\documentclass[10pt]{iopart}

\usepackage[utf8]{inputenc}

\usepackage{cite}
\usepackage{iopams}
\usepackage{siunitx}
\expandafter\let\csname equation*\endcsname\relax 
\expandafter\let\csname endequation*\endcsname\relax 

\usepackage{amsmath}
\usepackage{amssymb}
\usepackage{bm}

\usepackage{graphicx}
\usepackage{epsfig}
\usepackage{bm}
\usepackage{subcaption}

\usepackage{longtable}

\usepackage{url}
\usepackage{hyperref}
\usepackage{dcolumn}
\newcolumntype{d}{D{.}{.}{-1}}
\newcolumntype{f}[1]{D{.}{.}{#1}}

\newcommand{\eg}{{\textit{e.g., }}}
\newcommand{\ie}{{\textit{i.e., }}}

\begin{document}

\topical{QED tests with highly-charged ions}

\author{P Indelicato}

\address{Laboratoire Kastler Brossel, Sorbonne Universit\'{e},
CNRS, ENS-PSL Research University, Coll\`{e}ge de France,  Case\ 74;\ 4, place Jussieu, 75005 Paris, France}

\begin{abstract}
The current status of bound state quantum electrodynamics calculations of transition energies for few-electron ions is reviewed. Evaluation of one and two body QED correction is presented, as well as methods to evaluate many-body effects that cannot be
evaluated with present-day QED calculations. Experimental methods, their evolution over time, as well as progress in accuracy are presented. 
A detailed, quantitative, comparison between theory and experiment is presented for transition energies in few-electron ions. In particular the impact of the nuclear size correction on the quality of QED tests as a function of the atomic number is discussed.
The cases of hyperfine transition energies and of bound-electron Landé $g$-factor are also considered. 
\end{abstract}


\ead{paul.indelicato@lkb.upmc.fr}
\maketitle

\section{Introduction}
\label{sec:intro}

Quantum electrodynamics (QED), the \emph{theory of photons and electrons} \cite{jar1976}, was the first quantum  field theory, and the model for the other interactions included in the Standard Model of particle physics.
QED was born with the discovery by Lamb \& Retherford \cite{lar1947} that the degeneracy between the $2s_{1/2}$ and $2p_{1/2}$ states in hydrogen, as predicted by the Dirac equation, was not observed experimentally.
This fact, together with the discovery of the anomalous magnetic moment of the electron \cite{kaf1948}, led Bethe \cite{bet1947},   Feynman \cite{fey1949,fey1949a},  Schwinger \cite{sch1948,sch1948a,sch1949,sch1949a} and Tomonaga \cite{tom1946,ktt1947,ktt1947a} to propose QED as the physical description of the electromagnetic interaction, compatible with special relativity and quantum physics, which was followed by work from Dyson on bound states and convergence of the theory \cite{dys1949,dys1949a,dys1952}.

With the advent of lasers, atomic hydrogen spectroscopy has continually progressed, contributing to tests of QED and to regular improvements in the accuracy of the determination of the Rydberg constant. The increase in accuracy of the measurements has been constant and the number of lines measured rather large. Nowadays, the $1S-2S$ transition in hydrogen is known to \SI{10}{\hertz} \cite{pmab2011}, thanks to the use of frequency combs. The $1S-3S$ transition has been also measured with good accuracy \cite{fgtb2018}, as well as the $2S-4P$ \cite{bmmp2017}. 
Earlier measurements of  $2S-nl$ transitions have been summarized in \cite{bsaj2000}. Yet, progress in the accuracy of the determination of the Rydberg constant finally stopped following progress in the accuracy of hydrogen spectroscopy at the end of the 1990's.
The reason was that, assuming that all needed theoretical contributions are known, there are two unknowns in the theory-experiment comparison: the Rydberg constant and the proton charge radius. At the end of the 1990's, the uncertainty in the proton radius, which was then measured by electron scattering, became a limiting factor and it was decided to perform an independent, high-accuracy measurement using muonic hydrogen. A proposal was submitted and accepted at the Paul Scherrer Institute \cite{hpst1998}. This led to a very accurate measurement of the proton charge radius \cite{pana2010,ansa2013}, with a very large discrepancy of $7\sigma$ with respect to the value deduced from hydrogen spectroscopy in the 2010 CODATA evaluation  \cite{mtn2012,mnt2016}. This has become known as the \emph{proton size puzzle}, and is still not resolved today. The problem is the same for the deuterium charge radius, while the H-D isotopic shift is in good agreement \cite{pnfa2016}. There is also a discrepancy of $5\sigma$ for the muonic deuterium $2S$ hyperfine structure as pointed out in \cite{kpy2018}.
An alternative attempt to resolve this problem has also been pursued at NIST \cite{tbg2012,bgt2013}. It consists in measuring the Rydberg constant by doing the spectroscopy of circular, high-$n$ Rydberg states in one-electron medium-$Z$ ions \cite{jmtw2008,jmt2010}, which are insensitive to both QED and finite nuclear size corrections.

Helium has also been studied in great details. There are accurate measurements of both the $1s 2p\to 1s^2$ transitions \cite{euvh1993,euvh1996,euvh1997,rbsh2011} , and of the fine structure \cite{sdz1994,sdv1995,dnbj1997,mbcg1999,sah1998,sgh2000,cgnh2004,bgdw2009,ccgn2012,lpsw2013,lphf2016,zscj2017}.
Here again there are some discrepancies between measurements, \eg the isotopic shift between $^3$He and $^4$He \cite{zscj2017a}. The progress in accuracy, and the concomitant development of the theory of the helium fine structure, aims at providing an independent way of measuring the fine structure constant $\alpha$, and to compare the nuclear size of He isotopes through the isotopic shift. 

In parallel to this quest for a more accurate understanding of the simplest atoms, progress in experimental techniques has opened other ways of testing quantum electrodynamics in bound state systems (BSQED). In the non-relativistic approximation, the binding energy of an electron in an orbit of principal quantum number $n$ is given by
\begin{equation}
E_{\mathrm{NR}}(n)=-\frac {(Z \alpha)^2}{n^2} m_ec^2,
\end{equation}
where  $\alpha$ is the fine structure constant, $Z$ the atomic number, $m_e$ the electron mass and $c$ the speed of light The lowest-order (one loop) QED corrections behave as
\begin{equation}
E^{(1)}_{\mathrm{QED}}(n,\ell,j)=\frac{\alpha}{\pi}\frac {(Z \alpha)^4}{n^3}F^{(1)}_{(n,\ell,j)}\left(Z \alpha\right)  m_ec^2,
\label{eq:qed1st}
\end{equation}
where $F^{(1)}_{(n,\ell,j)}$ is a slowly varying function of $Z \alpha$, $\ell$ the angular momentum, and $j$ the total angular momentum. Going to high-$Z$ allows to have a ratio of QED contributions to transition energy that grows like $Z^2$. 

The progress in the spectroscopy of few-electron ions came though the use of complementary devices. The development of heavy-ion accelerators for nuclear physics led to the birth of a technique called  \emph{beam-foil spectroscopy} \cite{bas1968}. The original aim was to measure oscillator strengths for the determination of elemental abundances in astrophysics \cite{bas1985}. In this method, a beam of low-charged ions is passed through a thin foil (usually carbon). The electrons with orbital speeds slower or equal to the beam speed are ionized. A second, usually thiner, foil is used to recapture an electron in an excited state, and the de-excitation is observed with a suitable spectrometer. This method also allowed to measure lifetimes of metastable levels \cite{bic1968}. 

Beam-foil spectroscopy, with the use of more and more powerful accelerators like ALICE in Orsay, the Super-HILAC and BEVALAC at LBNL in Berkeley, the UNILAC and SIS at GSI, Darmstadt, the GANIL in Caen, and ATLAS at Argonne National Laboratory, has allowed to observe and measure transition energies and lifetimes in hydrogenlike, heliumlike and lithiumlike ions up to bismuth \cite{smgm1992} and  uranium \cite{mag1986,bciz1990,sbbc1991}.

The development of high-power lasers and tokamaks also enabled the spectroscopy of highly-charged ions. For example, the measurements of transition energies in hydrogenlike and heliumlike ions with $11\leq Z \leq 23$ were performed using laser-generated plasmas \cite{abzp1974}.
Vacuum spark devices were also used, providing transition energies in heliumlike ions with $16\leq Z \leq 39$ \cite{aamp1988}.
The TFR tokamak in Fontenay aux Roses near Paris \cite{tbbf1985,tcdl1985}, and the PLT and TFTR tokamaks at Princeton \cite{bhse1979,bfsb1982,bgch1984,bhzg1985,hbhg1987,dah1987,bbgh1989,bhbc1993}, ALCATOR C \cite{kkrs1984} and JET in England \cite{dhrs1989,dmj1989,htdj1989} were used to perform measurements of transition energies in few-electron ions, including transitions in core-excited ions, often called \emph{satellite} transitions. 

Astrophysical observations have also provided accurate relative measurements, for example in solar flares, for one, two \cite{saf1985} and three electron systems \cite{wap1976}.

After this initial phase of discovery measurements, new techniques emerged, designed to provide better accuracy and cleaner spectra at high-$Z$. These approaches were born from the realization, rather early, that the methods of non-relativistic QED (NRQED), which provides an expansion in $Z\alpha$ and $\log \left(Z\alpha\right)$ of the function $F\left(Z\alpha\right)$ defined in equation~\eqref{eq:qed1st}, do not converge at high-$Z$ \cite{moh1974,moh1974a}. It is also necessary to test the next order of QED corrections (the two-loop corrections), which are of order $\frac{\alpha}{\pi}\approx 2.3\times 10^{-3}$ times smaller than the one-loop corrections. Finally, more accuracy is needed to be able to disentangle QED corrections from nuclear effects, like nuclear polarization and the finite nuclear size correction, which becomes larger than QED effects for inner shells at high-$Z$ \cite{pmgs1989,pmgs1991,pas1995,pas1996,nlps1996,vap2014}.

To that aim, a variety of techniques have been used to reduce the uncertainty due to Doppler effect. For example, the accel-decel method where the ions are stripped after acceleration in a part of the accelerator and decelerated afterwards towards a gas cell where electron capture is performed. Combined with measurements at several decelerated beam energies, this method allowed to reduce the doppler effect uncertainty \cite{dsj1985,bifl1991}. 
There were also experiments comparing different lines of the same hydrogenlike ion, following the method developed for hydrogen \cite{hlww1975}. The iron \cite{smlr1987} and germanium  \cite{lcsd1987}  Lyman $\alpha$ $2P_{3/2}\to 1S_{1/2}$ and Balmer $4D_{5/2}\to2P_{3/2}$ energies have been compared, in fourth and first order respectively, using two curved crystal spectrometers symmetric with respect to the beam axis and 2D detectors. The Balmer line energy is very weakly dependent on QED and nuclear size, and is thus used as an internal reference, affected by the same Doppler effect as the calibrated line. In another method,  the impact of a fast uranium  beam on an argon gas target produced very slow recoil ions that could be studied by x-ray spectroscopy \cite{dbf1984,bdfl1985}. 

Another method developed for spectroscopy of highly charged ions is \emph{resonant coherent excitation} in thin crystals. This method uses the fact that the electric field due to regularly spaced atoms in a crystal is seen in the reference frame of a fast ion beam as a coherent source of light, which can excite transitions (see   \cite{gem1974} for an early review). The first  observation in light hydrogenlike ($5\leq Z \leq 9$ and heliumlike (F$^{7+}$) ions was performed  by Datz \etal \cite{dmck1978} and the method was then used in heavier elements thanks to high-energy accelerators \cite{aiky1999,ntik2013}. 

The main progress came from the use of storage rings with electron cooling, like CRYRING in Stockholm, TSR in Heidelberg or ESR in Darmstadt. The cooling allows to reduce the energy and momentum dispersion of the ions in the beam, reducing the Doppler broadening of the x-rays emitted by the ions. The cooling also forces the ions' speed to be the same as that of the electrons, thus enabling a better measurement of the Doppler shift. It is also possible to decelerate the ions to do measurements at different velocities and obtain a more accurate Doppler correction. The most accurate measurement of the hydrogenlike uranium $1s$ Lamb shift comes from a measurement at the ESR \cite{gsbb2005}. It also became possible to use radiative electron capture \cite{smbb1993,bfli1993,blbf1994,lbfb1994} or dielectronic recombination \cite{zslg1997,bkms2003,llos2008} to excite the ions.  This method has allowed to study, for example, $2p\to 2s$ transitions in lithiumlike gold, lead and uranium \cite{bkms2003}.

Another important progress came from the use of low-energy electron beam ion traps (EBIT). In these sources, a monoenergetic electron beam traveling along the axis of a Helmholtz coil ionizes injected atoms \cite{dla1969}. The space charge of the beam traps the ions radially and a set of electrodes traps them longitudinally. The atoms can be ionized in specific charge states by varying the electron beam energy. After the successful demonstration of dielectronic recombination in an electron beam ion source (an earlier version of this device with a much longer trapping area) \cite{bcal1984}, the EBIT became an instrument of choice for HCI spectroscopy. Up to now, however, they have not been able to perform spectroscopy of the heaviest hydrogenlike and heliumlike ions. The Super EBIT in Livermore has however allowed to do accurate measurements of the $1s^2 2p\; ^2P_{1/2}$ transition in lithiumlike uranium \cite{bctt2005}. 

In the last few years, EBITs have also been used to perform reference-free spectroscopy at the Max Planck institute of Nuclear Physics in Heidelberg. The use of a flat crystal spectrometer with a precision angular encoder and a laser system able to position very precisely the x-ray detector allows measurements at two angles, symmetric with respect to the optical axis. In this way it is possible to obtain directly the Bragg angle from the difference in angle between the two spectra \cite{kbbc2012}. This allowed reference free measurements of hydrogenlike and heliumlike $2p\to 1s$ transitions in sulfur, chlorine, argon and iron with few ppm accuracy \cite{bbkc2007,kmmu2014}. The use of EBITs, in combination with monocromatized x-rays from a synchrotron radiation facility (PETRA III) allowed to make very accurate measurements in heliumlike krypton relative to x-ray K-edges \cite{esbr2015}. A very promising new method, combining EBIT and x-ray free electron lasers (XFEL), has been demonstrated, allowing to measure $1s^2 2p\; ^2P_{1/2}\to 1s^2 2s\; ^2S_{1/2}$ transitions in lithiumlike iron and copper \cite{elbm2007,elsb2010}. This technique will probably become very important, once calibration and monochromator issues are resolved and  that reference-free measurements are made possible.   

At the same time, the use of electron-cyclotron ion sources (ECRIS), which are able to trap many more ions than EBITs, have been used with a vacuum double crystal spectrometer \cite{assg2014} to provide reference-free accurate measurements of $2p\to 1s$  transitions in heliumlike \cite{asgl2012,mssa2018} and core-excited lithiumlike \cite{sags2013} and berylliumlike \cite{mssa2018} argon ions. 

Other techniques have been proposed, which, although they cannot give the level of accuracy required in modern experiments, could have important applications. For example, the study of x-rays emitted in collisions between an ion with a K-hole  (bare or hydrogenlike) as a function of the impact parameter allows a measurement of atomic transition energies in the compound nucleus, when extrapolating to zero impact parameter. This method has been tested in Cl$^{17+}$ on Ar collisions, providing Br x-rays\cite{sstj1985,smjj1988}. Such experiments could allow for the spectroscopy of transitions between inner shells in superheavy elements around the critical $Z\approx 173$ when the $1s$ shell dives into the negative energy continuum \cite{blam1980}. 
Astrophysics applications are also very important, for example the analysis of the high-resolution X-ray observations of the Perseus cluster by the HITOMI satellite \cite{aaaa2018}.

Several reviews have focused on photon emission \cite{fis2005}, or on applications of highly-charged ions for precision physics and atomic clocks \cite{ksls2018}. Here I will present a detailed analysis of the agreement between theory and experiment for few-electron ion transition energies, and of the hyperfine energies and Landé $g$-factor as means to test QED. This review is structured as follows. In Sec. \ref{sec:ener}, I will present the principles of QED and relativistic many-body calculations of transition energies in few-electron heavy ions (Section \ref{sec:theo}). I will then discuss the experimental results and comparison with experiment in section \ref{sec:experiment} for one-, two- and three-electron systems. Then I will discuss other operators besides energy in Sec. \ref{sec:other-operators}. In Sec. \ref{sec:hfs} I will describe major results for the hyperfine structure measurements and theory in few electron systems and  in Sec. \ref{sec:lande} the Landé $g$-factors. Section \ref{sec:conc} is the conclusion.


\section{Transition energies}
\label{sec:ener}

\subsection{Theory for few-electron ions}
\label{sec:theo}

The relevant theory for calculating accurate transition energies in atoms is bound-state quantum electrodynamics (BSQED). 
This theory is based on the Furry bound picture \cite{fur1951}.  The calculation starts from the unperturbed Coulomb Dirac Hamiltonian $H_{D}$, which contains the field of 
the nucleus $V_{N}$,
\begin{equation}
 	H_{D}=c\bm \alpha \cdot \bm p+\beta mc^{2}+V_{N}\left(r\right)
 	\label{eq:dirac}
 \end{equation} 

This has the consequence that  this Coulomb field is included to all orders in the evaluation of all relevant quantities.
The electron-electron interaction is treated as a perturbation
 \begin{equation}
V_{\epsilon,g}=gH_{I}e^{-\epsilon|t|},
  \label{eq:vpot}
 \end{equation}
 where
 \begin{equation}
H_{I}=j^{\mu}A_{\mu}-\delta M(x).
\label{eq:hpot}
\end{equation}
Since the electromagnetic field can act at an infinite distance, the introduction of the $e^{-\epsilon|t|}$ parameter is needed  to turn off adiabatically the interaction at $t=\pm \infty$ to recover the unperturbed states. 
Here,  
\begin{equation}
j^{\mu}=-\frac{e}{2}\left[\bar \psi(x) \gamma^{\mu} ,\psi(x)\right]
\label{eq:4-current}
\end{equation}
is  the 4-current,  
\begin{equation}
\delta M(x)=\frac{\delta m}{2}\left[\bar \psi(x)  ,\psi(x)\right]
\end{equation}
is  the mass counter-term needed for a proper definition of perturbation theory, 
$g$ is a perturbation parameter, $e$ is the electron charge,
$A_{\mu}$ is the photon 4-vector operator, and $\psi(x)$  is the electron Dirac field operator. We also define $\bar 
\psi=\psi^{\dag}\gamma^{0}$. 
  
The evaluation of the perturbation contributions is done through the use of the adiabatic evolution operator
\begin{equation}
 U_{\epsilon,g}\left(t_{1},t_{2}\right)=T e^{-i \int^{t_{2}}_{t_{1}}dt 
 V_{\epsilon,g}(t)} \, ,
 \label{eq:evolop}
\end{equation}
where $T$ is the time ordering operator.  This operator enables one to define the adiabatic $S$-matrix  $S_{\epsilon,g}=\lim_{t \rightarrow \infty 
}U_{\epsilon,g}(-t,t)$. An unperturbed state with $p$ electrons and no real photons is denoted $\left|N_{p};0\right>=\left| 
n_{1},\ldots , n_{p};0\right>$, with an unperturbed energy given by  $E^{0}_{N_{p}}=\sum_{k=1}^{p}E_{n_{k}}$. Here $n_{k}$ represents the state of electron $k$.
The energy of this state including the effect of the perturbation from \eqref{eq:hpot} is given by the Gell-Mann and 
Low theorem \cite{gal1951,faw1971}, as symmetrized by Sucher \cite{suc1957}: 
\begin{equation}
 \Delta E_{N_{p}}=
  \lim_{
   \stackrel{\epsilon \rightarrow 0}{ g \rightarrow 1}
       }
 \frac{i\epsilon g}{2} \frac {\partial}{\partial g}\log \left< N_{p};0 \right|  
 S_{\epsilon,g}\left| N_{p};0\right>.
   \label{eq:gmlt}
\end{equation}

The $S$-matrix is then expanded in powers of $g$ as presented in \eg  \cite{moh1989,moh1996}). One obtains
\begin{eqnarray}
\left. g \frac {\partial}{\partial g}\log \left<S_{\epsilon,g} 
\right>_C\right|_{g=1}  & = & \frac{
 \left<S_{\epsilon,1}^{(1)} \right>_C+2\left<S_{\epsilon,1}^{(2)} 
 \right>_C+3\left<S_{\epsilon,1}^{(3)} \right>_C+\cdots}
 {1+\left<S_{\epsilon,1}^{(1)} \right>_C+\left<S_{\epsilon,1}^{(2)} 
 \right>_C+\left<S_{\epsilon,1}^{(3)} \right>_C+\cdots}
 \nonumber  \\
  & = & \left<S_{\epsilon,1}^{(1)} \right>_C+2\left<S_{\epsilon,1}^{(2)} 
 \right>_C-\left<S_{\epsilon,1}^{(1)} \right>_C^{2}
 \nonumber  \\
  & + & 3\left<S_{\epsilon,1}^{(3)} \right>_C
  -3\left<S_{\epsilon,1}^{(1)} \right>_C\left<S_{\epsilon,1}^{(2)} \right>_C
  +\left<S_{\epsilon,1}^{(1)} \right>_C^{3}
 \nonumber  \\
  & + & 4\left<S_{\epsilon,1}^{(4)} \right>_C
  -4\left<S_{\epsilon,1}^{(1)} \right>_C\left<S_{\epsilon,1}^{(3)} \right>_C
  -2\left<S_{\epsilon,1}^{(2)} \right>_C^{2}
    \nonumber  \\
  & + & 4\left<S_{\epsilon,1}^{(1)} \right>_C^{2}\left<S_{\epsilon,1}^{(2)} \right>_C
  -\left<S_{\epsilon,1}^{(1)} \right>_C^{4} \; ,
 \label{eq:smatexp}
\end{eqnarray}
where 
\begin{eqnarray}
 \left<S_{\epsilon,1}^{(j)} \right>_C & = & \left< N_{p};0 \right|  
 S_{\epsilon,1}^{(j)}\left| N_{p};0\right>_C
    \nonumber  \\
 \left<S_{\epsilon,g} \right>_C & = & \left< N_{p};0 \right|  
 S_{\epsilon,g}\left| N_{p};0\right>_C \; .
 \label{eq:spert}
\end{eqnarray}
At each order in \eqref{eq:smatexp}, the  terms of order $1/\epsilon^n$ for $n>1$ cancel out and the energy expression in \eqref{eq:gmlt} has thus a finite limit when $\epsilon\to 0$.

From the definition of the $S$-matrix and of the evolution operator 
\eqref{eq:evolop} one obtains thus
\begin{equation}
 S_{\epsilon,g}^{(j)}=\frac{\left(-ig\right)^{j}}{j!}\int d^{4}x_{j}\ldots \int d^{4}x_{1} 
 e^{-\epsilon \left|t_{j}\right|}\ldots
 e^{-\epsilon \left|t_{1}\right|}T\left[H_{I}\left(x_{j}\right) \ldots 
 H_{I}\left(x_{1}\right)\right] \; .
 \label{eq:sjdef}
\end{equation}

At each order, the term  in \eqref{eq:sjdef} can be expressed as an integral of  products of Dirac Green's functions, the photon Green's function, and eigenfunctions of the Coulomb Dirac equation.
The Dirac Green's functions are given by
\begin{equation}
G\left(\bm x_2,\bm x_1,z\right)=\sum_n \frac{\phi_n\left(x_2\right)\phi_n^{\dag}\left(x_1\right)}{E_n-z},
\label{eq:green-sum-poles}
\end{equation}
where $\phi_n\left(x_2\right)$ are the eigenfunctions of the Dirac equation, and $E_n$ the corresponding eigenvalue, including the two continua with $E_n\leq-mc^2$ and $E_n\ge mc^2$.
The photon Green's function is given by 
\begin{eqnarray}
H\left(\bm x_{2}-\bm x_{1},q_0\right)&=&-\frac{e^{-bx_{21}}}{4 \pi x_{21}} \nonumber \\
x_{21}=\left| \bm x_{2}-\bm x_{1}\right|;&&   b=-i\left( q_0^2+i\delta\right)^{\frac{1}{2}}, \Re(b)>0.
\label{eq:photgreen}
\end{eqnarray}

In the rest of this section, I will give some examples of the expression of the energy shift for several cases and present the results.
For a complete derivation of these results, and more details, the reader is referred to, \eg \cite{moh1989,moh1989a,lkd1993,mps1998,bei2000,iam2017,art2017,lai2017}.

\subsubsection{Bound state QED for one-electron systems}
\label{sec:hlike-bsqed}

\begin{figure*}[htbp]
\centering

        \includegraphics[width=0.3\textwidth]{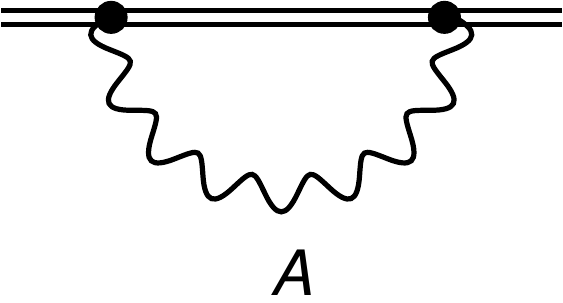}
        \includegraphics[width=0.12\textwidth]{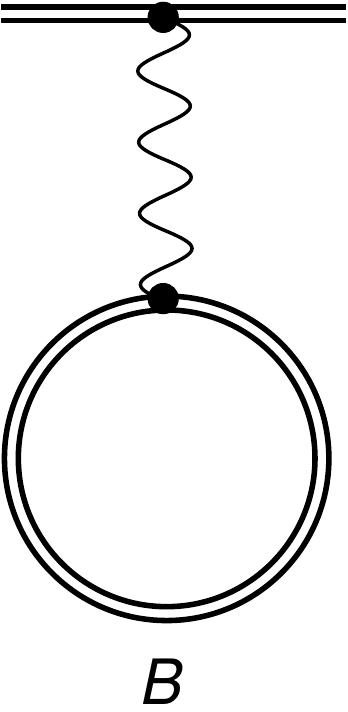}
\caption{One-loop QED corrections. A: Self-energy. B: Vacuum polarization. Dots represent vertices. The double lines with a free end represent a bound electron wave function, and the double lines between two vertices represent a bound-electron Dirac propagator.
\label{fig:one-loop-qed}
}
\end{figure*}

\begin{figure*}[htbp]
\centering

        \includegraphics[width=\textwidth]{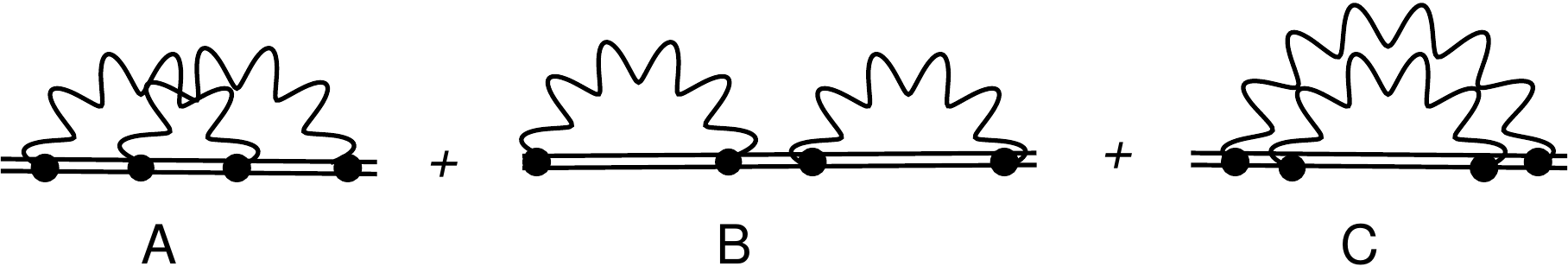}
        \includegraphics[width=0.6\textwidth]{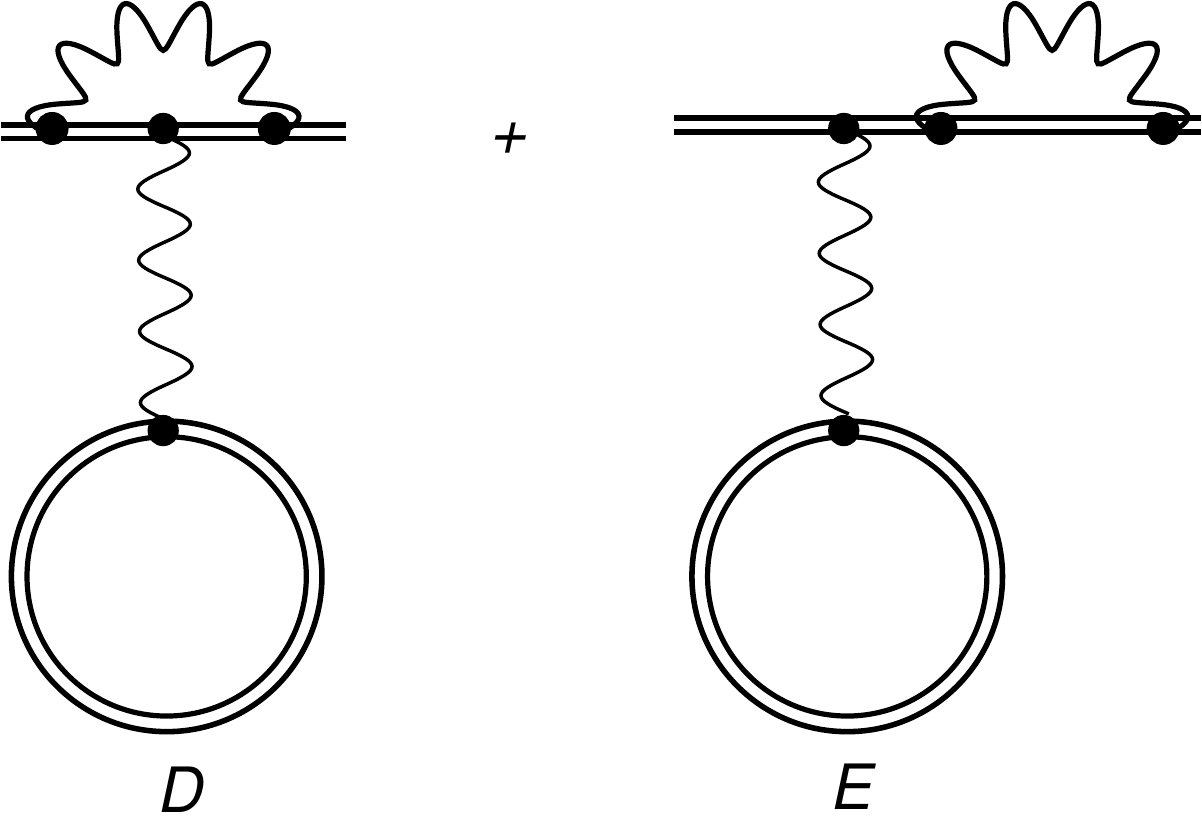}
        \includegraphics[width=\textwidth]{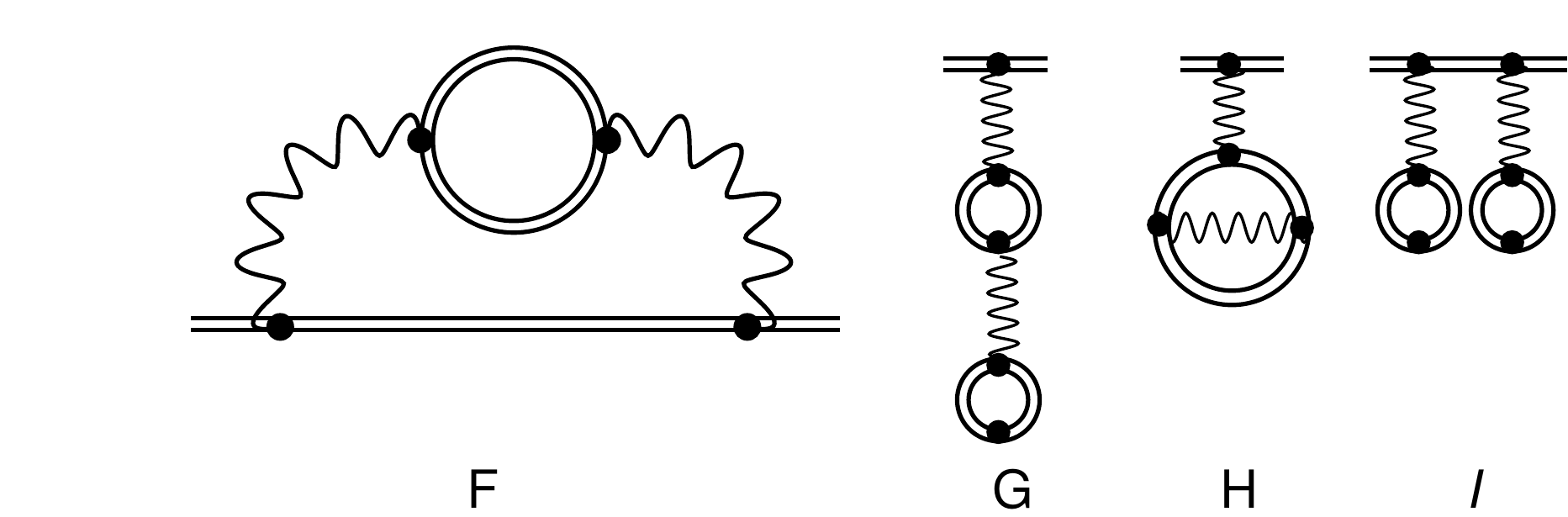}
\caption{Two-loop QED corrections. A, B, C: two-loop self-energy. D, E: SE-VP corrections. F: S(VP)E correction. G, H: Källén and Sabry correction. I: loop-after-loop vacuum polarization.
\label{fig:two-loop-qed}
}
\end{figure*}

The energy of a given level in a one electron-system is given by the Dirac equation, with finite size nuclear corrections, QED corrections and recoil corrections.
The solution of the Dirac equation for a point nucleus can be written as
\begin{eqnarray}
         E^{\mathrm{D}}_{n\kappa}(Z)& = & \frac{m_e c^{2}}{\sqrt{1+\frac{(Z\alpha)^{2}}
         {(n-|\kappa|+\sqrt{\kappa^{2}-(Z\alpha)^{2}})^{2}}}},
        \label{eq:direigen}  \\
         & = &  \frac{m_e c^{2}}{\sqrt{1+\frac{(Z\alpha)^{2}}
         {(n-j-\frac{1}{2}+\sqrt{\left(j+\frac{1}{2}\right)^{2}-(Z\alpha)^{2}})^{2}}}},
        \label{eq:direigenj}
\end{eqnarray}
where $|\kappa|=l$ when $\kappa>0$ and $|\kappa|=l+1$ when $\kappa<0$. The total angular momentum quantum number is given by $j=|\kappa|-\frac{1}{2}$.
The level energy is thus given by 
\begin{eqnarray}
E_{n\kappa}(Z,A)& =& E^{\mathrm{D}}_{n,\kappa} (Z)+  E^{(1)}_{\mathrm{QED}}(n,\kappa,Z) + E^{(2)}_{\mathrm{QED}}(n,\kappa,Z) + E^{\mathrm{Nuc.}}_{n,\kappa} \left(Z,A\right) \nonumber \\ 
&& +  E^{\mathrm{Rec.}}_{n,\kappa} \left(Z,M_{A}\right) \\
&=&E^{\mathrm{D}}_{n,\kappa} (Z)+ \frac{\alpha}{\pi}\frac {(Z \alpha)^4}{n^3}F^{(1)}_{n,\kappa}\left(Z \alpha\right)m_e c^{2} \nonumber \\
&&+ \left(\frac{\alpha}{\pi}\right)^2\frac {(Z \alpha)^4}{n^3}F^{(2)}_{n,\kappa}\left(Z \alpha\right)m_e c^{2}\nonumber \\
&&+ E^{\mathrm{Nuc}}_{n,\kappa} \left(Z,M_{A}\right) +  E^{\mathrm{Rec.}}_{n,\kappa} \left(Z,M_{A}\right),
\label{eq:hyd_like_level_ener}
\end{eqnarray}
where $ E^{(1)}_{\mathrm{QED}}(n,\kappa,Z)$ represents the first order QED corrections, self-energy and vacuum polarization presented in figure \ref{fig:one-loop-qed}, while $ E^{(2)}_{\mathrm{QED}}(n,\kappa,Z)$ is the sum of the two-loop corrections
from figure \ref{fig:two-loop-qed}. The   $E^{\mathrm{Nuc}}_{n,\kappa} \left(Z,M_{A}\right)$  correction represents the finite nuclear correction and possibly nuclear polarization correction. The other correction, $E^{\mathrm{Rec.}}_{n,\kappa} \left(Z,M_{A}\right)$ represents the recoil effect. The nuclear polarization will be discussed in Sec. \ref{sec:nuclear} and the recoil correction in Sec. \ref{sec:nuclear-recoil}. For low $Z$, the two QED corrections in \eqref{eq:hyd_like_level_ener} are often represented by a low-$Z$ approximation, reviewed in detail in  \cite{egs2001,egs2007} and in the 2010 CODATA paper \cite{mtn2012}.  More recently two- and three-photon nuclear-polarization corrections have been evaluated in detail for hydrogen and deuterium \cite{ppy2018}. In this work, the Friar moment correction and the corresponding inelastic contribution to the two-photon exchange, which partially cancel each other, have been studied in detail.  Very recently Yerokhin \etal \cite{ypp2019} have provided detailed calculations of the transition energies in one-electron atoms for elements with $1\leq Z \leq 5$, with all available corrections known to date, including the results from \cite{ppy2018}. There are a number of recent works, in the framework of the low $Z\alpha$ expansion and NRQED, which provide one and two-loop QED corrections, useful for low-$Z$ atoms and excited states \cite{jcp2005}. The evaluation of the one- and two-loop Bethe logarithm provides the lowest order contribution to the $F^{(1)}_{n,\kappa}\left(Z \alpha\right) $ and $F^{(2)}_{n,\kappa}\left(Z \alpha\right) $  functions. The one-electron, one-loop Bethe logarithm has been evaluated with very high accuracy for $n\leq 20$ by Drake \& Swainson \cite{das1990}. This was extended to Rydberg states for $n\leq 200$ in \cite{jam2005}. The two-loop self-energy Bethe logarithm has been evaluated \cite{paj2003,jen2006}, as well as the one-loop self-energy, two-electron Bethe logarithm \cite{gad1983,das1990,kor2004}.

The basis for a direct evaluation to all orders in $Z\alpha$ of the diagrams has been set by Wichmann \& Kroll\cite{wak1956} for vacuum polarization and in \cite{bls1959,bam1959} for the self-energy. The first high-precision evaluation of the self-energy was performed by Mohr \cite{moh1974,moh1974a} for the $1s$ level at medium and high-$Z$. It was extended to super-heavy elements by Cheng \& Johnson \cite{caj1976} and to different $n$, $\ell$, $j$ levels in \cite{moh1982,moh1992,mak1992,iam1998a,iam1998,bim2001}. All-order calculations at low-$Z$ are very difficult as the $E^{(1)}_{\mathrm{QED}}(n,\kappa,Z)$ correction from QED is formally of order $\frac{\alpha}{\pi}mc^2$, and thus terms of order \num{1}, $Z\alpha$, $(Z\alpha)^2$ and $(Z\alpha)^3$ have to be cancelled, requiring very large accuracy. The calculation was thus performed for $1\leq Z\leq 5$ in \cite{jms1999,jms2001,jam2004,jam2005} using highly-efficient resummation techniques \cite{jmsw1999} for $S$ and $P$ states with principal quantum numbers up to \num{4}. It should be noted as shown by Mohr \cite{moh1974} that the expansion in $Z\alpha$ is only asymptotic, and does not converge even for relatively small values of $Z$.

\begin{figure*}[htbp]
\centering
        \includegraphics[width=0.5\textwidth]{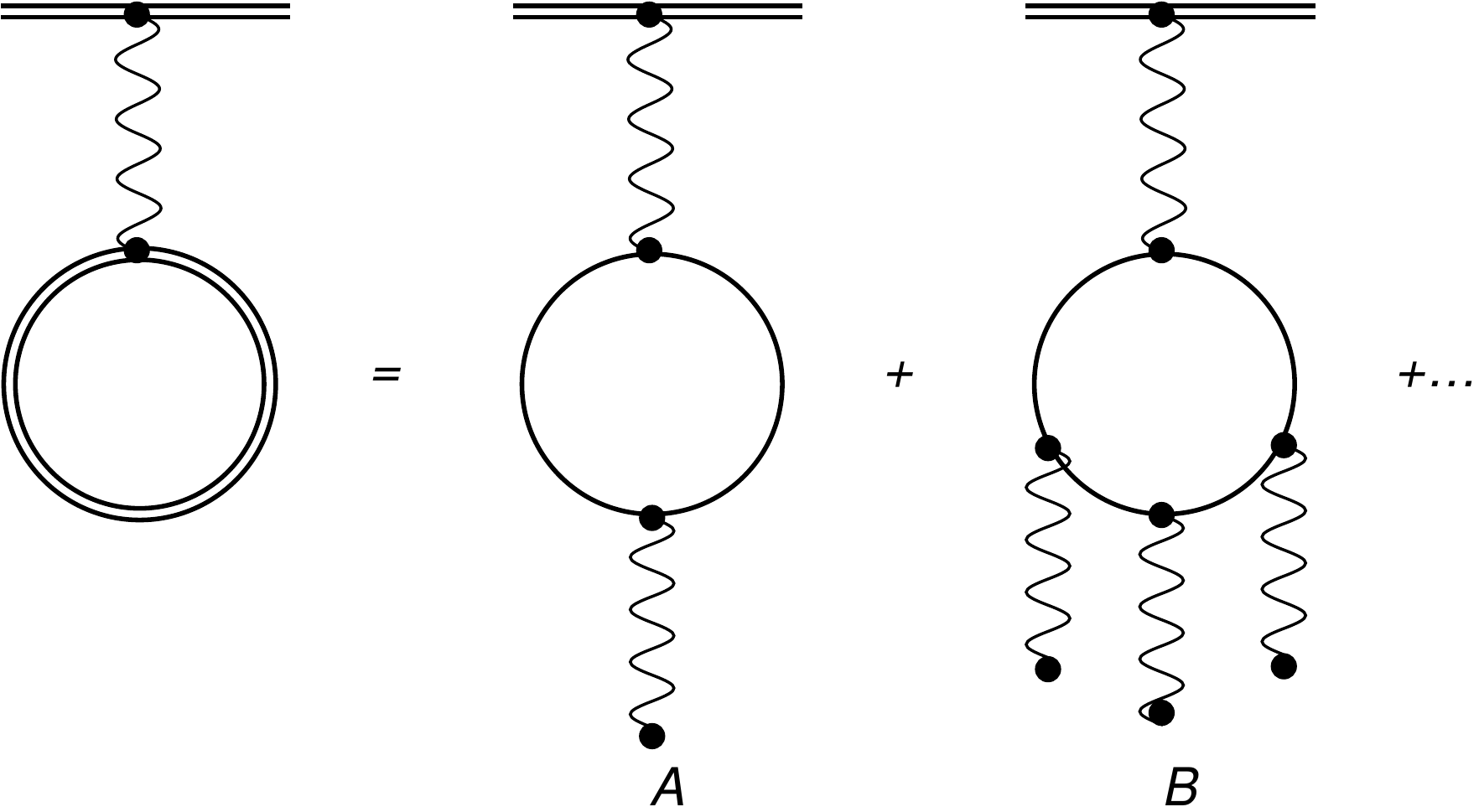}
\caption{Expansion of the vacuum polarization in the number of loop interactions with the nucleus.  A represents the Uehling potential and B and higher orders the Wichmann and Kroll correction.
\label{fig:vp-loop-exp}
}
\end{figure*}

The vacuum polarization is usually expanded in powers of $Z\alpha$, corresponding to the number of interactions with the nucleus in the electron-positron loop as shown in figure \ref{fig:vp-loop-exp}. Since there is only one electron vertex for the bound electron, each order in the vacuum polarization expansion can be represented by a potential. The only non-zero contributions have an odd number of interactions with the nucleus. The first contribution of order $Z\alpha$ is evaluated using the Uehling potential \cite{ueh1935} and the second one, of order $(Z\alpha)^3$  and higher by the Wichmann and Kroll potential \cite{wak1956}. Accurate expressions to evaluate the Uehling potential can be found in \cite{far1976,blom1972}.  Calculations of the vacuum polarization for the contributions to all orders in $Z\alpha$ can be found in \cite{sam1988,mnf1989,plss1993} for the $n=1$ and $n=2$ levels. The sum of Wichmann and Kroll contributions of order $(Z\alpha)^n$, with $n=3, \, 5, \, 7$, has also been evaluated in Beier \etal \cite{bpgs1997}.

It should be noted that besides the diagram with the electron-positron loop in figure \ref{fig:one-loop-qed} B, there are equivalent ones with  a muon-antimuon loop. This effect is completely negligible for electronic atoms.

The situation for the second-order QED diagrams from figure \ref{fig:two-loop-qed} is not yet as complete as what has been done for the first order diagrams. The lowest order in the number of interactions between the electron-position loops and the nucleus corresponding to the contributions from figure  \ref{fig:two-loop-qed} G and H is called the Källén and Sabry potential \cite{kas1955}. It can be evaluated using accurate numerical formulas from Fullerton \& Rinker \cite{far1976}. 
The loop-after-loop contribution (figure  \ref{fig:two-loop-qed} I) has been evaluated for finite-size nuclei by solving the Dirac equation numerically with and without the Uehling potential \cite{plss1993,pllp1996,ind2013}. It has been also calculated with standard QED methods \cite{mllp1995,bpgs1997}. The next contributions that have been evaluated are the crossed vacuum-polarization self-energy diagrams named SE-VP  in figure \ref{fig:two-loop-qed} D and E. They were first evaluated in \cite{plss1993,lpsk1993} by including the Uehling potential in the
evaluation of the self-energy.  More accurate values can be found in \cite{pllp1996}. The next term, the S(VP)E  diagram  (figure \ref{fig:two-loop-qed}  F) was evaluated in   \cite{pllp1996,mas1996}. More recently very accurate evaluations of the SEVP, VPVP, Källén and Sabry, and S(VP)E contributions for the $1s$, $2s$, $2p_{1/2}$ and $2p_{3/2}$ levels for $1\leq Z \leq 100$ have been performed by Yerokhin \etal \cite{yis2008}.  

The SESE correction has been the most difficult to evaluate. The set of three diagrams  in figure \ref{fig:two-loop-qed} A, B and C must be evaluated together, and renormalization is difficult. A first step was performed by Mallampalli \& Sapirstein \cite{mas1998}, who calculated a number of contributions except the most difficult terms to renormalize. The full SESE contribution calculation for the $1s$ state and high-$Z$ elements was finally performed by Yerokhin \etal \cite{yis2003}.  Improvements in accuracy and extension to lower $Z$ and the $n=2$ levels can be found in several works \cite{yis2003a,yis2005,yer2009,yer2018}.

\begin{figure*}[htbp]
\centering
        \includegraphics[width=0.8\textwidth]{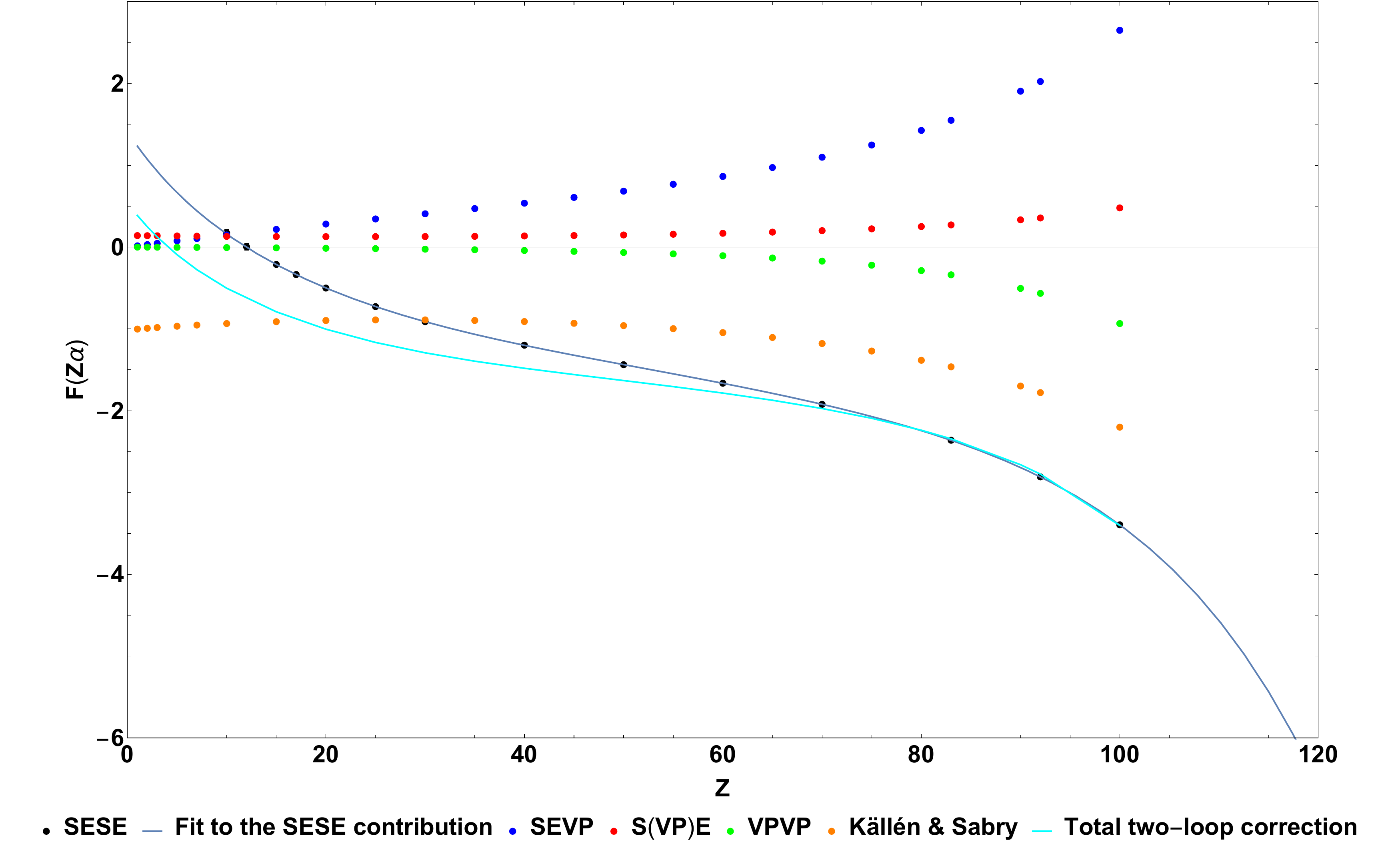}
\caption{Comparison of the most accurate calculations of $F\left(Z\alpha\right)$ of the SESE correction from  Yerokhin \etal \protect\cite{yis2003,yis2003a,yis2005},  and Yerokhin \protect\cite{yer2018} and  SEVP, S(VP)E, VPVP and the Källén and Sabry corrections from  Yerokhin \etal  \protect\cite{yis2008} for the $1s$ level. The fitted function is provided by  \protect\eqref{eq:sese1s-series} with the parameters given in the text.
\label{fig:1s-wto-loop-comp}
}
\end{figure*}

A comparison of the latest calculations for all the two-loop contributions from figure \ref{fig:two-loop-qed} are presented in figure \ref{fig:1s-wto-loop-comp}  for the $1s$ level. Up to now it has not been possible to evaluate directly the SESE contribution for low-$Z$. The low-$Z$ part has been evaluated using NRQED calculations which provide a series in $Z\alpha$ and $\log\left(\frac{1}{(Z\alpha)^2}\right)$:
\begin{eqnarray}
         F^{\mathrm{SESE}}_{1s}\left(Z \alpha\right) & = & B_{40}+B_{50} Z\alpha \nonumber \\
         && +\bigg[B_{60}+B_{61} \log\left(\frac{1}{(Z\alpha)^2}\right)+B_{62} \left\{\log\left(\frac{1}{(Z\alpha)^2}\right)\right\}^2 \nonumber \\
         &&             \qquad \qquad+B_{63} \left\{\log\left(\frac{1}{(Z\alpha)^2}\right)\right\}^3\bigg] (Z\alpha)^2 \nonumber \\
         && + \bigg[B_{70}+B_{71} \log\left(\frac{1}{(Z\alpha)^2}\right)+B_{72} \left\{\log\left(\frac{1}{(Z\alpha)^2}\right)\right\}^2\bigg] (Z\alpha)^3 \nonumber \\
        && + \bigg[B_{80}+B_{81} \log\left(\frac{1}{(Z\alpha)^2}\right)+B_{82} \left\{\log\left(\frac{1}{(Z\alpha)^2}\right)\right\}^2\bigg] (Z\alpha)^4 \nonumber \\
         && + B_{90}(Z\alpha)^5.
        \label{eq:sese1s-series}
\end{eqnarray}
In this expression, only $B_{40}$, $B_{50}$, $B_{61}$, $B_{62}$, $B_{63}$, $B_{71}$ and $B_{72}$ have been evaluated. The values are summarized in Yerokhin \& Shabaev \cite{yas2015}.
One has $B_{40}=1.409244$, $B_{50}=-24.26506$, $B_{61}=48.388913$, $B_{62}=\frac{16}{27}-\frac{16}{9}\log 2=-0.639669$, $B_{63}=-\frac{8}{27}$. Recently the $B_{72}$ coefficient value was obtained as $B_{72}=-\frac{2}{3}A_{50}=-6.19408$, where $A_{50}=\pi\left(\frac{139}{32}-2\log 2\right)$ is the equivalent coefficient for the expansion of the one-loop self-energy \cite{kai2018}. 

I performed a weighted fit of \eqref{eq:sese1s-series} with the most accurate  values from \cite{yis2003,yis2003a,yis2005,yer2009,yer2018}, using the unknown $B_{ij}$ coefficients as free parameters. This fit has a reduced $\chi^2=0.22$, which means that the theoretical error bars are somewhat pessimistic.
I get $B_{60}= -95.467$, $B_{70}= -193.497$, $B_{71}=27.0276$, $B_{80} = 1085.57$, $B_{81} = -228.220$, $B_{82} =-116.29$ and $B_{90} = -787.4255$. A previous fit with less accurate data provided  $B_{60}=-84(15)$ \cite{yer2009}. Extrapolation to $Z=1$ of the higher-order remainder as defined in equation (7) of \cite{yer2009} gives $G_{\mathrm{SESE}}^{\mathrm{h.o.}}(Z=1)=-98.8$ against \num{-86+-15}.

\subsubsection{Nuclear corrections}
\label{sec:nuclear}

The contribution from the nucleus in \eqref{eq:hyd_like_level_ener} is the sum of four effects
\begin{eqnarray}
E^{\mathrm{Nuc}}_{n,\kappa} \left(Z,M_{A}\right) & =& E^{\mathrm{Vol}}_{n,\kappa} \left(Z,RMS_{Z,A}\right) + E^{\mathrm{Shape}}_{n,\kappa} \left(Z,A\right) + E^{\mathrm{Def}}_{n,\kappa} \left(Z,A\right) \nonumber \\
  && + E^{\mathrm{Pol}}_{n,\kappa} \left(Z,A\right)
  \label{eq:nuc_corr}
\end{eqnarray}
The first contribution is the most obvious one and is due to the finite nuclear size. For electronic atoms, it depends mostly on the RMS radius of the nucleus. Most calculations use available compilations for these radii, which collect measurements performed with muonic atoms, electron scattering and atomic x-ray or laser spectroscopy \cite{aam2013,ang2004}. For the shape contribution, three models of charge distribution in the nucleus are widely used:  the uniform model and the 2 or 3-parameter Fermi model.  Other models have been used, like sum of gaussians (SOG) \cite{sic1974} or Fourier-Bessel \cite{vjv1987,fbhs1995} to better reproduce the nuclear charge density distribution. For the Fermi model, the charge distribution is given by 
\begin{equation}
\rho (r) = \frac{\rho_0}{1+e^{\frac{(r-c)}{a}}},
\label{eq:fermi-model}
\end{equation}
where $c$ is the half-density radius. The surface thickness parameter $t$ is given by $t=(4 \ln 3) a$. 
The different moments can be defined by
\begin{equation}
<r^n>=\int_0^{\infty} r^2 r^n\rho (r) dr \,.
\end{equation}
The RMS radius is obtained for $n=2$. 
I use as an example  hydrogenlike $^{208}$Pb, which is the heaviest element where the Fourier-Bessel and SOG coefficients are known from earlier compilations \cite{vjv1987,fbhs1995}. The results are presented in Table \ref{tab:nucl-dep}, together with the moments of the different charge distributions. It is clear that the Fourrier-Bessel and Fermi model with the same RMS radius give very close binding energy values. The SOG model, even though it gives an identical RMS radius value, leads to a larger shift. The table also shows the large differences between the higher moments of the distribution.

\begin{table}
\begin{center}
\caption{Comparison between different nuclear models for the $1s$ level binding energy in $^{208}$Pb. The last column represents differences in energy between two adjacent lines. The sum of Gaussians and Fourrier-Bessel parameter are taken from \cite{vjv1987}. All lengths are in fermi.}
\label{tab:nucl-dep}
\begin{tabular}{cD{.}{.}{3}D{.}{.}{3}D{.}{.}{1}D{.}{.}{3}D{.}{.}{3}D{.}{.}{3}D{.}{.}{3}}
\hline
Model	&	\multicolumn{1}{c}{RMS} 	&	\multicolumn{1}{c}{$c$} 	&	\multicolumn{1}{c}{$t$} 	&	\multicolumn{1}{c}{$\sqrt[\frac[4]{<r^4>}$ } 	&	\multicolumn{1}{c}{$\sqrt[\frac[6]{<r^6>}$ } 	&	\multicolumn{1}{c}{Ener. (eV)} 	&	\multicolumn{1}{c}{$\Delta E$ (eV)} 	\\											\hline															
SOG	               &	5.5030	&	         	&		&	0.6404	&	1.4020	&	-101336.480	&		\\
Fourrier-Bessel	&	5.5031	&	         	&		&	0.6457	&	1.4099	&	-101336.642	&	0.162	\\
Fermi	       &	5.5030	&	6.6455	&	2.3	&	5.8551	&	6.1421	&	-101336.650	&	0.008	\\
Fermi	       &	5.5030	&	6.7240	&	2.2	&	5.8368	&	6.1635	&	-101336.627	&	-0.023	\\
Fermi	       &	5.5030	&	6.7603	&	2.0	&	5.8283	&	6.2600	&	-101336.616	&	-0.011	\\
\hline
\end{tabular}
\end{center}
\end{table}

More difficult to evaluate are corrections connected to internal nuclear structure. The first one is nuclear deformation. It has been studied most precisely by means of muonic atom spectroscopy. For a deformed nucleus, one writes the parameter $c$ as
\begin{equation}
c=R_0\left[1+\beta_2 Y_{20}\left(\theta,\phi\right)+\beta_4 Y_{40}\left(\theta,\phi\right)\right].
\label{eq:def_nuc}
\end{equation} 
This equation provides an effective, angular-dependent charge distribution. It is then averaged over all directions to provide either an average charge distribution, or higher-order multipole contributions as described, \eg in \cite{tssr1984}. 
Measurements of the $\beta_2$ and $\beta_4$ parameters for several isotopes of thorium, uranium and americium, using muonic atom spectroscopy, can be found in \cite{zstb1984,jshn1985,znhr1986}. Quadrupole parameters and deformation for rare earth elements can be found in \cite{tssr1984}. The effect of the nuclear deformation has been taken into account for the first time in highly-charged ions in the evaluation of the lithiumlike $2p_{j}\to 2s_{1/2}$ transition in $^{238}$U by Blundell \etal \cite{bjs1990}. Since the nuclear spin of $^{238}$U is zero, the charge distribution cannot have nuclear moments beyond the monopole term, and Blundell \etal \cite{bjs1990} performed a spherical average of the charge distribution.  Kozhedub \etal \cite{kast2008}, although they also work with spin-0 nuclei, have provided a general formalism (Eqs. (4) to (11)) for the evaluation of this effect. This effect was also taken into account in the evaluation of x-ray transition energies in core-excited neutral atoms \cite{ial1992,ibl1998}. In  \cite{kast2008} one can find specific calculations for hydrogenlike and lithiumlike  uranium and two isotopes of neodymium.

\begin{figure*}[htbp]
\centering
        \includegraphics[width=\textwidth]{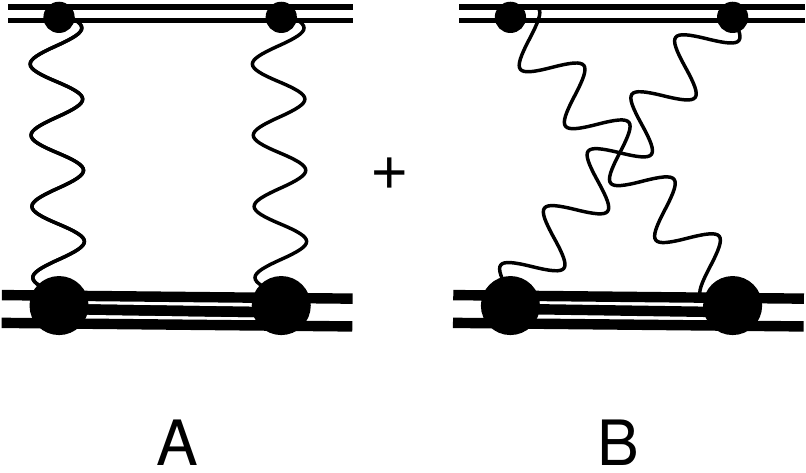}
\caption{Nuclear polarization diagrams. The nucleus is represented by the heavy double line. The triple line represents the excited nucleus in the intermediate state.
\label{fig:nuc-polar}
}
\end{figure*}

The last contribution to the transition energy in highly-charged ions is due to the interaction of the nucleus with the bound electrons, and in particular the $1s$ ones, which perturb the nucleus and lead to a small change in the nuclear charge distribution. This corresponds to the diagram of figure \ref{fig:nuc-polar}. The contribution is complex to evaluate, as it requires knowledge of the internal nuclear structure, \ie excited states, resonances, etc. The effect has been evaluated in heavy few electron ions \cite{pmgs1989,pmgs1991,pas1995}. In these papers, a factor of $\frac{1}{2\pi}$ was omitted. This was first noted by Nefiodov \etal \cite{nlps1996}, who were evaluating the nuclear polarization in lead and uranium and corrected the calculation \cite{pas1996}. The size of this contribution to the $1s$ binding energy is of the order of \SI{0.1}{\electronvolt}. A more recent calculation performed for lead and uranium takes into account also the transverse interaction between the electron and the nucleus \cite{yhhi2001}. It gives results similar to the ones from earlier works.

The magnitudes of all the one-body contributions discussed up to now to the Lyman $\alpha_1$ transition energy in hydrogenlike ions are plotted in figure \ref{fig:size-contrib} as a function of $Z$, together with the uncertainties of the transition energy measurements.

\subsubsection{Recoil corrections}
\label{sec:nuclear-recoil}

Up to now, we have treated the nucleus as infinitely heavy, and done all the calculations in a central potential. Of course the nucleus has a finite mass $M_{A,Z}$. For one-electron ions, the nuclear recoil can be taken into account at the lowest order by replacing \cite{bab1948,moh1983,jas1985}  \eqref{eq:direigen}, by
\begin{equation}
E_{D,M_{A,Z} }= \left(\frac{M_{A,Z} }{M_{A,Z} +m_e}-1\right)\left(E^{\mathrm{D}}_{n\kappa}(Z)-m_ec^2\right)
\label{eq:red_dirac}
\end{equation}
and adding the relativistic reduced mass correction
\begin{equation}
E^{1}_{M_{A,Z} }=-\frac{\left(Z\alpha\right)^4}{8n^4}\frac{m_e}{M_{A,Z}}m_ec^2.
\end{equation}

For multi-electron systems, the one-body recoil correction  can be evaluated from the relativistic normal mass shift (RNMS) Hamiltonian \cite{sto1961,sto1963,sha1985,pal1987,sha1988,saa1994}
\begin{equation}
H^1_{\mathrm{RNMS}} = \frac{1}{2M_{A,Z}}\left\{ \sum_i \bm{p}_i^2 -\frac{\alpha Z}{r_i}   \left[ \bm{\alpha}_i +\frac{ \left(\bm{\alpha}_i\cdot \bm{r}_i\right) \bm{r}_i} {r_i^2}\right]  \cdot \bm{p}_i \right\}.
\label{eq:hrnms}
\end{equation}
In \cite{sha1985}, \eqref{eq:hrnms} and all higher-order one-body QED contributions were derived from the QED formalism, which was later extended in \cite{sha1988} to many-body 
systems. In \cite{pal1987}, Stone's theory was reformulated, independently from the work of \cite{sha1985,sha1988}, to a form allowing one to use the Dirac approach.

The  two-body contribution is obtained from the relativistic specific mass shift Hamiltonian 
\begin{equation}
H^2_{\mathrm{RSMS}} = \frac{1}{2M_{A,Z}} \left\{\sum_{i\neq j} \bm{p}_i \cdot \bm{p}_j   -\frac{\alpha Z}{r_i} \left[\bm{\alpha}_i +\frac{ \left(\bm{\alpha}_i\cdot \bm{r}_i\right) \bm{r}_i} {r_i^2}\right] \cdot   \bm{p}_j\right\}.
\label{eq:rsms}
\end{equation}
 
 Both can be treated in perturbation theory as they give small contributions. Higher-order relativistic corrections can be evaluated following the work of V. Shabaev \cite{sha1998}. Application to hydrogenlike  atoms have been performed in \cite{sha1998a,yas2015a} and in two-electron atoms in
 \cite{asyp2005,mpsz2018}. For lithiumlike ions, this calculation has been performed in  \cite{asy1995}. For the ground state energy of an hydrogenlike uranium ion the recoil contribution is   \SI{0.46}{\electronvolt} for a total energy of \SI{132}{\kilo\electronvolt}.
 Very recently Malyshev \etal \cite{mamsp2019} calculated the recoil corrections from QED, up to order $1/Z$ for the $1s^2$ level of two-electron ions and for the $1s^2 2s$ and $1s^2 2p$ levels of three-electron ions. These QED results  were compared to the one obtained with the Hamiltonian in \eqref{eq:rsms}.

\subsubsection{Two-electron bound state QED corrections}
\label{sec:helike-bsqed}

\begin{figure*}[htbp]
\centering
        \includegraphics[width=0.8\textwidth]{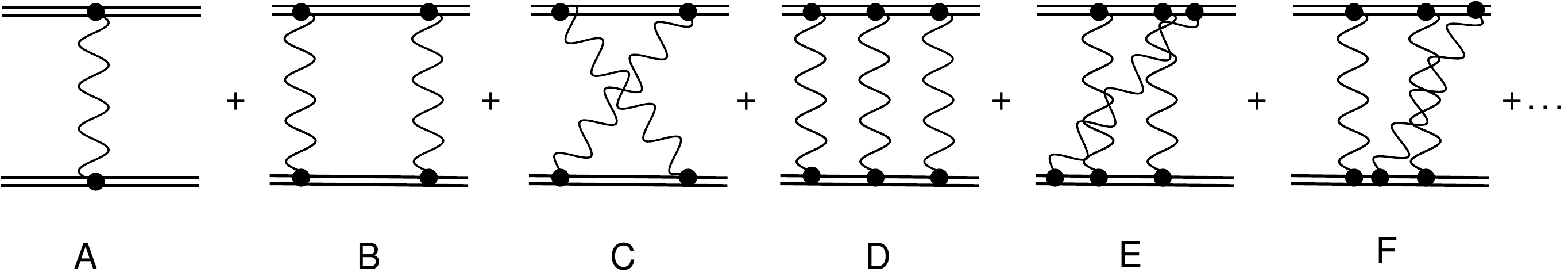}
\caption{The two-electron interaction QED diagram. A: electron-electron interaction. B and C: second order electron-electron contribution. Diagram A, B, D constitute the first orders of the ladder approximation.
\label{fig:two-electron-interaction}
}
\end{figure*}

All corrections provided in the previous section are obviously directly applicable to the two-body problem. One can add the one-body energies defined in \eqref{eq:direigenj} for each individual electron. Two-body corrections must then be added.
 These two-body corrections can be separated into two kinds, those involving only the electron-electron interaction and those with loop corrections to the electron-electron interaction.
 The first kind is represented in figure \ref{fig:two-electron-interaction} up to third order. 
 The first order diagram represents the electron-electron interaction mean-value between the electron wavefunctions connected at the two vertices. From these diagrams one can derive the expression of the electron-electron interaction
\begin{eqnarray}
    V_{ij} & = & \frac{1}{r_{ij} } -\frac{\bm{\alpha}_i \cdot
\bm{\alpha}_j}{r_{ij}} - \frac{\bm{\alpha}_i \cdot
\bm{\alpha}_j}{r_{ij}}
\left(\cos\left(\alpha\omega_{ij}r_{ij}\right)-1\right) \nonumber\\ & + &
\left( \bm{\alpha}_i \cdot \bm{\nabla}_i \right)\left(
\bm{\alpha}_j \cdot \bm{\nabla}_j
\right)\frac{\cos\left(\alpha\omega_{ij}r_{ij}\right)-1} {(\alpha\omega_{ij})^2
r_{ij}},
\label{eq:interact}
\end{eqnarray}
given here in the Coulomb gauge and in a.u. In this expression $r_{ij}= \left| \bm{r}_i-\bm{r}_j \right|$ is the inter-electronic distance, $\omega_{ij} $ is the energy of the photon exchanged between the electrons and $\bm{\alpha}_i$ are the Dirac matrices. It should be noted that  the $\nabla$ operators act on the $r_{ij}$ and not on  the following wave  functions. In this expression the first term is the Coulomb interaction, the second the magnetic interaction, and the two  $\omega_{ij} $-dependent terms represent the retardation interaction, connected to the time it takes for the exchanged photon to travel the distance between the interacting electrons.

 Even though each order has two more vertices than the previous one, the expansion parameter for these diagrams is not $\alpha$ but $1/Z$. The convergence of the series at low-$Z$ is thus very slow.
 This sequence of diagrams can be separated in two parts. The first one, called the ladder approximation, corresponds to diagrams with \num{1}, \num{2},\ldots , $n$  photons, all parallel (as  diagrams A, B, D on figure \ref{fig:two-electron-interaction}).
 In QED, these diagrams must be evaluated order by order simultaneously with the crossed diagram of the same order to conserve gauge invariance and remove divergent $1/\epsilon$ terms in \eqref{eq:spert}.
 In these diagrams, if one retains only the part of the Dirac propagator corresponding to the sum over electronic states and the Coulomb part of the interaction, the second-order result is very close to what can be obtained from standard many-body techniques like RMBPT \cite{mas2000}. 
 The crossed diagrams, like C and E on figure \ref{fig:two-electron-interaction}, and the part of the propagator corresponding to positrons represent a purely non-radiative QED correction. A more detailed discussion about the role of the negative-energy state is given below in section \ref{sec:many-body}. 
 
 The second order correction has been evaluated for two-electron ions for the $1s^2 \, ^1S_0$ state  \cite{bmjs1993,pssl1996} and for the $1s 2p \,^3P_J$ states \cite{mas2000,lasm2001}.  A similar calculation involving the ground state and all $1s 2l$ states of heliumlike ions has also been done by the Saint Petersburg group \cite{abps2000}.  The calculation of these diagrams has been extended to the  $2p_{j}\to 2s_{1/2}$ transition in lithiumlike ions \cite{yass2001}.
 
\begin{figure*}[htbp]
\centering
        \includegraphics[width=0.5\textwidth]{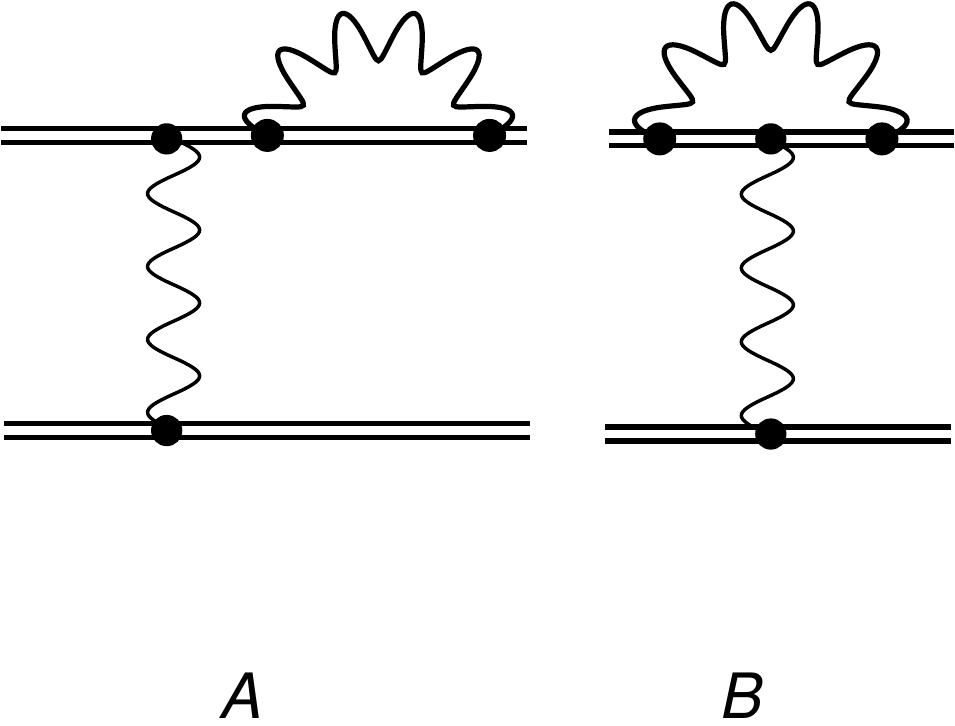}
\caption{Self-energy screening diagrams.
\label{fig:self-screen}
}
\end{figure*}

\begin{figure*}[htbp]
\centering
        \includegraphics[width=0.5\textwidth]{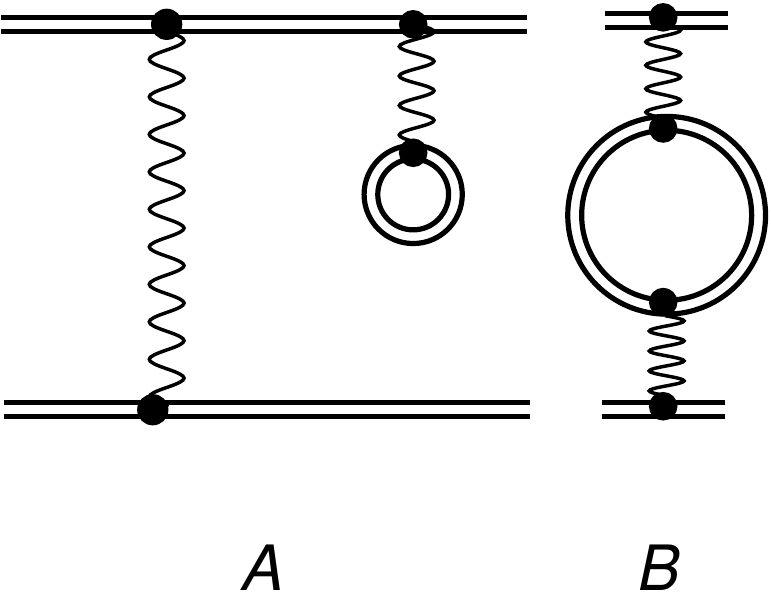}
\caption{Vacuum polarization screening diagrams.
\label{fig:vp-screen}
}
\end{figure*}

 The second kind of diagrams shown in Figs. \ref{fig:self-screen} and \ref{fig:vp-screen} are radiative corrections. The A part in each figure represents the self-energy and vacuum polarization correction to the external leg wavefunction. The B part is a radiative correction to the electron-electron interaction. These corrections are often referred to as screening corrections as the presence of the electron without radiative corrections provides an effect corresponding to replacing $Z$ in the expression of the self-energy in \eqref{eq:qed1st} by $Z-\sigma$ where $\sigma \approx 1$ is a screening coefficient, and keeping only the term of order $\sigma$. The self-energy screening then behaves as $\alpha^2(Z\alpha)^3$ in first approximation.  It shows very simply that the electron-electron interaction provides a sequence of perturbations in $1/Z$.
 The self-energy screening correction has been evaluated for the ground state of heliumlike ions in \cite{pssl1996,yas1997}. A more general scheme has been proposed to evaluate all possible self-energy screening contributions between $1s$, $2s$, $2p_{1/2}$ and $2p_{3/2}$ states \cite{iam2001} using the method proposed in Indelicato \& Mohr \cite{iam1991}. Specific calculations of the QED corrections to  the ground state  self-energy screening and two-photon corrections in He-like ions have also been performed in Holmberg \etal \cite{hsl2015}.

 The vacuum polarization screening correction has been evaluated for the ground state and $1s 2l$ states of  two-electron ions by Artemyev \etal \cite{abps2000}. 
 The vacuum polarization correction was evaluated for the $2p_{j}\to 2s_{1/2}$ transition in lithiumlike ions \cite{abps1999}. The self-energy screening correction has been  evaluated for the $2p_{1/2}\to 2s_{1/2}$ transition in three-electron ions by Yerokhin \etal \cite{yabp1999}. 
 
 It should be noted that the case of quasi-degenerate states like $1s 2p_{1/2}\,J=1$ and $1s 2p_{3/2}\, J=1$ requires a specific procedure and the evaluation of non-diagonal matrix elements as well, for both the radiative and non-radiative diagrams  \cite{abps2000,lis2001,lasm2001}. 

\begin{figure}
\centering
        \includegraphics[width=0.3\textwidth]{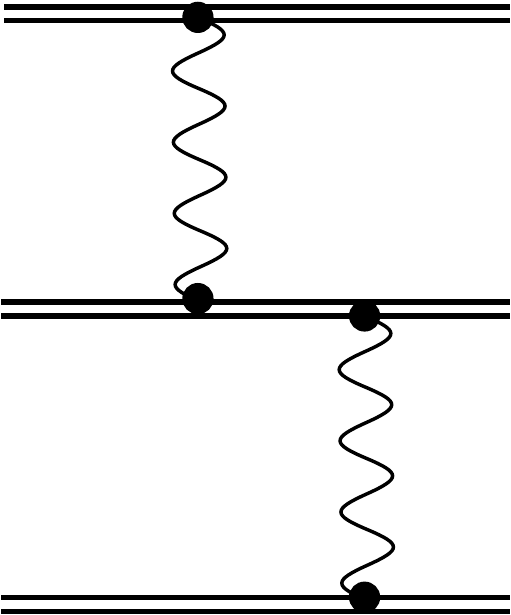}
\caption{Three-body interaction in QED.
\label{fig:three-body}
}
\end{figure}

When performing calculations for atoms with more than two electrons, there are additional diagrams to be considered. For example for three and more-electron atoms,  three-body QED corrections, represented on figure \ref{fig:three-body} should be taken into account \cite{mit1971}.

\subsubsection{Relativistic many-body effects}
\label{sec:many-body}

The slow convergence in $Z$ of the successive terms of the ladder approximation and the difficulty of performing full QED calculations beyond the two-photon exchange requires the development of methods able to include higher orders.
Over the years, many methods have been developed to calculate correlation energy with high-accuracy, some purely non-relativistic, and some fully relativistic. On the relativistic side, one can mention $1/Z$ expansion, Relativistic Configuration Interaction, Multiconfiguration Dirac-Fock \cite{gra1970,des1975}, Relativistic Many-Body Perturbation theory \cite{jas1986,jbs1988,qgw1990,lin1994,sap1988}, coupled-clusters or S-matrix \cite{sap1988,sac2011}. There is however a difficulty in these calculations connected to the fact that they are not directly QED-based, except for the S-matrix method. This is because the solutions of the Dirac equation for a Coulomb potential have bound-states and a positive energy continuum as the Schrödinger equation, but also a negative energy continuum. In QED, this continuum is reinterpreted as a positron component of the operators. But in many-body techniques this is not possible and there are states with energy $-m_ec^2$ to which the bound states can decay. When adding the electron-electron interaction, this leads to unstable atoms. This was identified as early as 1951 by Brown \& Ravenhall \cite{bar1951} and is now known as the \emph{Brown and Ravenhall disease} or continuum dissolution. This was later worked out in more detail in  \cite{suc1980,mit1981,has1984} where it was shown how to implement projection operators to avoid this problem. Projection operators were then implemented in RMBPT \cite{jas1986,hllm1986,lhlm1987,jbs1988} and MCDF \cite{iad1993,ind1995}.  Grant \& Quiney \cite{gaq2000} discussed the fact that projection operators are not unique and depend on the specific method being used. Yet, it is a direct consequence of QED that cannot be circumvented. There have been however recent attempts to include the negative energy continuum without a QED-compatible method  \cite{wnt2007,taw2011}.
The basis of the different many-body techniques is thus the no-pair Hamiltonian
\begin{equation}
{\cal H}^{\mbox{no pair}}=\sum_{i=1}^m H_{D}(r_i) + \sum_{i<j} {\cal
V}\left( \left| {\bf r}_i-{\bf r}_j \right| \right) ,
\label{eq:no-pair-j}
\end{equation}
where $H_{D}$ is a one-electron Dirac operator defined in \eqref{eq:dirac} and ${\cal V}$ is an operator representing the two-body interaction.
This operator has the expression
\begin{equation}
   { \cal V}_{ij}=\Lambda^{++}_{ij} V_{ij} \Lambda^{++}_{ij}.
\label{eq:vproj}
\end{equation}
where $\Lambda^{++}_{ij}=\Lambda^{+}_{i}\Lambda^{+}_{j}$ is an operator projecting onto the one-electron positive energy states and $V_{ij}$ is given in \eqref{eq:interact}.

Another difficulty with many-body methods comes from the gauge dependance of the electron-electron interaction \eqref{eq:interact}. In QED, a diagram or set of diagrams provides gauge-invariant results. For example, the two diagrams B and C in figure \ref{fig:two-electron-interaction} must be evaluated together, and provide a gauge-independent result. But when the operator \eqref{eq:interact} is used with \eqref{eq:no-pair-j}  the gauge invariance is lost. This issue has been discussed in several works \cite{gid1987,gra1987,suc1988,gai1988,lam1989}.
Sucher \cite{suc1988} has shown that for calculations of the MCDF type, the Lorentz gauge adds spurious terms that do not appear in the Coulomb gauge. Lindgren \cite{lin1990} has shown that these contributions come from the fact that relativistic many-body calculations take only into account the reducible part of the QED diagrams, and that the gauge dependence is canceled when the irreducible part is taken into account (see also \cite{lov1978}). A more correct derivation, using the two-time Green's function formalism has been performed by Shabaev \cite{sha1993}.

The other difficulty associated with the use of all-order methods lies in the energy dependance of the exchanged photon. The value of $\omega_{ij} $ in \eqref{eq:interact} is perfectly well defined in the independent particle approximation, but it is not clear what value to use when the interaction operators are evaluated between correlation orbitals, which can have energies much more negative than the bound orbitals. There is thus always a question of the use of the magnetic and retardation part of the operator in the evaluation of the correlation energy \cite{gid1987,gra1987}. The MCDFGME code allows to do this, which can lead to a very large contribution to the correlation energy at high-$Z$. Following the method described in  Indelicato \cite{ind1988}, one finds at $Z=92$, for the $1s^2 \, ^1S_0$ level, that the Coulomb non-relativistic contribution to the correlation energy, obtained by $1/Z$ expansion \cite{aps1971,aps1975,bla1976} is \SI{-1.26}{\electronvolt}, while the relativistic part is \SI{1.38}{\electronvolt}, the magnetic interaction part is \SI{-6.31}{\electronvolt} and the retardation part is \SI{1.62}{\electronvolt}  \cite{ind1988,ind1992}.

In view of these difficulties, and to be able to perform QED-based many-body calculations including complex cases with degenerate levels, several methods have been proposed and used. Besides the S-matrix formalism already mentioned \cite{sap1988}, one can mention the covariant evolution operator method \cite{lsa2004,lsh2006,lsh2011}. The alternative method is the \emph{two-time Green's function} approach \cite{sha2002,art2017a}.

The other difficulty in the many-body approach is connected to the evaluation of the screened self-energy corrections of figure \ref{fig:self-screen}. In the past, these corrections have been evaluated using an effective-$Z$ parameter in the self-energy part of \eqref{eq:qed1st}. This effective-$Z$ was derived by comparing the mean radius of the orbital to the radius of the hydrogenlike orbital with the same $n, \,\kappa$. Then it was proposed \cite{igd1987,iad1990} to use the Welton approximation \cite{wel1948}, which correctly represents the lowest order of the self-energy.
More recently Shabaev \etal \cite{sty2013} proposed a model operator, derived from QED, which allows to calculate the screened self energy using a potential. A computer code which implements this  model operator is available \cite{sty2015}.
Other approaches use direct evaluation of the QED diagrams, replacing the electron-electron interaction with different effective electronic potentials. Such a method was developed by Blundell \& Snyderman \cite{bas1991} and used to evaluate the self-energy contribution to transition energies in few-electron ions, Li-like, Na-like and Cu-like \cite{blu1992} or for several perturbing potentials \cite{bcs1997}.

\subsection{Experiments and comparison with theory}
\label{sec:experiment}
The different contributions and techniques summarized in the previous section are complicated to evaluate, and require many approximations. The quality of these approximations becomes more and mode difficult to assess when the number of electrons increases.
Moreover, testing QED at the present time cannot come from the comparison between theory and experiment for a single ion. Of course hydrogenlike ions constitute a very favorable testing ground for QED contributions. Yet the data at very high-$Z$ are scarce, and not very accurate except for a few cases like the measurement of the Lyman $\alpha$ lines in uranium \cite{gsbb2005}, where the QED contribution was measured with an accuracy of $\approx$1\%. Much data also exists for two and three-electron ions. In the case of heliumlike ions there are both $n=2\to n=1$ and $n=2, \, \Delta n=0$ transitions which have been measured for a broad range of $Z$. Again the number of data and precision at very high-$Z$ are not sufficient. In Li-like ions, the fine structure transitions $1s^2 2p \, ^2P_{J} \to 1s^2 2s \, ^2S_{1/2}$ have been accurately measured across a broad range of $Z$, with good accuracy, even at high-$Z$, and thus  provide useful comparison for the fine-structure, at the cost of an increased complexity for the theory.
 
In this section I present a detailed discussion of available experimental data and the comparison with theory. In particular I perform weighted fits  to the experiment-theory differences with functions $b Z^n$. In order to do a consistent treatment, I use in each case the most recent and accurate calculation that has values for all the elements which have been measured. In the case of three-electron ions, for which the transition energies are much more difficult to calculate, I also present a detailed comparison of available calculations. 
In the different tables presented here,  and whenever possible, the published values have been corrected for the latest values of the fundamental constants \cite{mnt2016}, although the difference is usually negligible compared to the error bars.

\subsubsection{Hydrogenlike ions }
\label{sec:hlike}

\begin{figure*}[htbp]
\centering
        \includegraphics[width=\textwidth]{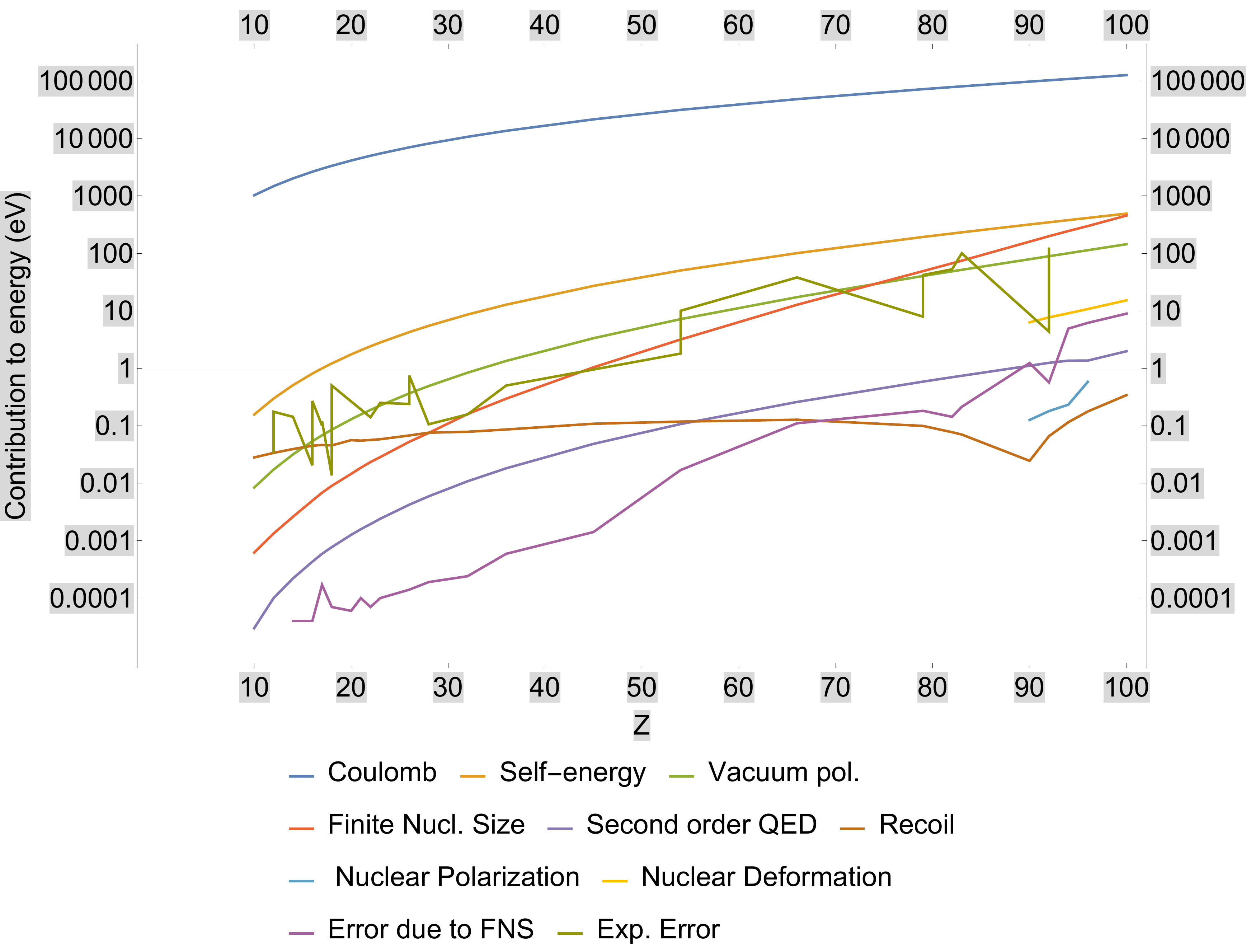}
\caption{Size of all the one-electron contributions to the Lyman $\alpha_1$ transition energy as a function of $Z$. The second order QED corrections from figure \protect \ref{fig:two-loop-qed} are summed together.
The uncertainties for all the available measurements of the Lyman $\alpha_1$ are also plotted for comparison, together with uncertainties connected to the errors on the nuclear size.
\label{fig:size-contrib}
}
\end{figure*}

\begin{figure*}[htbp]
\centering
        \includegraphics[width=\textwidth]{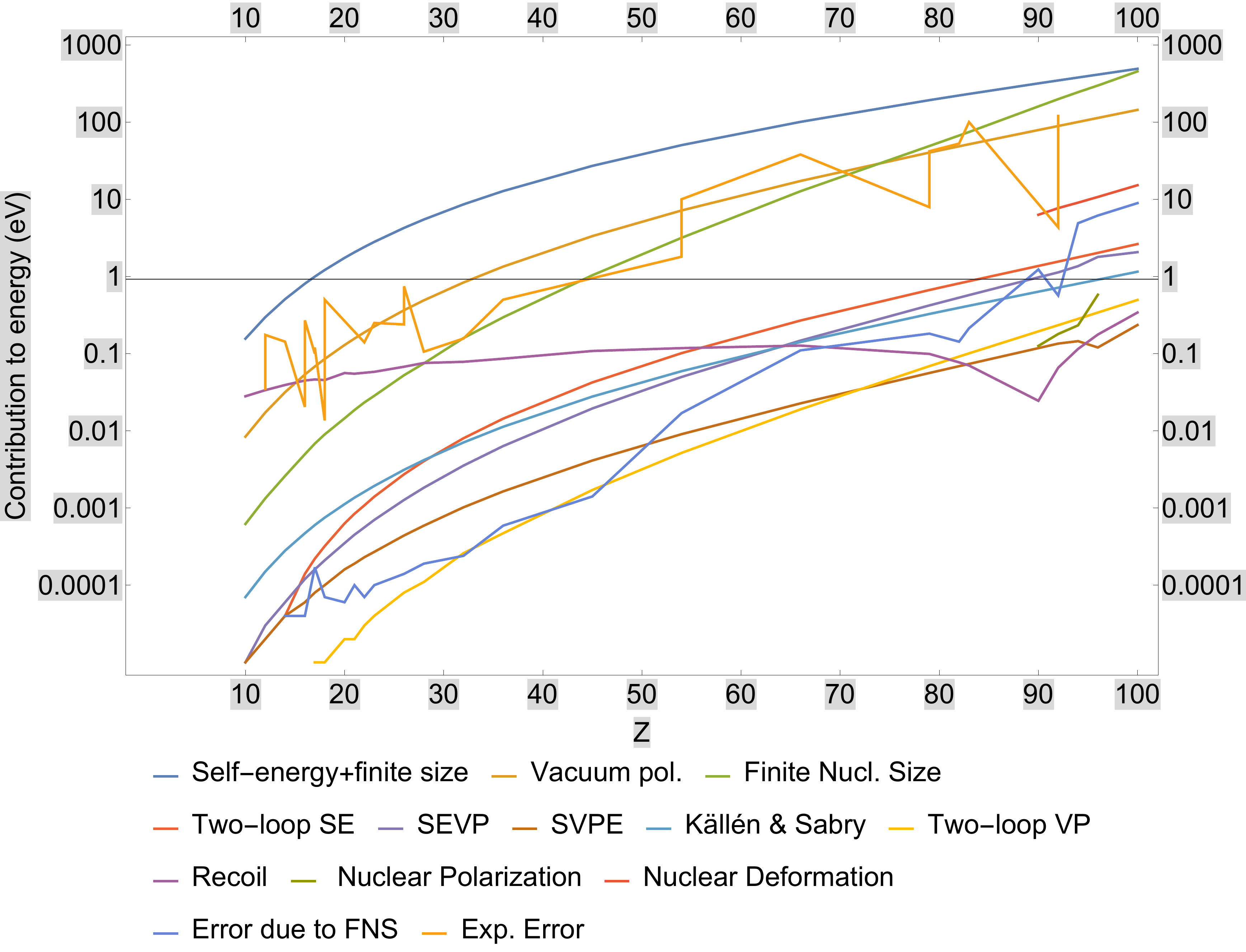}
\caption{Size of all the QED contributions to the Lyman $\alpha_1$ transition energy as a function of $Z$, compared to nuclear contributions and experimental errors.
\label{fig:h-qed-contrib}
}
\end{figure*}

The study of one-electron ions is indeed expected to be the best way to test the different corrections presented in Secs. \ref{sec:hlike-bsqed}. Yet it is not as simple as it seems because of the different nuclear corrections, the accuracy of which is limited by our knowledge of the nuclear structure. This may be seen in Table \ref{tab:nucl-dep} where the dependance on the nuclear shape is shown to be sizable in some cases. To illustrate this, I have plotted in figure \ref{fig:size-contrib} the different QED, recoil and nuclear contributions as a function of $Z$ for the Lyman $\alpha_1$ transition. The effect of the uncertainties on the nuclear size are also plotted, using the nuclear radii from the latest tabulation of Angeli \& Marinova \cite{aam2013}. These uncertainties can be large when compared to QED corrections, and have a strong impact at high-$Z$. 
The calculations were performed with the 2018 version of the MDFGME Multiconfiguration Dirac-Fock code \cite{iad2005}, which contains up to date values for  one and two-loop QED corrections, recoil and finite nuclear size from the references listed in \ref{sec:hlike-bsqed}. All the second-order QED corrections are plotted together.
The nuclear polarization corrections mentioned in \ref{sec:nuclear} are also shown. Figure  \ref{fig:h-qed-contrib} zooms on the QED and nuclear corrections. The different contributions to second-order QED from  figure \protect \ref{fig:two-loop-qed} are plotted separately. 

Over the years there have been several tabulations of one-electron energy levels. The tables from Erickson \cite{eri1977} were used for many years. It was shown later \cite{moh1974,moh1974a} that the higher-order contributions to the self-energy used in these tables were not as accurate as claimed at the time.
Subsequently, a tabulation of level energies for $10\leq Z \leq 40$ was published by Mohr \cite{moh1983}. Then Johnson \& Soff \cite{jas1985} provided an up-to-date table for $1 \leq Z \leq 110$, which was used for \SI{30}{years}, until a new tabulation taking into account all two-loop QED corrections and nuclear size corrections from  \cite{aam2013} was published  \cite{yas2015}. I use this recent publication as the reference calculation to compare theory and experiment.

\begin{figure*}[htbp]
\centering
        \includegraphics[width=\textwidth]{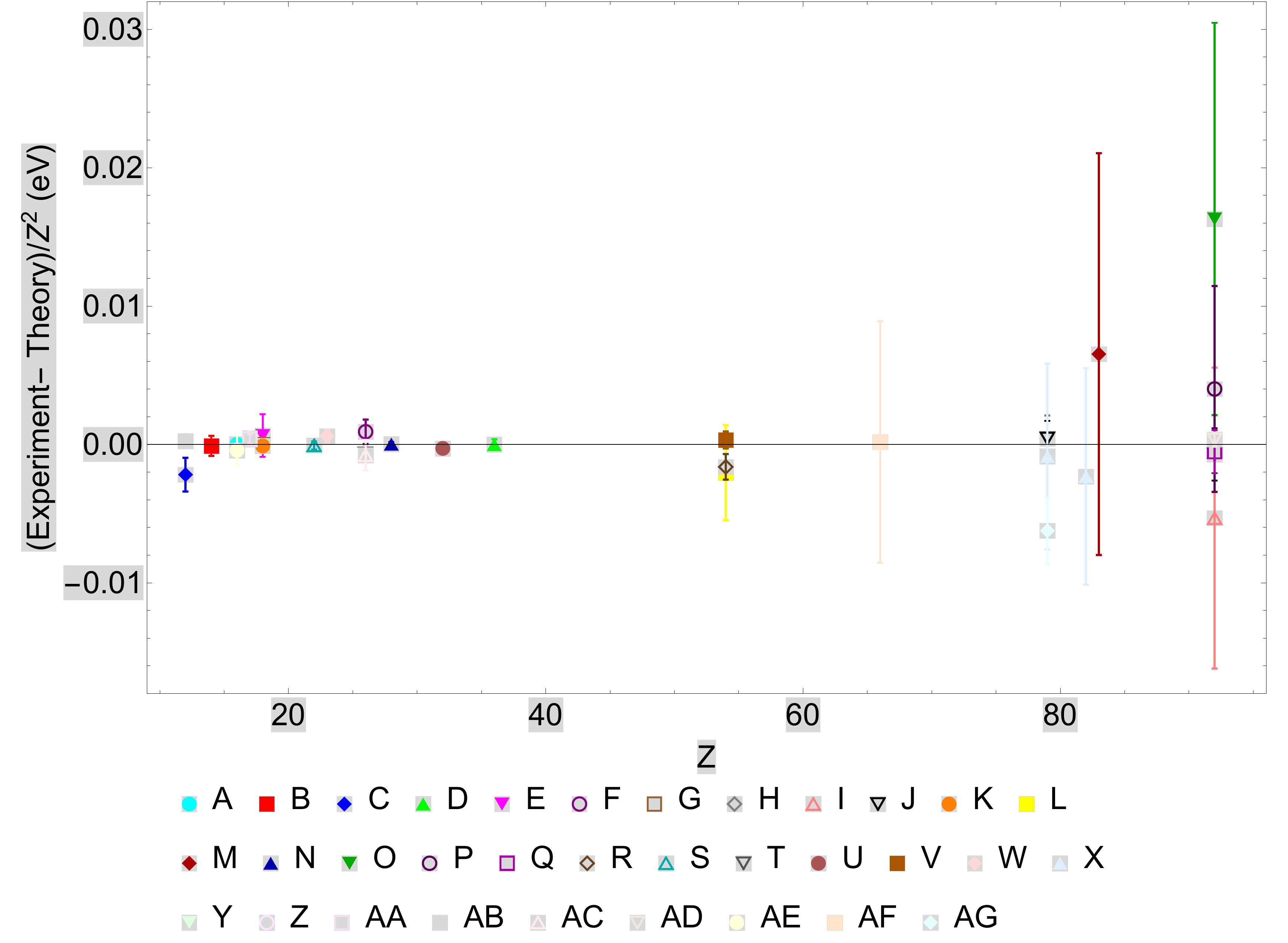}
\caption{
Comparison between theory  \protect\cite{yas2015} and experiment for the Lyman $\alpha 1$ line in hydrogenlike ions. The data are from Table \protect  \ref{tab:hlikelya1-comp}.
References:
A: \protect\cite{kmmu2014},
B: \protect\cite{tkbb2002},
C: \protect\cite{astf1980},
D: \protect\cite{tbil1985},
E: \protect\cite{bmic1983},
F: \protect\cite{btim1983},
G: \protect\cite{bdfl1985},
H: \protect\cite{blbf1994},
I: \protect\cite{ldhs1994},
J: \protect\cite{bmlg1995},
K: \protect\cite{mrkk1986},
L: \protect\cite{biss1989},
M: \protect\cite{smgm1992},
N: \protect\cite{bifl1991},
O: \protect\cite{bciz1990},
P: \protect\cite{smbb1993},
Q: \protect\cite{smbd2000},
R: \protect\cite{wbbl2000},
S: \protect\cite{tas2002},
T: \protect\cite{cldh2007},
U: \protect\cite{clsd2009},
V: \protect\cite{tgbb2009},
W: \protect\cite{gcph2010},
X: \protect\cite{kabe2017},
Y: \protect\cite{kkrs1984},
Z: \protect\cite{dsj1985},
AA: \protect\cite{rsdc1984},
AB: \protect\cite{hfkb1998},
AC: \protect\cite{smlr1987},
AD: \protect\cite{gsbb2005},
AE: \protect\cite{sbbt1982},
AF: \protect\cite{bfli1993},
AG: \protect\cite{gths2018}.
\label{fig:h-z-vs-exp-lya1}
}
\end{figure*}

\begin{figure*}[htbp]
\centering
        \includegraphics[width=\textwidth]{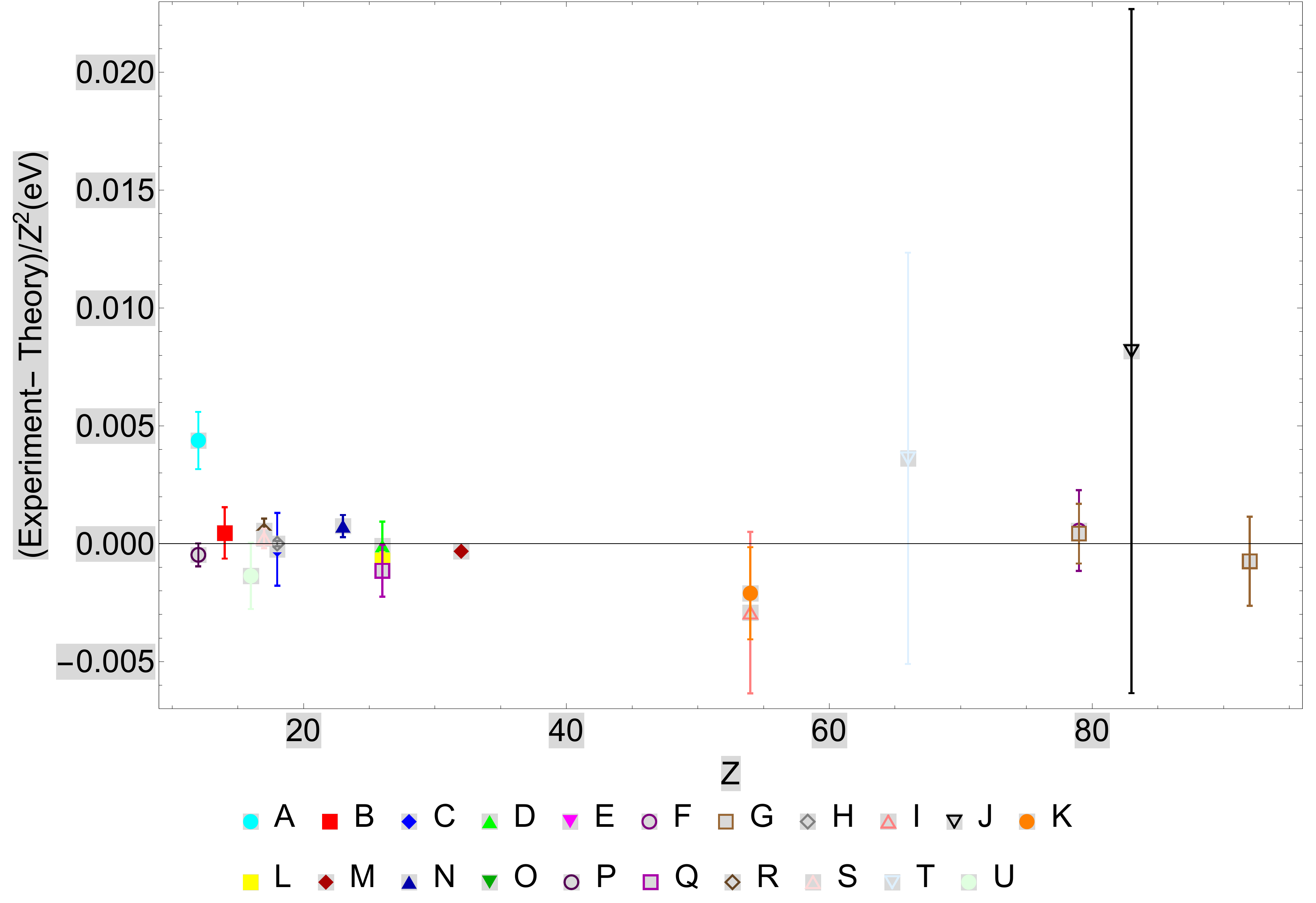}
\caption{
Comparison between theory  \protect\cite{yas2015} and experiment for the Lyman $\alpha 2$ line in hydrogenlike ions. The data are those from Table \protect  \ref{tab:hlikelya2-comp}.
References:
A: \protect\cite{astf1980},
B: \protect\cite{tkbb2002},
C: \protect\cite{bmic1983},
D: \protect\cite{btim1983},
E: \protect\cite{bdfl1985},
F: \protect\cite{blbf1994},
G: \protect\cite{bmlg1995},
H: \protect\cite{mrkk1986},
I: \protect\cite{biss1989},
J: \protect\cite{smgm1992},
K: \protect\cite{wbbl2000},
L: \protect\cite{cldh2007},
M: \protect\cite{clsd2009},
N: \protect\cite{gcph2010},
O: \protect\cite{kkrs1984},
P: \protect\cite{hfkb1998},
Q: \protect\cite{smlr1987},
R: \protect\cite{dsj1985},
S: \protect\cite{rsdc1984},
T: \protect\cite{bfli1993},
U: \protect\cite{sbbt1982}.
\label{fig:h-z-vs-exp-lya2}
}
\end{figure*}

\begin{figure*}[htbp]
\centering
        \includegraphics[width=\textwidth]{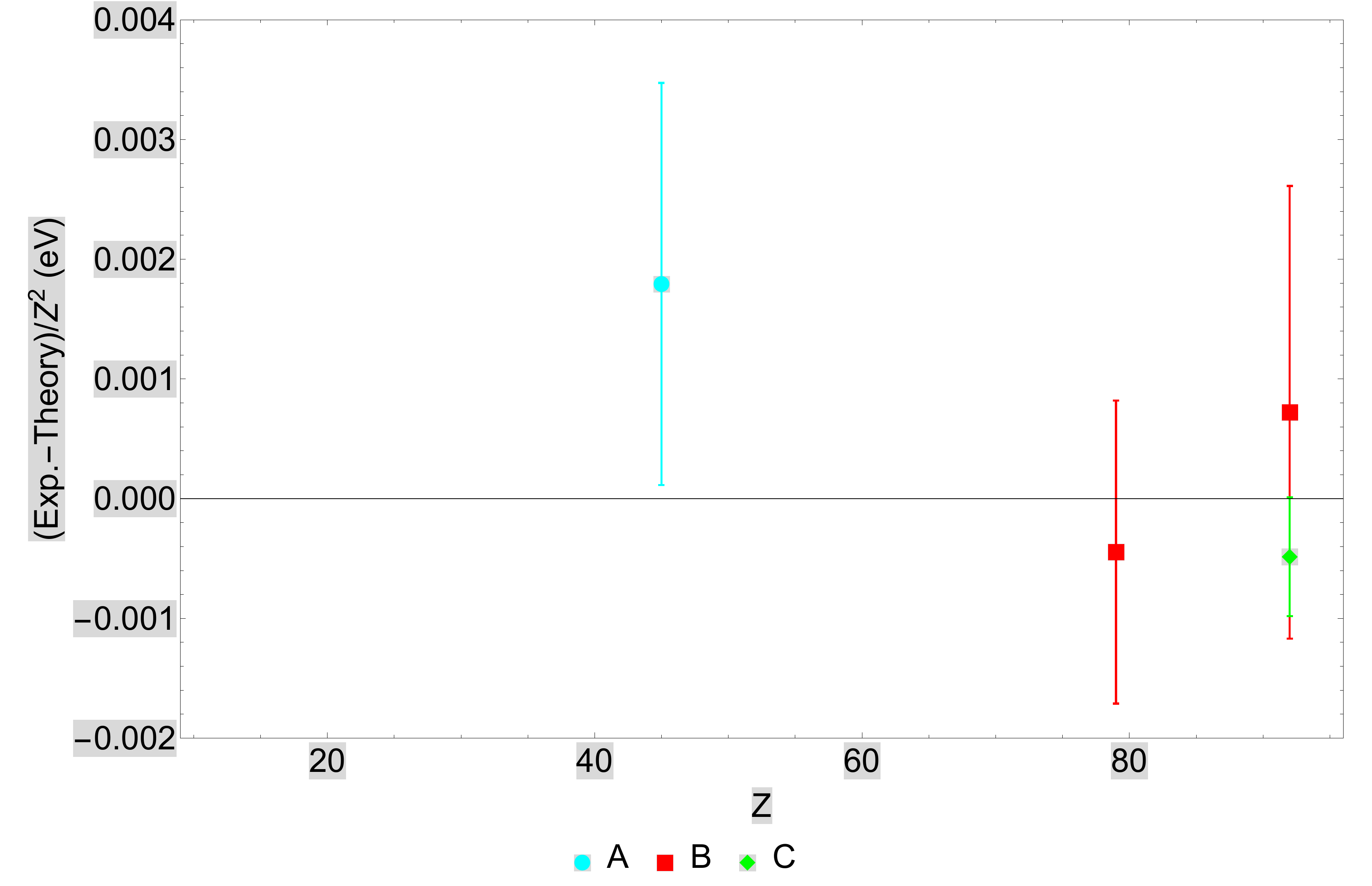}
\caption{
Comparison between theory  \protect\cite{yas2015} and experiment for the $1s$ binding energy in hydrogenlike ions. The data are those from Table \protect  \ref{tab:h-bind-comp}.
References:
A: \protect\cite{nno2003},
B: \protect\cite{bmlg1995},
C: \protect\cite{gsbb2005}.
\label{fig:h-z-vs-exp-grd}
}
\end{figure*}

\begin{figure*}[htbp]
\centering
        \includegraphics[width=\textwidth]{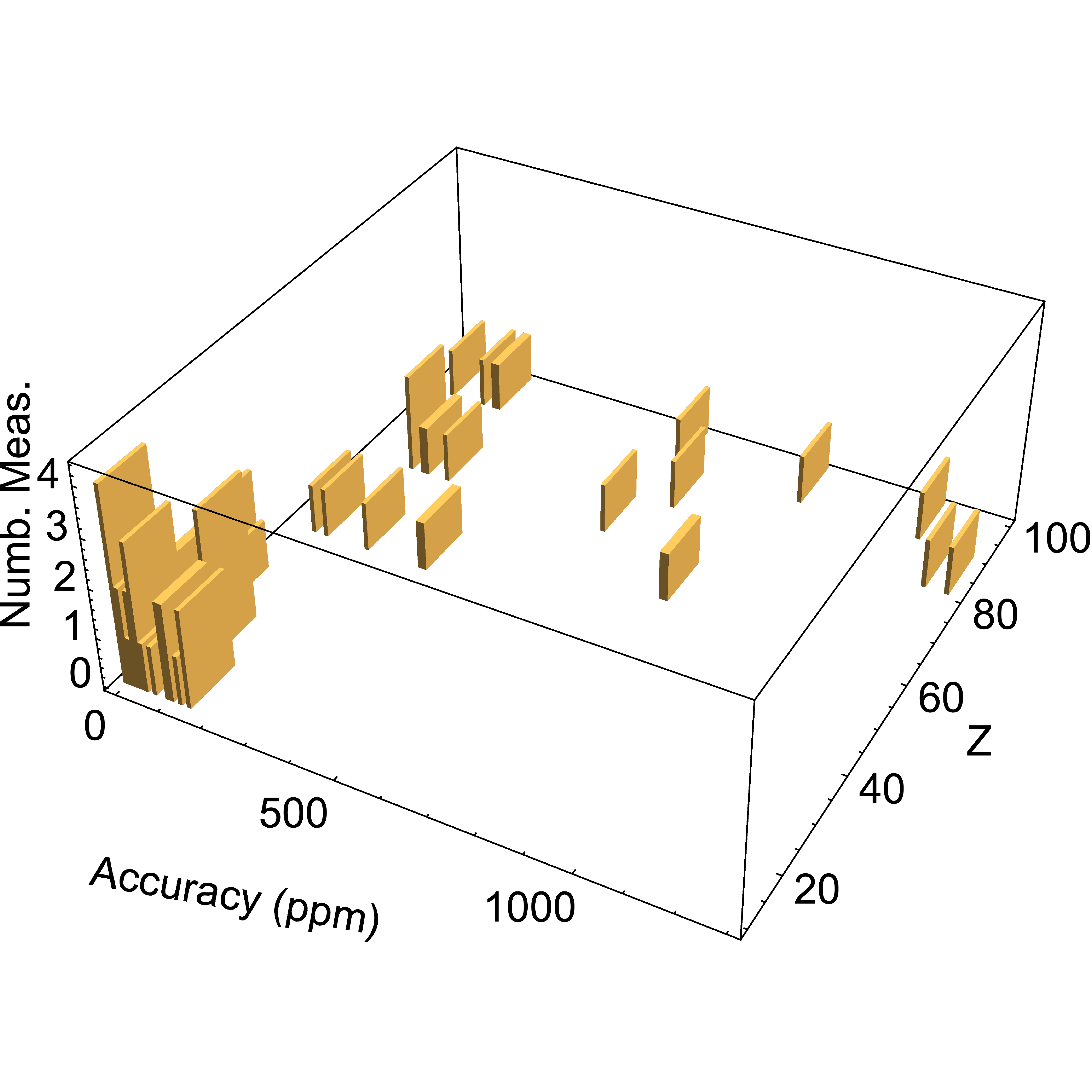}
\caption{
Histogram of the number of measurements for a given accuracy in parts per million and the atomic number $Z$ for Ly$\alpha_1$ and Ly$\alpha_2$. All measurements from Tables   \protect \ref{tab:hlikelya1-comp} and \protect \ref{tab:hlikelya2-comp} are taken into account \cite{dijk2019}.
\label{fig:h-z-vs-ppm-histo}
}
\end{figure*}

\begin{figure*}[htbp]
\centering
        \includegraphics[width=\textwidth]{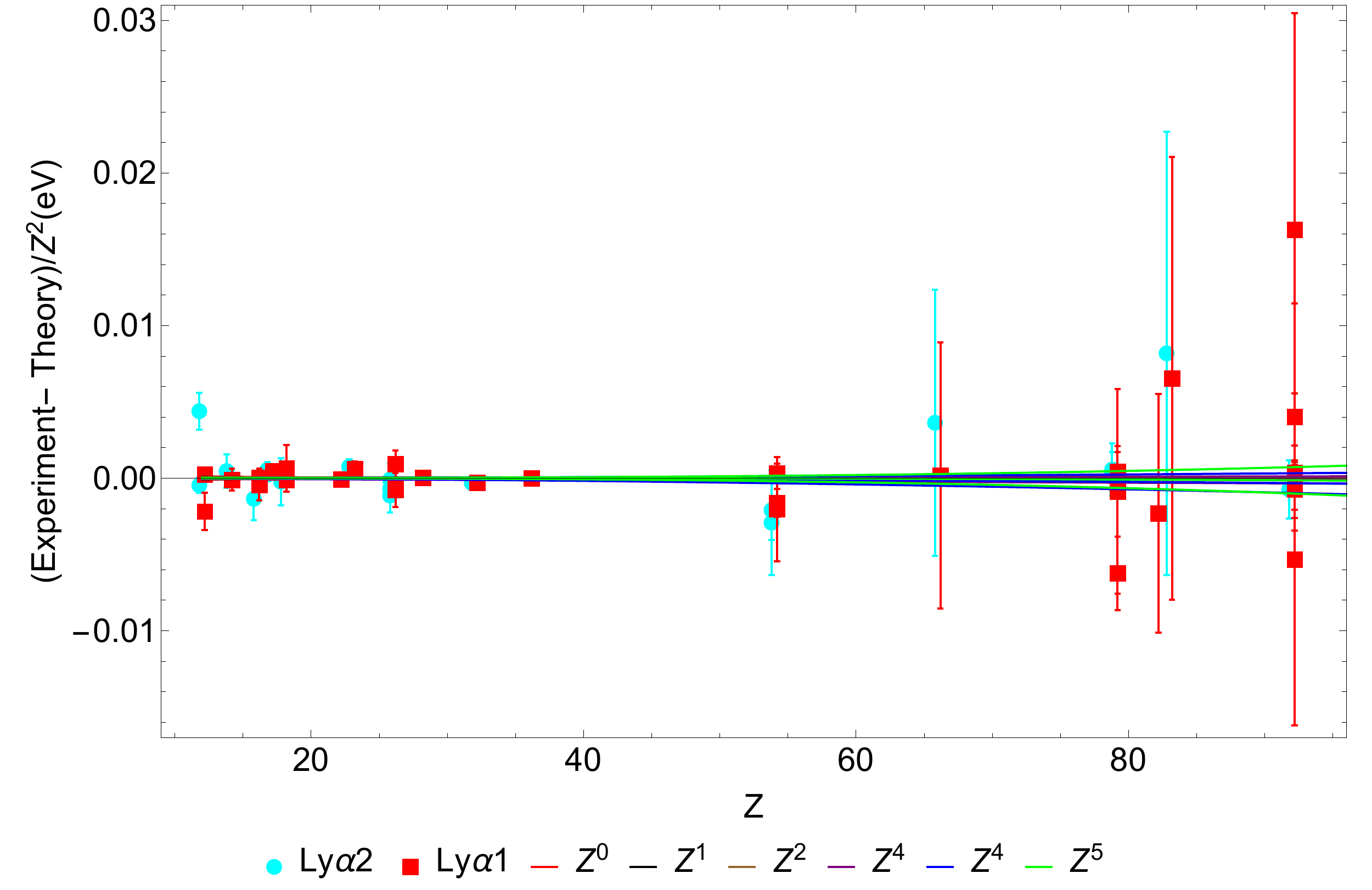}
\caption{
Fit of the  differences between  experiments and the calculations of Yerokhin \& Shabaev \protect \cite{yas2015} for the Lyman $\alpha$ transitions in hydrogenlike ions with different functions $b Z^n$, $n=0,\ldots,\, 5$.
The error bands corresponding to $(b\pm \delta b) Z^n$, where $\delta b$ is the $1\sigma$ error bar on the fitted constant $b$, are also shown. The Lyman $\alpha_1$ and  $\alpha_2$ points have been shifted by $\delta Z=\pm 0.2$ to make the figure more readable.
\label{fig:lh-ly--fit-yero-diff-exp}
}
\end{figure*}

\begin{figure*}[htbp]
\centering
        \includegraphics[width=\textwidth]{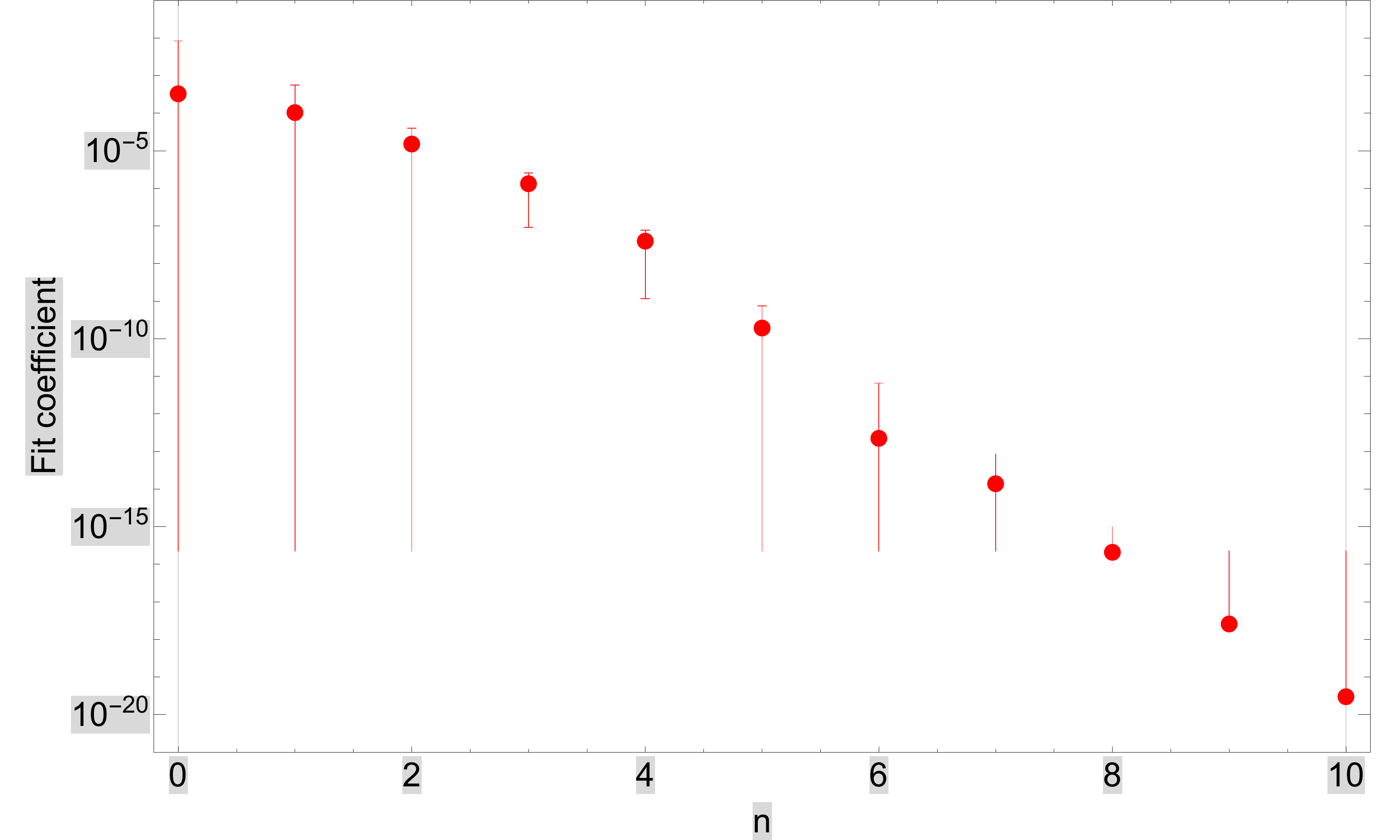}
\caption{
Absolute values of the fit coefficient  $b$ for $n=0,\ldots,\, 10$ with error bars for the differences between  experiments and the calculation of Yerokhin \& Shabaev \protect \cite{yas2015} for Lyman $\alpha$ transitions in hydrogenlike ions.
\label{fig:ly-h-diff-exp-coeff}
}
\end{figure*}

The comparison between experiment and theory is presented in Tables  \ref{tab:hlikelya1-comp} and \ref{tab:hlikelya2-comp} for the Lyman $\alpha 1$ and $\alpha 2$ respectively for all $Z\ge 12$. The few cases where the binding energy has been obtained directly by using radiative recombination are presented in Table \ref{tab:h-bind-comp}. I chose $Z=12$ as the minimum value, since, except for hydrogen, there are no measurements of the Lyman $\alpha$ lines for lower $Z$, and there are very few measurements of $n=2$ intrashell transitions at high-$Z$. A comparison for the intrashell transitions can be found in Table 5 of \cite{yas2015}. In a number of cases, the result given in the experimental papers is only the Lamb-shift, \ie the difference between the measured energy and the Dirac energy, even though the transition energy is what has been measured.  In this case I added back the theoretical $2p_{j}$ energy used in the paper to get the transition energy. In the case of \cite{gsbb2005}, the authors provide both the binding energy due to radiative recombination and the measurement of the Lyman $\alpha_1$ transition energy. The difference between experiment and theory for the Lyman $\alpha1$ and Lyman $\alpha2$ lines is plotted in Figs. \ref{fig:h-z-vs-exp-lya1} and \ref{fig:h-z-vs-exp-lya2} respectively. The values for the $1s$ binding energy are plotted in figure \ref{fig:h-z-vs-exp-grd}. One can see that the overall agreement is quite good, but that there is a lack of very accurate measurements at medium and high-$Z$, except for the \SI{4.5}{\electronvolt} measurement in uranium \cite{gsbb2005}. This is confirmed by figure \ref{fig:h-z-vs-ppm-histo} in which I plot the histogram of the number of measurements with a given accuracy in parts-per-million (ppm) as a function of $Z$ and of the accuracy.  
In order to quantify the quality of the agreement and look for possible systematic deviations, I have fitted simultaneously all the experiment-theory differences for the   Lyman $\alpha 1$ and $\alpha 2$ with functions of the type $b Z^n$,  $n=0$ to \num{10}.
The comparison between the fitted functions and the data is presented in figure \ref{fig:lh-ly--fit-yero-diff-exp} for $n=0$ to \num{5}. The error bands, corresponding to the functions $(b\pm\delta b)Z^n$, where $\delta b$  is the uncertainty on the fitted coefficient $b$, are also plotted. The coefficients $b$ are plotted in figure \ref{fig:ly-h-diff-exp-coeff}, with the corresponding error bars. This analysis and all the subsequent ones have been performed with Mathematica \num{11.3} \cite{Mathematica11-2018}.
One can see that the agreement is excellent, and that all the coefficients are compatible with \num{0} within their error bars. The error bands are quite symmetrical around the horizontal axis, which shows that within the present experimental accuracy and uncertainty on the calculation connected to the uncertainty in nuclear charge radius, there is no global effect. With the available data, the error bands correspond to energy differences of \SI{+-5}{\electronvolt} to \SI{+-8}{\electronvolt} at $Z=92$. 

\begin{figure*}[htbp]
\centering
        \includegraphics[width=\textwidth]{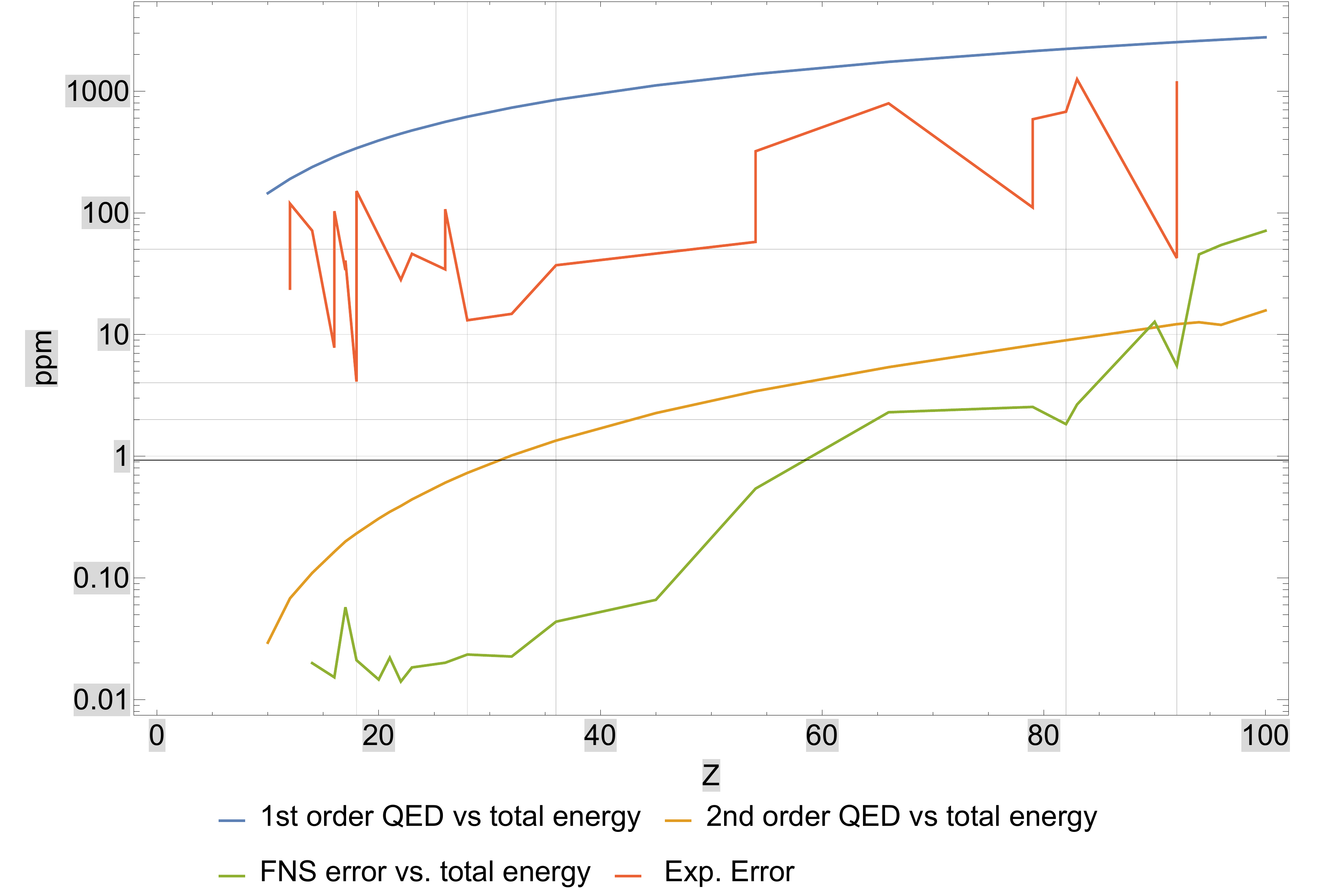}
\caption{
Comparison between one and two-loop QED corrections,  experimental uncertainties, and  uncertainties on the theoretical energy of the ion due to the uncertainties on the nuclear size, in parts per million, for the Lyman $\alpha 1$ transitions.  
All measurements presented in Table   \protect \ref{tab:hlikelya1-comp}  are taken into account.
The vertical lines represent the best measured elements, Ar, Ni, Kr, Pb and U. The horizontal line are drawn at  \SI{1}{ppm},  \SI{2}{ppm}, \SI{4}{ppm}, \SI{10}{ppm} and  \SI{50}{ppm}.
\label{fig:h-qed-vs-uncert-nuc-exp}
}
\end{figure*}

It is now of interest, to define future experimental strategies, to compare the size of the first and second order QED contributions to the uncertainties from available experiments and nuclear size uncertainties. Using the data from Tables  \ref{tab:hlikelya1-comp} and \ref{tab:hlikelya2-comp} and theoretical values for QED, I show in figure \ref{fig:h-qed-vs-uncert-nuc-exp} the present status of experimental tests of QED. Clearly the high-precision, low-$Z$ experiments performed in Heidelberg on argon do not provide information on the second-order QED contribution.
The best uranium measurement uncertainty, which is right now  \SI{\approx 50}{ppm} would need to be improved by at least one order of magnitude, to around  \SI{0.4}{\electronvolt} to significantly test the size of second order QED corrections. This would be at the limit of uncertainties due to nuclear effects. In the long term, it would probably be best to focus on $^{208}$Pb, where the nucleus is very insensitive to nuclear polarization and deformation because of its double-magic property and has a well-known radius, which provides an uncertainty in the theoretical evaluation of only \SI{2}{ppm}. Yet performing a  test of second-order QED to 20\% accuracy would require to reach an experimental accuracy of \SI{\approx 2}{ppm}, which is a very difficult task.

\begin{table}
\begin{center}
\caption{Comparison between experimental and theoretical \protect\cite{yas2015}  energies for the $2p_{3/2}\to 1s$ (Lyman $\alpha 1$) transition in hydrogenlike ions}
\label{tab:hlikelya1-comp}
\begin{tabular}{cD{.}{.}{3}D{.}{.}{3}D{.}{.}{3}l}
\hline
$Z$	&	\multicolumn{1}{c}{Experiment }	&	\multicolumn{1}{c}{Error }	&	\multicolumn{1}{c}{Diff. Th.}	&	\multicolumn{1}{c}{Reference} \\
	\hline
12	&	1472.32	&	0.17	&	-0.31	&	\cite{astf1980}	\\
12	&	1472.67	&	0.03	&	0.04	&	\cite{hfkb1998}	\\
14	&	2006.06	&	0.14	&	-0.02	&	\cite{tkbb2002}	\\
16	&	2622.704	&	0.020	&	0.004	&	\cite{kmmu2014}	\\
16	&	2622.59	&	0.27	&	-0.11	&	\cite{sbbt1982}	\\
17	&	2962.46	&	0.10	&	0.11	&	\cite{kkrs1984}	\\
17	&	2962.49	&	0.12	&	0.13	&	\cite{dsj1985}	\\
17	&	2962.47	&	0.10	&	0.12	&	\cite{rsdc1984}	\\
18	&	3322.993	&	0.014	&	0.000	&	\cite{kmmu2014}	\\
18	&	3323.20	&	0.50	&	0.21	&	\cite{bmic1983}	\\
18	&	3322.989	&	0.017	&	-0.004	&	\cite{bdfl1985}	\\
18	&	3322.96	&	0.06	&	-0.04	&	\cite{mrkk1986}	\\
22	&	4976.86	&	0.14	&	-0.03	&	\cite{tas2002}	\\
23	&	5443.95	&	0.25	&	0.32	&	\cite{gcph2010}	\\
26	&	6973.80	&	0.60	&	0.62	&	\cite{btim1983}	\\
26	&	6972.73	&	0.24	&	-0.45	&	\cite{cldh2007}	\\
26	&	6972.65	&	0.75	&	-0.53	&	\cite{smlr1987}	\\
28	&	8101.76	&	0.11	&	0.01	&	\cite{bifl1991}	\\
32	&	10624.24	&	0.16	&	-0.30	&	\cite{clsd2009}	\\
36	&	13508.95	&	0.50	&	-0.01	&	\cite{tbil1985}	\\
54	&	31278.00	&	10.00	&	-5.95	&	\cite{biss1989}	\\
54	&	31279.20	&	2.70	&	-4.75	&	\cite{wbbl2000}	\\
54	&	31284.90	&	1.80	&	0.95	&	\cite{tgbb2009}	\\
66	&	48038.00	&	38.00	&	0.74	&	\cite{bfli1993}	\\
79	&	71573.05	&	10.43	&	2.62	&	\cite{blbf1994}	\\
79	&	71573.11	&	7.90	&	2.68	&	\cite{bmlg1995}	\\
79	&	71565.00	&	41.90	&	-5.43	&	\cite{kabe2017}	\\
79	&	71531.50	&	15.00	&	-38.93	&	\cite{gths2018}	\\
82	&	77919.00	&	52.60	&	-15.58	&	\cite{kabe2017}	\\
83	&	80188.00	&	100.00	&	44.96	&	\cite{smgm1992}	\\
92	&	102130.00	&	92.00	&	-45.10	&	\cite{ldhs1994}	\\
92	&	102168.99	&	16.00	&	-6.11	&	\cite{bmlg1995}	\\
92	&	102313.00	&	120.00	&	137.90	&	\cite{bciz1990}	\\
92	&	102209.00	&	63.00	&	33.90	&	\cite{smbb1993}	\\
92	&	102170.70	&	13.20	&	-4.40	&	\cite{smbd2000}	\\
92	&	102178.12	&	4.33	&	3.02	&	\cite{gsbb2005}	\\\hline
\end{tabular}
\end{center}
\end{table}

\begin{table}
\begin{center}
\caption{Comparison between experimental and theoretical \protect\cite{yas2015}  energies for the $2p_{1/2}\to 1s$ (Lyman $\alpha 2$) transition in hydrogenlike ions}
\label{tab:hlikelya2-comp}
\begin{tabular}{cD{.}{.}{3}D{.}{.}{3}D{.}{.}{3}l}
\hline
$Z$	&	\multicolumn{1}{c}{Experiment }	&	\multicolumn{1}{c}{Error }	&	\multicolumn{1}{c}{Diff. Th.}	&	\multicolumn{1}{c}{Reference} \\
	\hline
12	&	1472.32	&	0.17	&	0.63	&	\cite{astf1980}	\\
12	&	1471.62	&	0.07	&	-0.07	&	\cite{hfkb1998}	\\
14	&	2004.41	&	0.21	&	0.09	&	\cite{tkbb2002}	\\
16	&	2619.35	&	0.36	&	-0.35	&	\cite{sbbt1982}	\\
17	&	2958.62	&	0.10	&	0.09	&	\cite{kkrs1984}	\\
17	&	2958.68	&	0.15	&	0.16	&	\cite{dsj1985}	\\
17	&	2958.59	&	0.12	&	0.06	&	\cite{rsdc1984}	\\
18	&	3318.10	&	0.50	&	-0.08	&	\cite{bmic1983}	\\
18	&	3318.17	&	0.02	&	-0.01	&	\cite{bdfl1985}	\\
18	&	3318.18	&	0.04	&	0.00	&	\cite{mrkk1986}	\\
23	&	5431.10	&	0.25	&	0.40	&	\cite{gcph2010}	\\
26	&	6951.90	&	0.70	&	-0.07	&	\cite{btim1983}	\\
26	&	6951.49	&	0.23	&	-0.48	&	\cite{cldh2007}	\\
26	&	6951.19	&	0.74	&	-0.78	&	\cite{smlr1987}	\\
32	&	10574.94	&	0.16	&	-0.33	&	\cite{clsd2009}	\\
54	&	30848.00	&	10.00	&	-8.53	&	\cite{biss1989}	\\
54	&	30850.40	&	5.70	&	-6.13	&	\cite{wbbl2000}	\\
66	&	47045.00	&	38.00	&	15.78	&	\cite{bfli1993}	\\
79	&	69335.96	&	10.69	&	3.48	&	\cite{blbf1994}	\\
79	&	69335.15	&	7.90	&	2.67	&	\cite{bmlg1995}	\\
83	&	77393.00	&	100.00	&	56.29	&	\cite{smgm1992}	\\
92	&	97605.61	&	16.00	&	-6.33	&	\cite{bmlg1995}	\\\hline
\end{tabular}
\end{center}
\end{table}

\begin{table}
\begin{center}
\caption{Comparison between experimental and theoretical \protect\cite{yas2015}  energies for the $1s$ binding energy in hydrogenlike ions.
\protect \cite{gsbb2005} a:  average of the six measurements on Table II of \protect \cite{gsbb2005} (three using Lyman $\alpha$1 and three using the K-RR spectra).
\protect \cite{gsbb2005} b:  average of the three measurements using K-RR spectra  Table II \protect \cite{gsbb2005}, to avoid using the $2p_{3/2}$ QED contribution to deduce the $1s$ binding energy.}
\label{tab:h-bind-comp}
\begin{tabular}{cD{.}{.}{3}D{.}{.}{3}D{.}{.}{3}l}
\hline
\hline
Z	&	\multicolumn{3}{c}{$1s$ binding energy }					&	\multicolumn{1}{c}{Reference}	\\
	&	\multicolumn{1}{c}{Experiment }	&	\multicolumn{1}{c}{Error }	&	\multicolumn{1}{c}{Diff. Th.}	&		\\
\hline									
45	&	-28308.4	&	3.4	&	3.6	&	\cite{nno2003}	\\
79	&	-93257.4	&	7.9	&	-2.8	&	\cite{bmlg1995}	\\
92	&	-131820.2	&	4.2	&	-4.1	&	\cite{gsbb2005} a	\\
92	&	-131825.1	&	6.4	&	-9.0	&	\cite{gsbb2005} b	\\
\hline									
\end{tabular}
\end{center}
\end{table}

\subsubsection{Heliumlike ions}
\label{sec:helike}
In this section, I study the present status of the comparison between theory and experiment for He-like ions. There has been some controversy after the claims in  \cite{ckgh2012,cpgh2014}  of a possible systematic deviation between theory and experiments for the $n=2 \to n=1$ transitions in two-electron ions. It was later shown, with the help of new high-accuracy experimental data and more detailed statistical analysis, that this was not the case \cite{epp2013,bab2015,esbr2015,mssa2018}. Yet it remains important to extend this analysis to more data to disentangle possible effects.
For this reason, I present here a study of both the $n=2 \to n=1$ and the $n=2$, $\Delta n=0$ transitions with measurements extending up to uranium. In the case of He-like ions, there are also a few direct measurements of the ground state energy \cite{mes1995,sem1996}.
There have been many theoretical calculations of two-electron ion energies, as they constitute a mandatory test of many-body techniques. Among them I can cite the work of Drake, who developed the \emph{unified relativistic theory}  to calculate energy levels in He-like ions \cite{dra1979}, with a tabulation for all elements \cite{dra1988}. The MCDF method was also used \cite{gid1987,igd1987,ind1988}. 

\begin{figure*}[htbp]
\centering
        \includegraphics[width=\textwidth]{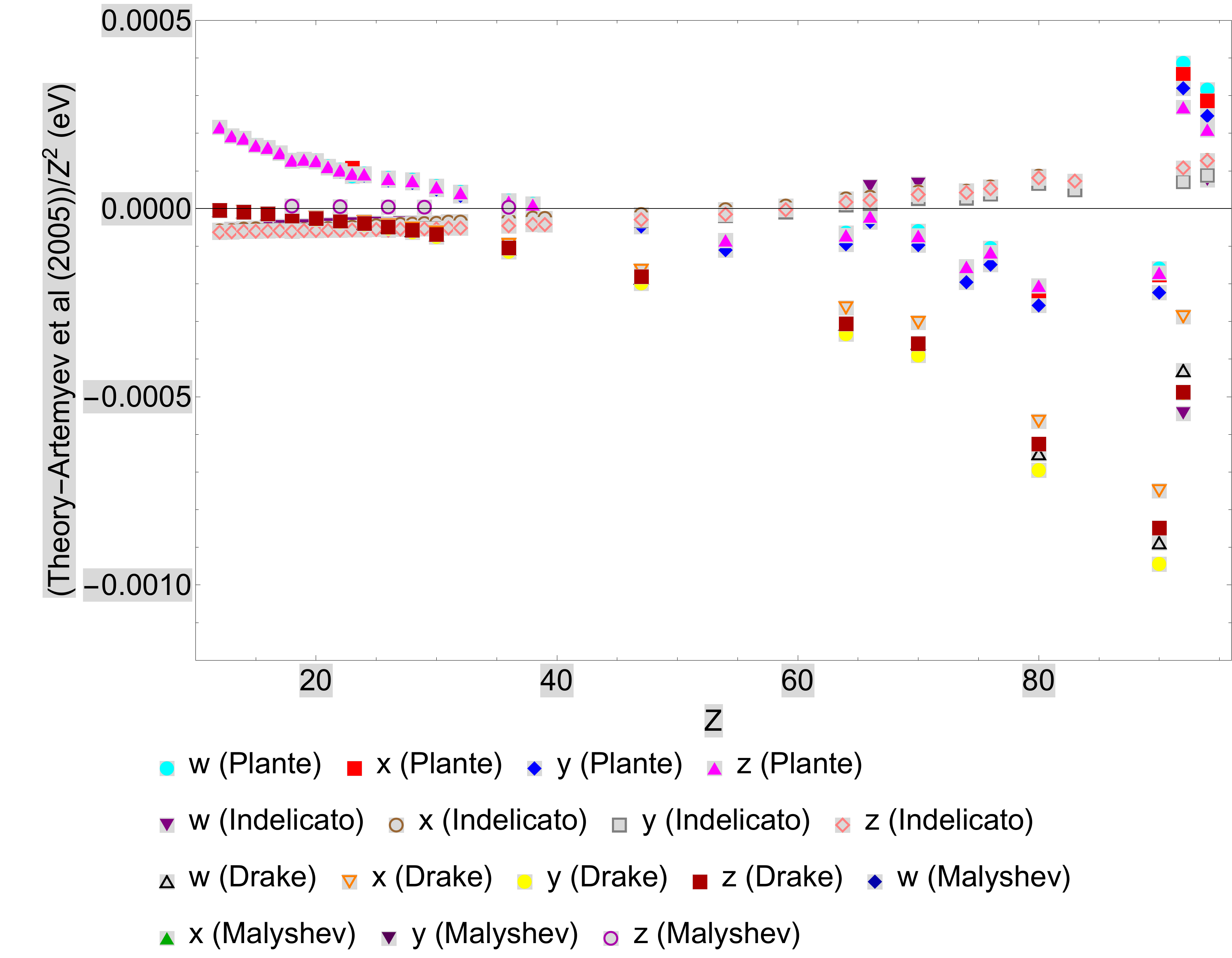}
\caption{Difference between the calculations using the \emph{unified relativistic theory}  \protect \cite{dra1988}, RMBPT, \protect \cite{pjs1994}, the MCDF method \protect \cite{ind1988,ind1992} and  recent improved values  \protect \cite{mkgt2019}, and the calculations of Artemyev \etal \protect  \cite{asyp2005} used as a reference. Differences are divided by $Z^2$.  
\label{fig:he-plante-vs-artemyev}
}
\end{figure*}

As can be seen in figure \ref{fig:he-plante-vs-artemyev}, there are sizable differences at high-$Z$ between the calculations of   \cite{dra1988,ind1988,pjs1994,asyp2005}. The most recent calculations \cite{mkgt2019}, performed for $Z=18$, \num{22}, \num{26}, \num{29}, \num{36} differ only by \SIrange{2}{4}{\milli\electronvolt} from the calculations  of Artemyev \etal \cite{asyp2005}. The MCDF results \cite{ind1992} correspond to the calculation described in \cite{ind1988}, with exact non-relativistic Coulomb correlation energy, relativistic corrections, and magnetic and retardation corrections. The self-energy screening is evaluated using the Welton method. The differences in the most recent calculations come from the changes  in nuclear size, from different treatments of the magnetic and retardation correlations and from the presence of two-loop QED corrections. All these contributions have magnitudes of a few \si{\electronvolt}. Cheng \& Chen \cite{cac2000} have published large scale relativistic configuration interaction (RCI) calculations in which the QED screening and relaxation effects are  evaluated by using Dirac-Kohn-Sham (DKS)  potentials. Yet these calculations have been performed only for $22\leq Z \leq 36$ and cannot be used as a reference calculations for the whole range of $Z$. 

In the case of heliumlike ions, the metastability of some of the excited levels has allowed measurements of $n=2$, $\Delta n=0$ transitions up to uranium. The use of hyperfine quenching \cite{ipm1989}, where the mixing of the adjacent $1s2p\,^3P_0$ and $1s2p\,^3P_1$ levels allows the $1s2p\,^3P_0$, whose transition to the ground state is totally forbidden, to acquire part of the lifetime of the $1s2p\,^3P_1$ with a rate that depends on the square of the energy splitting, has enabled measurements of the $1s2p\,^3P_1-1s2p\,^3P_0$ energy splitting for $Z=28$, \num{47} and \num{64}.  Calculations of this effect can be found in  \cite{ipm1989,agjf1995,jcp1997}. 

All available experimental results for  $n=2$, $\Delta n=0$ transitions or splittings are presented in Table  \ref{tab:helike-exc-3p0-3s1} for the $1s2p\,^3P_0 \to 1s 2s\, ^3S_1$ transition, in Table  \ref{tab:helike-exc-3p2-3s1} for the $1s2p\,^3P_2 \to 1s 2s\, ^3S_1$ transition and in Table  \ref{tab:helike-exc-3p1-3s1} for the $1s2p\,^3P_1 \to 1s 2s\, ^3S_1$  and $1s2p\,^3P_1 \to 1s 2s\, ^1S_0$ transitions and for the  $1s2p\,^3P_1 \to 1s 2s\, ^3P_0$ energy splitting.  In each Table the difference between the experimental energy and theory is displayed. I started these tables at $Z=7$, the minimum value of $Z$ for which the $1s2p\,^1P_1$ level calculations converged in Plante \etal \cite{pjs1994}. For $Z\ge 12$ the difference is evaluated using values from Artemyev \etal \cite{asyp2005} . 

There are a number of cases where relative measurements of transition energies were performed on $n=2 \to n=1$ transitions using one of the w, x, y or z lines as a reference. In that case, it would not make sense to include them in the global analysis of the agreement between theory and experiment for the $n=2 \to n=1$ transitions. However the energy differences are significant as they represent the fine structure interval,  in a range of $Z$ not so well covered by direct measurements. A number of measurements, performed with Tokamaks, ECRIS, or astrophysics plasmas, which have been calibrated against the w or z lines are thus included in Table  \ref{tab:helike-exc-1p1-3s1} for the $1s2p\,^1P_1 \to 1s 2p\, ^3P_J$ and $1s2p\,^1P_1 \to 1s 2s\, ^3S_1$ splitting.

\begin{table}
\begin{center}
\caption{Comparison between experimental and theoretical energies for the $1s2p\,^3P_0 \to 1s 2s\, ^3S_1$ transitions in heliumlike ions. Theory is from Plante
\etal \protect\cite{pjs1994} for $Z<12$ and from  Artemyev \etal \protect\cite{asyp2005} for $Z\geq 12$.}
\label{tab:helike-exc-3p0-3s1}
\begin{tabular}{cD{.}{.}{4}D{.}{.}{4}D{.}{.}{4}l}
\hline
\hline
Z	&	\multicolumn{3}{c}{ $1s2p\,^3P_0 \to 1s 2s\, ^3S_1$}					&	\multicolumn{1}{c}{Reference}	\\
	&	\multicolumn{1}{c}{Exp. (eV)}	&	\multicolumn{1}{c}{Err.}	& 	\multicolumn{1}{c}{Diff. Th.}	&		\\
\hline									
7	&	6.49925	&	0.00014	& 	-0.00030	&	\cite{pss1984}	\\
8	&	7.56033	&	0.00018	& 	-0.00030	&	\cite{saps1981}	\\
8	&	7.56036	&	0.00006	& 	-0.00027	&	\cite{pss1984}	\\
8	&	7.56061	&	0.00037	& 	-0.00003	&	\cite{eas1971}	\\
9	&	8.62799	&	0.00060	& 	-0.00038	&	\cite{saps1981}	\\
9	&	8.62817	&	0.00042	& 	-0.00020	&	\cite{kmmp1985}	\\
9	&	8.62757	&	0.00036	& 	-0.00080	&	\cite{eas1971}	\\
10	&	9.70363	&	0.00015	& 	-0.00004	&	\cite{pss1984}	\\
10	&	9.70340	&	0.00030	& 	-0.00026	&	\cite{hds1993}	\\
10	&	9.70385	&	0.00030	& 	0.00019	&	\cite{eas1971}	\\
12	&	11.88396	&	0.00091	& 	0.00026	&	\cite{kmmp1985}	\\
12	&	11.88404	&	0.00011	& 	0.00034	&	\cite{clwf2000}	\\
13	&	12.9908	&	0.0014	& 	-0.0001	&	\cite{clwf2000}	\\
14	&	14.11028	&	0.00048	& 	-0.00032	&	\cite{hhmd1994}	\\
14	&	14.11124	&	0.00048	& 	0.00064	&	\cite{dab1981}	\\
14	&	14.1082	&	0.0031	& 	-0.0024	&	\cite{lhpd1980}	\\
14	&	14.11023	&	0.00047	& 	-0.00037	&	\cite{clwf2000}	\\
15	&	15.2442	&	0.0011	& 	-0.0005	&	\cite{hsm1995}	\\
16	&	16.3905	&	0.0013	& 	-0.0027	&	\cite{dab1981}	\\
16	&	16.39244	&	0.00087	& 	-0.00076	&	\cite{hsm1996}	\\
16	&	16.3907	&	0.0050	& 	-0.0025	&	\cite{lhpd1980}	\\
17	&	17.5615	&	0.0050	& 	0.0031	&	\cite{dab1981}	\\
17	&	17.5651	&	0.0016	& 	0.0067	&	\cite{lhpd1980}	\\
17	&	17.5651	&	0.0019	& 	0.0067	&	\cite{bdl1978}	\\
18	&	18.766	&	0.031	& 	0.024	&	\cite{dam1977}	\\
18	&	18.74195	&	0.00051	& 	0.00075	&	\cite{klsb1995}	\\
18	&	18.7469	&	0.0011	& 	0.0057	&	\cite{bfs1986}	\\
36	&	44.312	&	0.032	& 	0.063	&	\cite{mdbb1990}	\\
92	&	260.0	&	7.9	& 	8.0	&	\cite{mag1986}	\\
\hline
\end{tabular}
\end{center}
\end{table}

\begin{center}
\begin{longtable}{cD{.}{.}{8}D{.}{.}{8}D{.}{.}{8}l}
\caption{Comparison between experimental and theoretical energies for the $1s2p\,^3P_2 \to 1s 2s\, ^3S_1$ transition in heliumlike ions. Theory is from Plante \etal \protect\cite{pjs1994} for $Z<12$ and from Artemyev \etal \protect\cite{asyp2005} for $Z\geq 12$.}
\label{tab:helike-exc-3p2-3s1}\\
\hline
\hline
	&	\multicolumn{3}{c}{ $1s2p\,^3P_2 \to 1s 2s\, ^3S_1$}					&		\\
$Z$	&	\multicolumn{1}{c}{Exp. (eV)}	&	\multicolumn{1}{c}{Err.}	&	\multicolumn{1}{c}{Exp.-Th.}	& 	\multicolumn{1}{c}{Reference}	\\
\hline			
\endfirsthead
\caption{  $1s2p\,^3P_2 \to 1s 2s\, ^3S_1$  transition  (continued)}\\
\hline
\hline
	&	\multicolumn{3}{c}{ $1s2p\,^3P_2 \to 1s 2s\, ^3S_1$}					&		\\
$Z$	&	\multicolumn{1}{c}{Exp. (eV)}	&	\multicolumn{1}{c}{Err.}	&	\multicolumn{1}{c}{Exp.-Th.}	& 	\multicolumn{1}{c}{Reference}	\\
\hline									
\endhead
\hline 
\endfoot

\hline 
\endlastfoot
7	&	6.536475	&	0.000086	& 	-0.00023	&	\cite{pss1984}	\\
8	&	7.63595	&	0.00019	& 	-0.00037	&	\cite{saps1981}	\\
8	&	7.63615	&	0.00007	& 	-0.00017	&	\cite{pss1984}	\\
8	&	7.63623	&	0.00038	& 	-0.00008	&	\cite{eas1971}	\\
9	&	8.76641	&	0.00037	& 	0.00052	&	\cite{saps1981}	\\
9	&	8.76542	&	0.00043	& 	-0.00047	&	\cite{kmmp1985}	\\
9	&	8.76573	&	0.00037	& 	-0.00016	&	\cite{eas1971}	\\
10	&	9.93403	&	0.00010	& 	-0.00009	&	\cite{pss1984}	\\
10	&	9.933835	&	0.000080	& 	-0.00028	&	\cite{hds1993}	\\
10	&	9.93407	&	0.00016	& 	-0.00004	&	\cite{bfs1986}	\\
10	&	9.93383	&	0.00010	& 	-0.00029	&	\cite{clwf2000}	\\
10	&	9.93368	&	0.00016	& 	-0.00044	&	\cite{eas1971}	\\
11	&	11.15196	&	0.00017	& 	-0.00005	&	\cite{clwf2000}	\\
12	&	12.43099	&	0.00075	& 	0.00109	&	\cite{kmmp1985}	\\
12	&	12.43014	&	0.00024	& 	0.00024	&	\cite{clwf2000}	\\
13	&	13.78169	&	0.00077	& 	0.00049	&	\cite{clwf2000}	\\
14	&	15.21820	&	0.00037	& 	-0.00020	&	\cite{hhmd1994}	\\
14	&	15.21857	&	0.00037	& 	0.00017	&	\cite{dab1981}	\\
14	&	15.2184	&	0.0019	& 	0.0000	&	\cite{lhpd1980}	\\
14	&	15.21786	&	0.00045	& 	-0.00054	&	\cite{clwf2000}	\\
15	&	16.75643	&	0.00057	& 	-0.00047	&	\cite{hsm1995}	\\
16	&	18.41085	&	0.00055	& 	-0.00075	&	\cite{dab1981}	\\
16	&	18.41140	&	0.00055	& 	-0.00020	&	\cite{hsm1996}	\\
16	&	18.4092	&	0.0025	& 	-0.0024	&	\cite{lhpd1980}	\\
16	&	18.388	&	0.020	& 	-0.0236	&	\cite{sbcs2013}	\\
17	&	20.19977	&	0.00066	& 	-0.00053	&	\cite{dab1981}	\\
17	&	20.1995	&	0.0031	& 	-0.0008	&	\cite{lhpd1980}	\\
17	&	20.1986	&	0.0004	& 	-0.0017	&	\cite{bdl1978}	\\
18	&	22.132	&	0.036	& 	-0.009	&	\cite{dam1977}	\\
18	&	22.14225	&	0.00063	& 	0.00095	&	\cite{klsb1995}	\\
18	&	22.1424	&	0.0040	& 	0.0011	&	\cite{bfs1986}	\\
18	&	22.1430	&	0.0110	& 	0.0017	&	\cite{sbcs2013}	\\
22	&	31.8324	&	0.0057	& 	0.0062	&	\cite{glmb1986}	\\
26	&	45.747	&	0.015	& 	0.026	&	\cite{bbdd1981}	\\
28	&	54.7948	&	0.0097	& 	0.0009	&	\cite{zllw1988}	\\
29	&	59.997	&	0.023	& 	0.026	&	\cite{bbdd1985}	\\
36	&	111.596	&	0.030	& 	-0.004	&	\cite{mdbb1990}	\\
54	&	465.06	&	1.74	& 	-0.16	&	\cite{mbbd1989}	\\
92	&	4509.71	&	0.99	& 	-0.32	&	\cite{tkbi2009}	\\
\hline
\end{longtable}
\end{center}

\begin{table}
\begin{center}
\caption{Comparison between experimental and theoretical  energies for the $1s2p\,^3P_1 \to 1s 2s\, ^3S_1$, $1s2p\,^1P_1 \to 1s 2s\, ^3S_1$ and $1s2p\,^3P_1 \to 1s 2s\, ^1S_0$ transitions and for the   $1s2p\,^3P_1 \to 1s 2s\, ^3P_0$ energy splitting in heliumlike ions. Theory is from Plante \etal \protect\cite{pjs1994} for $Z<12$ and from Artemyev \etal \protect\cite{asyp2005}for $Z\geq 12$.}
\label{tab:helike-exc-3p1-3s1}
\begin{tabular}{cD{.}{.}{8}D{.}{.}{8}D{.}{.}{8}l}
\hline
\hline
Z	&	\multicolumn{3}{c}{ $1s2p\,^3P_1 \to 1s 2s\, ^3S_1$}					&	\multicolumn{1}{c}{Reference}	\\
	&	\multicolumn{1}{c}{Exp. (eV)}	&	\multicolumn{1}{c}{Err.}	& 	\multicolumn{1}{c}{Diff. Th.}	&		\\
\hline									
7	&	6.50027	&	0.00014	& 	-0.00036	&	\cite{pss1984}	\\
8	&	7.56758	&	0.00014	& 	-0.00034	&	\cite{saps1981}	\\
8	&	7.56770	&	0.00012	& 	-0.00022	&	\cite{pss1984}	\\
8	&	7.56786	&	0.00037	& 	-0.00007	&	\cite{eas1971}	\\
9	&	8.64712	&	0.00036	& 	0.00000	&	\cite{eas1971}	\\
10	&	9.74091	&	0.00023	& 	-0.00004	&	\cite{pss1984}	\\
10	&	9.74098	&	0.00031	& 		&	\cite{eas1971}	\\
12	&	11.98714	&	0.00070	& 	0.00014	&	\cite{kmmp1985}	\\
13	&	13.14520	&	0.00084	& 	-0.00030	&	\cite{clwf2000}	\\
16	&	16.798	&	0.009	& 	0.005	&	\cite{sbcs2013}	\\
18	&	19.341	&	0.012	& 	-0.045	&	\cite{sbcs2013}	\\
\hline									
	&	\multicolumn{3}{c}{ $1s2p\,^3P_1 \to 1s 2p\, ^3P_2$}					&	\multicolumn{1}{c}{Reference}	\\
\hline									
9	&	0.11873568	&	0.00000012	& 	-0.000028	&	\cite{mmtf1999}	\\
\hline									
	&	\multicolumn{3}{c}{ $1s2p\,^3P_1 \to 1s 2s\, ^3P_0$}					&	\multicolumn{1}{c}{Reference}	\\
	&	\multicolumn{1}{c}{Exp. (eV)}	&	\multicolumn{1}{c}{Err.}	& 	\multicolumn{1}{c}{Diff. Th.}	&		\\
\hline									
7	&	0.00104886	&	0.00000010	& 	-0.00003416	&	\cite{thm1998}	\\
7	&	0.00107505	&	0.00000012	& 	-0.00000796	&	\cite{mhts1996}	\\
12	&	0.1032953	&	0.0000019	& 	-0.0000047	&	\cite{mat2000}	\\
28	&	-2.33	&	0.15	& 	-0.006	&	\cite{dllb1991}	\\
47	&	0.790	&	0.040	& 	-0.013	&	\cite{bbcd1993}	\\
64	&	18.57	&	0.19	& 	-0.02	&	\cite{ibbc1992}	\\
\hline									
	&	\multicolumn{3}{c}{ $1s2p\,^3P_1 \to 1s 2s\, ^1S_0$}					&	\multicolumn{1}{c}{Reference}	\\
\hline									
7	&	0.12228784	&	0.00000012	& 	0.00019852	&	\cite{thm1998}	\\
14	&	0.89647825	&	0.00000074	& 	0.00017825	&	\cite{dcm2008}	\\
14	&	0.896468	&	0.000025	& 	0.000168	&	\cite{ram2002}	\\
\hline															
\end{tabular}
\end{center}
\end{table}

\begin{table}
\begin{center}
\caption{Comparison between experimental and theoretical  energies for the  $1s2p\,^1P_1 \to 1s 2s\, ^3S_1$,  $1s2p\,^1P_1 \to 1s 2p\, ^3P_1$ and $1s2p\,^1P_1 \to 1s 2p\, ^3P_2$  energy splitting in heliumlike ions. Theory is from Artemyev \etal \protect\cite{asyp2005}.}
\label{tab:helike-exc-1p1-3s1}
\begin{tabular}{cD{.}{.}{8}D{.}{.}{8}D{.}{.}{4}l}
\hline
\hline
Z	&	\multicolumn{3}{c}{ $1s2p\,^1P_1 \to 1s 2s\, ^3S_1$}					&	\multicolumn{1}{c}{Reference}	\\
	&	\multicolumn{1}{c}{Exp. (eV)}	&	\multicolumn{1}{c}{Err.}	& 	\multicolumn{1}{c}{Diff. Th.}	&		\\
\hline									
\hline									
18	&	35.419	&	0.011	& 	-0.015	&	\cite{sbcs2013}	\\
18	&	35.37	&	0.16	& 	-0.06	&	\cite{tbbf1985}	\\
21	&	43.11	&	0.29	& 	-1.20	&	\cite{tcdl1985} (blend)	\\
22	&	47.76	&	0.18	& 	0.09	&	\cite{bhzg1985}	\\
23	&	51.60	&	0.20	& 	0.33	&	\cite{tcdl1985}	\\
24	&	55.11	&	0.20	& 	-0.03	&	\cite{tcdl1985}	\\
25	&	59.46	&	0.20	& 	0.15	&	\cite{tcdl1985}	\\
26	&	63.69	&	0.36	& 	-0.13	&	\cite{saf1985}	\\
28	&	73.07	&	0.48	& 	-0.91	&	\cite{hbhg1987}	\\
Z	&	\multicolumn{3}{c}{ $1s2p\,^1P_1 \to 1s 2s\, ^3P_1$}					&	\\
18	&	15.978	&	0.079	& 	-0.070	&	\cite{tbbf1985}	\\
21	&	20.78	&	0.30	& 	-0.01	&	\cite{tcdl1985}	\\
22	&	22.83	&	0.18	& 	0.12	&	\cite{bhzg1985}	\\
23	&	25.50	&	0.20	& 	0.66	&	\cite{tcdl1985}	\\
24	&	27.31	&	0.20	& 	0.09	&	\cite{tcdl1985}	\\
25	&	30.18	&	0.20	& 	0.30	&	\cite{tcdl1985}	\\
26	&	32.79	&	0.79	& 	-0.07	&	\cite{saf1985}	\\
28	&	40.60	&	0.49	& 	0.70	&	\cite{hbhg1987}	\\
Z	&	\multicolumn{3}{c}{ $1s2p\,^1P_1 \to 1s 2s\, ^3P_2$}					&		\\
18	&	13.142	&	0.079	& 	-0.150	&	\cite{tbbf1985}	\\
21	&	15.307	&	0.296	& 	0.067	&	\cite{tcdl1985}	\\
22	&	15.786	&	0.181	& 	-0.058	&	\cite{bhzg1985}	\\
23	&	16.882	&	0.200	& 	0.455	&	\cite{tcdl1985}	\\
24	&	17.290	&	0.200	& 	0.293	&	\cite{tcdl1985}	\\
25	&	17.107	&	0.200	& 	-0.446	&	\cite{tcdl1985}	\\
26	&	17.911	&	0.288	& 	-0.190	&	\cite{saf1985}	\\
28	&	18.641	&	0.489	& 	-0.540	&	\cite{hbhg1987}	\\
\hline															
\end{tabular}
\end{center}
\end{table}

\begin{figure*}[htbp]
\centering
      \includegraphics[width=\textwidth]{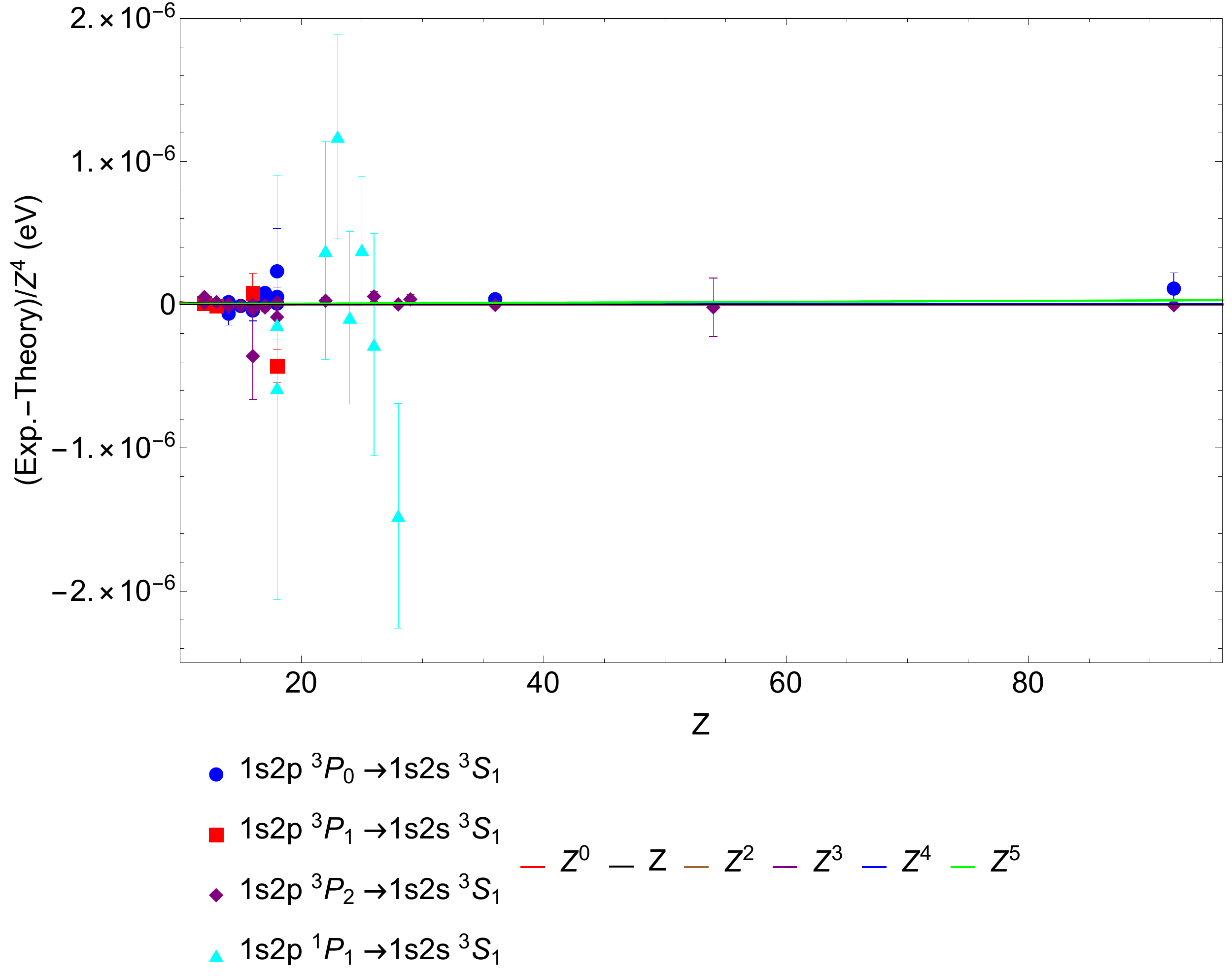}
\caption{Experiment-Theory  \protect \cite{asyp2005}  differences for the $1s 2p \; ^{2S+1}P_J\to 1s 2s \;^3S_1$ intra shell transitions or splittings in two-electron ions. All measurements from Tables  \ref{tab:helike-exc-3p0-3s1},  \ref{tab:helike-exc-3p2-3s1}, \ref{tab:helike-exc-3p1-3s1} and  \ref{tab:helike-exc-1p1-3s1} are shown, except the  $1s 2p \; ^1P_1\to 1s 2s \;^3S_1$ transition in V from \protect \cite{tcdl1985}, which is a blend. The fits to the  differences between  experiments and the calculation of Artemyev \etal \protect \cite{asyp2005} with different functions $b Z^n$, $n=0,\ldots,\, 5$ are also shown. 
The error bands corresponding to $(b\pm \delta b) Z^n$, where $\delta b$ is the $1\sigma$ error on $b$, are also plotted. The fits have been performed on all eight transitions presented in this figure and in figure \ref{fig:he-n2-diff-exp-th-fit2}, which have been separated for better readability.
\label{fig:he-n2-diff-exp-th-fit1}
}
\end{figure*}

\begin{figure*}[htbp]
\centering
       \includegraphics[width=\textwidth]{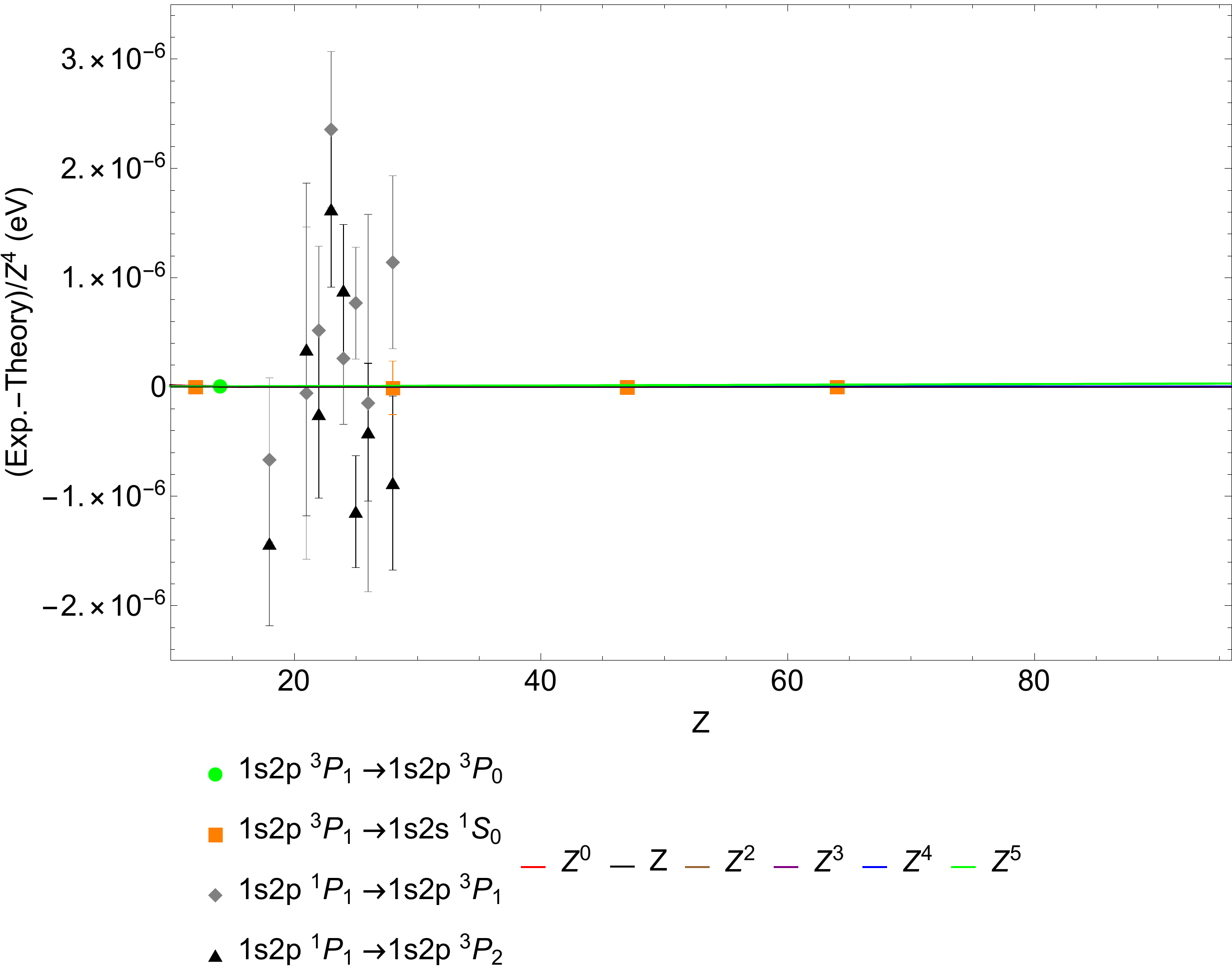}
\caption{Experiment-Theory  \protect \cite{asyp2005}  difference for the $1s 2p \; ^3P_1\to 1s 2p \;^3P_0$,  $1s 2p \; ^3P_1\to 1s 2s \;^1S_0$,  $1s 2p \; ^1P_1\to 1s 2p \;^3P_1$ and  $1s 2p \; ^1P_1\to 1s 2p \;^3P_2$ intra shell transitions or splittings in two-electron ions. All measurements from Tables  \ref{tab:helike-exc-3p0-3s1},  \ref{tab:helike-exc-3p2-3s1}, \ref{tab:helike-exc-3p1-3s1} and  \ref{tab:helike-exc-1p1-3s1} are shown.The fits to the  differences between  experiments and the calculation of Artemyev \etal \protect \cite{asyp2005} with different functions $b Z^n$, $n=0,\ldots,\, 5$ are also shown. 
The error bands corresponding to $(b\pm \delta b) Z^n$, where $\delta b$ is the $1\sigma$ error on $b$, are also plotted.
The fits have been performed on all eight transitions presented in this figure and in figure \ref{fig:he-n2-diff-exp-th-fit1}, which have been separated for better readability.
\label{fig:he-n2-diff-exp-th-fit2}
}
\end{figure*}

\begin{figure*}[htbp]
\centering
        \includegraphics[width=\textwidth]{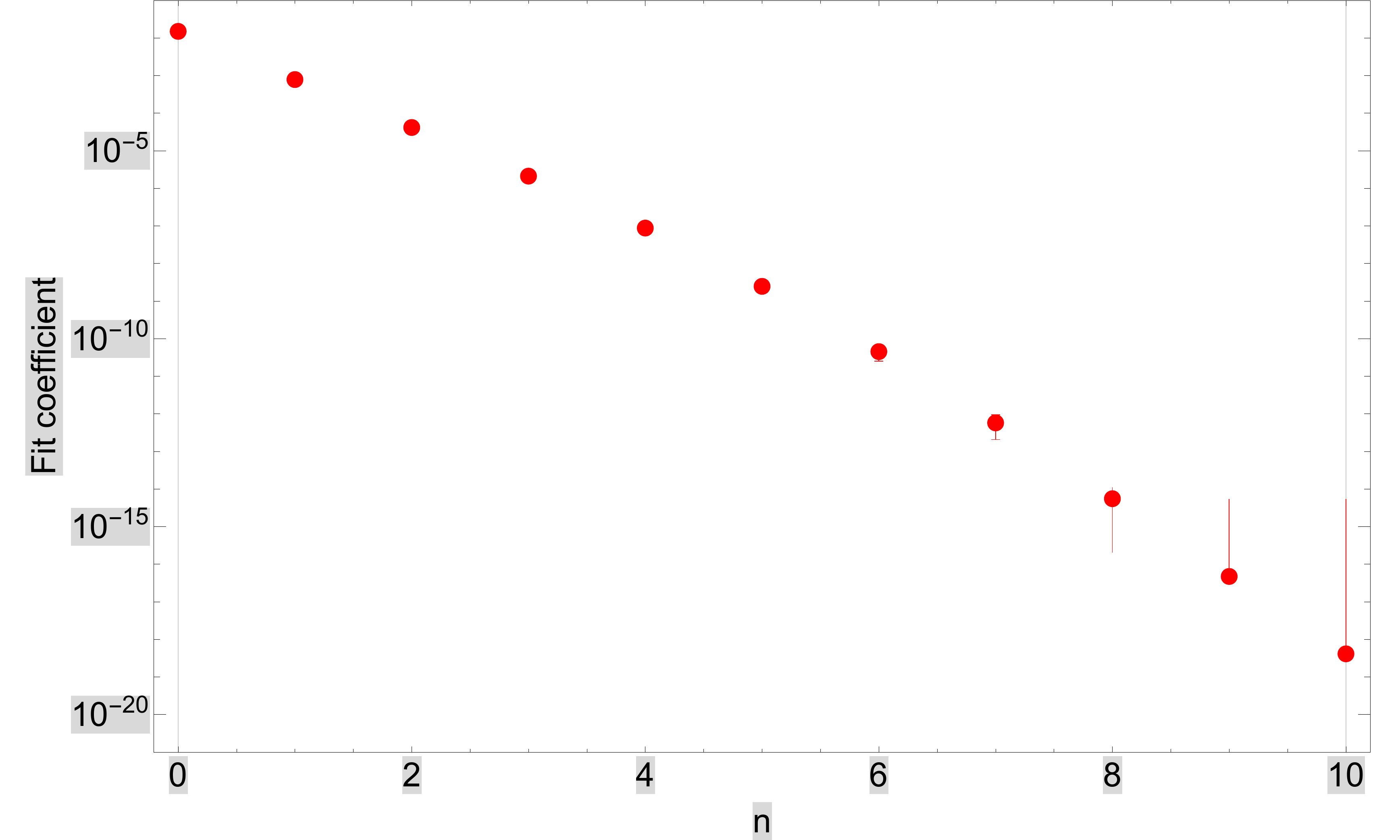}
\caption{
Absolute values of the fit coefficient  $b$ for $n=0,\ldots,\, 10$ with error bars for the differences between  experiments and the calculation of Artemyev \etal \protect \cite{asyp2005} for the  $n=2$ intra shell transitions or splittings in heliumlike ions.
\label{fig:he-n=2-fit-artemyev-diff-exp-coeff}
}
\end{figure*}

The experiment-theory energy differences are displayed in figures \ref{fig:he-n2-diff-exp-th-fit1} and \ref{fig:he-n2-diff-exp-th-fit2} for the various lines and splittings from  Tables  \ref{tab:helike-exc-3p0-3s1},  \ref{tab:helike-exc-3p2-3s1},  \ref{tab:helike-exc-3p1-3s1} and  \ref{tab:helike-exc-1p1-3s1}. 
The figures also show the fits with functions of the type $f(Z)=b Z^n$, to all eight lines or splittings, but each figure shows only four of them for better readability. The errors on the $b$ values for $n>0$ are so small that the band cannot be distinguished from the fitted functions. The values of the coefficients $b$ with their error bars are plotted in figure \ref{fig:he-n=2-fit-artemyev-diff-exp-coeff}.

\begin{figure*}[htbp]
\centering
    \begin{subfigure}{0.8\textwidth}
       \includegraphics[clip, trim=0.1cm 3.5cm 0.0cm 3.5cm, width=\textwidth]{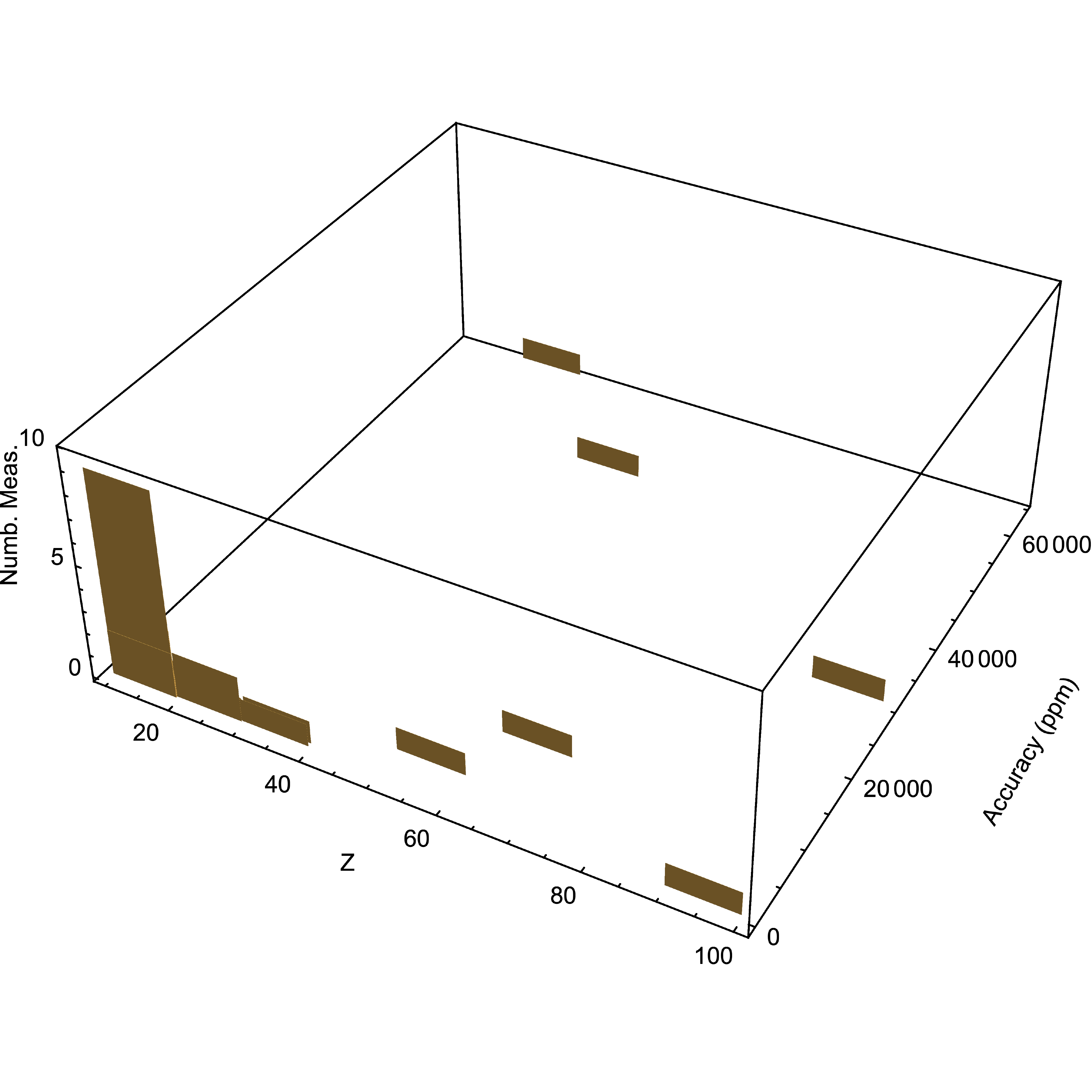}
       \subcaption{Complete view}
    \end{subfigure}
     \begin{subfigure}{0.8\textwidth}
       \includegraphics[clip, trim=0.1cm 4cm 0.1cm 4cm, width=\textwidth]{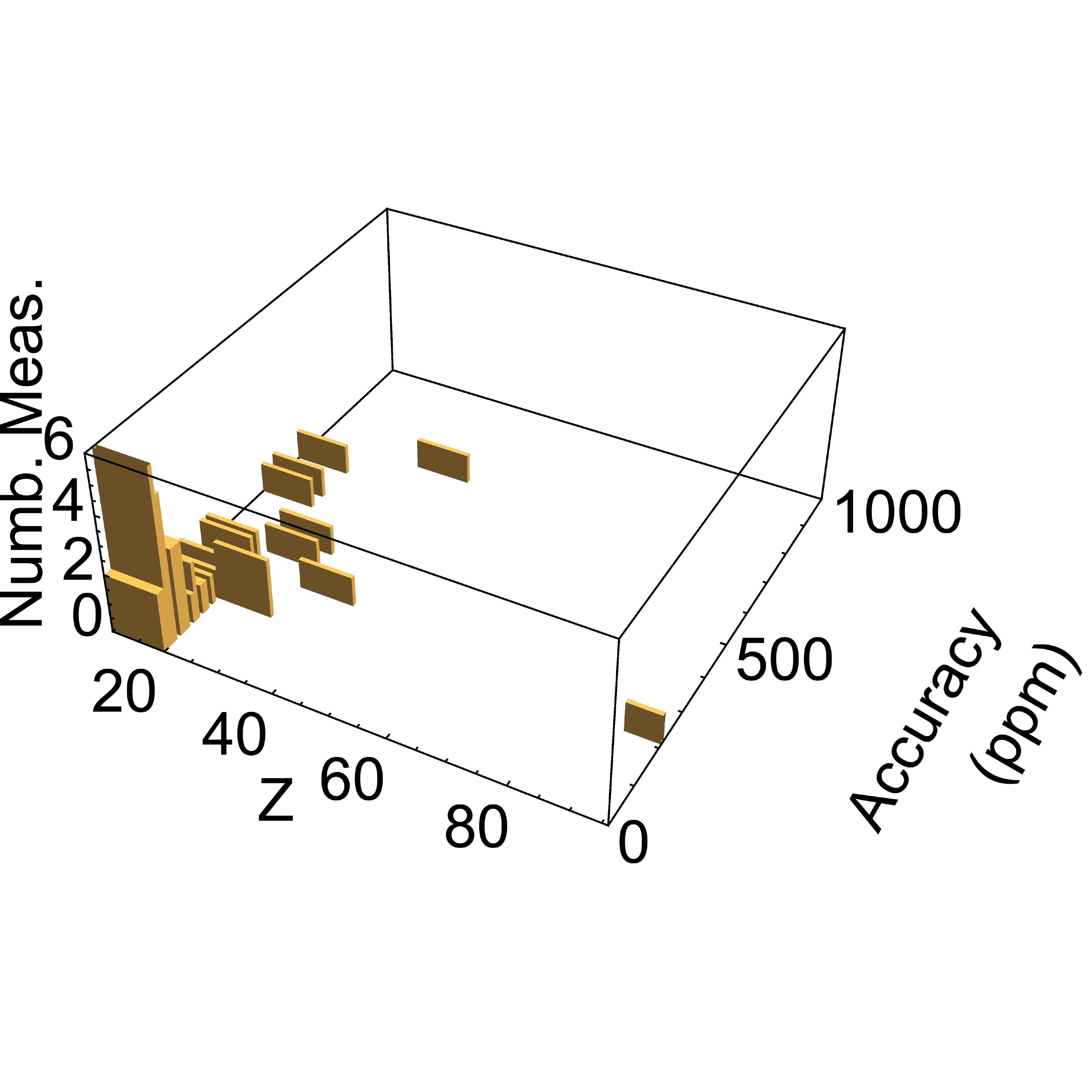}
       \subcaption{Zoomed view on most accurate values}
    \end{subfigure}
\caption{
Histogram of the number of measurements for a given accuracy in parts-per-million and the atomic number $Z$ for the  $n=2$ intra shell transitions or splittings in two-electron ions.
\label{fig:he-z-vs--ppm-histo-n=2}
}
\end{figure*}

All fitted functions tend towards positive energies, contrary to what is observed for the hydrogenlike ions in figure \ref{fig:lh-ly--fit-yero-diff-exp}, though the maximum deviation from \num{0} is only  \SI{2}{\electronvolt} for $Z=100$ and $n=5$.  
This is of the same order of magnitude as the best experimental error bars for uranium measurements. We can then conclude that there is no significant systematic deviation that can be identified with current data and theory.
The histogram of number of measurements of a given accuracy for a given atomic number is plotted in figure \ref{fig:he-z-vs--ppm-histo-n=2}. The situation is quite different from the $n=2 \to n=1$ transitions in H-like ions.
At low-$Z$ there are many very accurate results, some measurements being even performed by laser spectroscopy. But for medium-$Z$ and high-$Z$ elements the relative accuracy can be quite bad, up to approximately 6\%.
There is also a wide range of $Z$ without any accurate measurement available. There is thus a real need to improve on the latest bent-crystal spectroscopy results on uranium \cite{tkbi2009} and extend them to elements in the rare-earth region of the periodic table.

\begin{figure*}[htbp]
\centering
        \includegraphics[width=\textwidth]{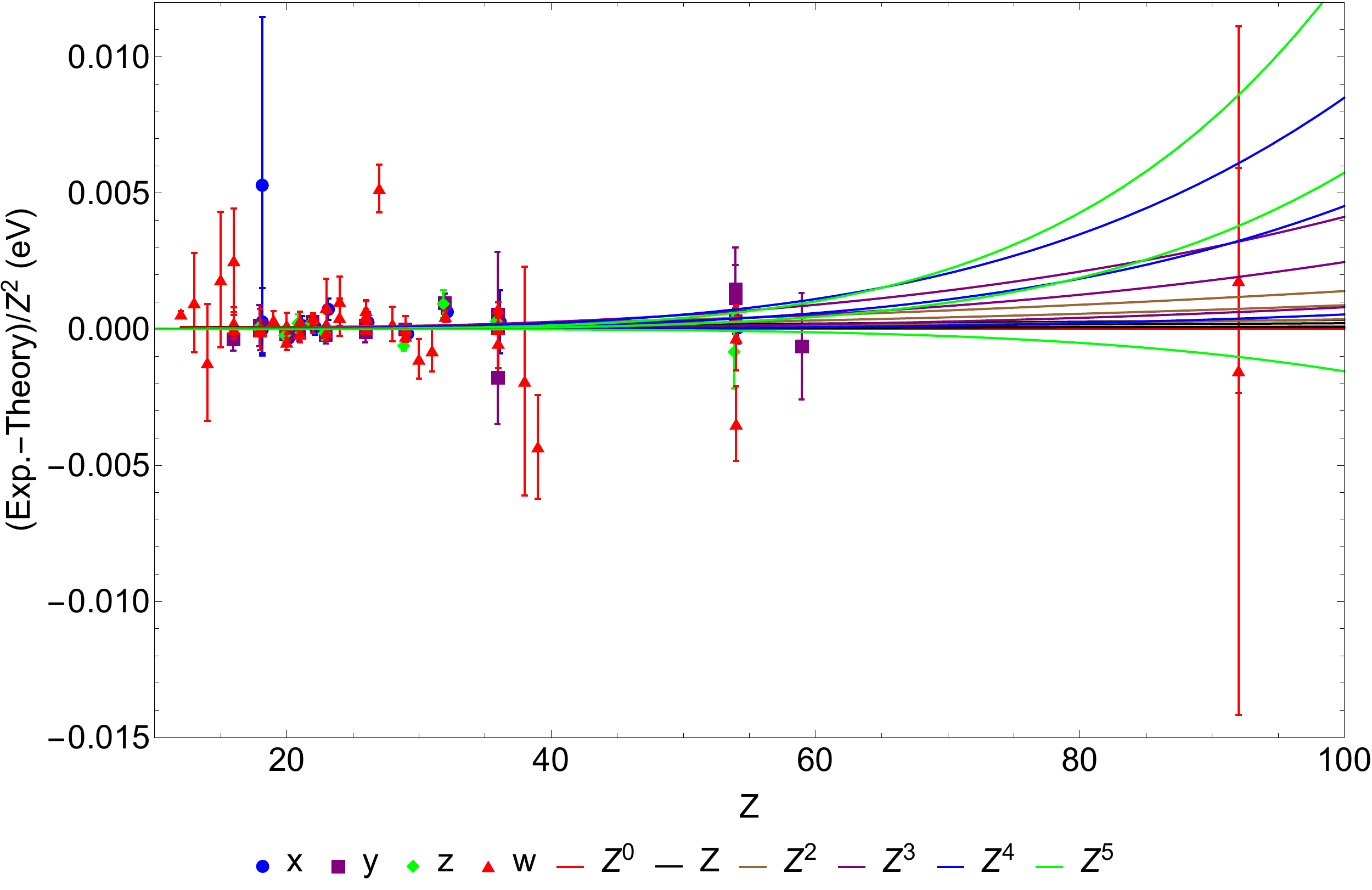}
\caption{
Fit of the  differences between  experiments and the calculations of  Artemyev \etal \protect\cite{asyp2005}  for the w, x, y and z transitions in heliumlike ions with different functions $b Z^n$, $n=0,\ldots,\, 5$.
The error bands corresponding to $(b\pm \delta b) Z^n$, where $\delta b$ is the $1\sigma$ error bar on $b$, are also shown. All the data from Tables \protect \ref{tab:helike-w},  \protect \ref{tab:helike-x}, \protect \ref{tab:helike-y} and  \protect \ref{tab:helike-z} are included.
All energy differences are divided by $Z^2$.
\label{fig:he-wxyz-fit-artemyev-diff-exp}
}
\end{figure*}

\begin{figure*}[htbp]
\centering
        \includegraphics[width=\textwidth]{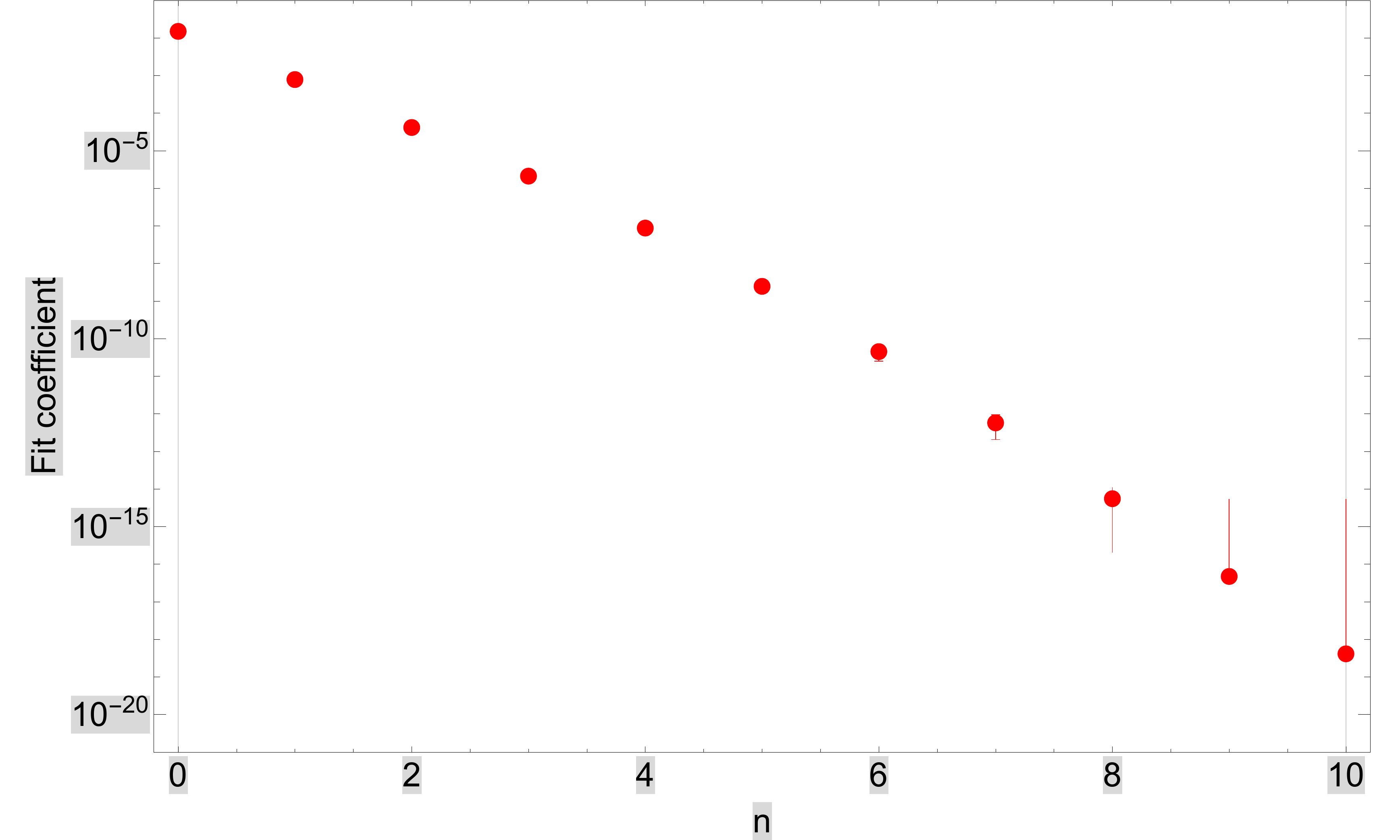}
\caption{
Absolute values of the fit coefficient  $b$ for $n=0,\ldots,\, 10$ with error bars for the differences between  experiments and the calculations of  Artemyev \etal \protect\cite{asyp2005}  for the w, x, y, z transitions in heliumlike ions.
\label{fig:he-wxyz-fit-artemyev-diff-exp-coeff}
}
\end{figure*}

I then show the equivalent analysis for the  $n=2 \to n=1$ transitions. The experimental results for the $1s2p\,^1P_1 \to 1s^2\, ^1S_0$ (w) transition are presented in Table \ref{tab:helike-w}, those for  the $1s2p\,^3P_1 \to 1s^2\, ^1S_0$ (x) transition in Table  \ref{tab:helike-x}, those for  the $1s2p\,^3P_2 \to 1s^2\, ^1S_0$ (y) transition in Table  \ref{tab:helike-y} and those for  the $1s2s\,^3S_1 \to 1s^2\, ^1S_0$ (x) transition in Table  \ref{tab:helike-z}. The label between parentheses represents the conventional notation from Gabriel \cite{gab1972}.
Figure \ref{fig:he-wxyz-fit-artemyev-diff-exp} represents the different data sets displayed in Tables \ref{tab:helike-w} to \ref{tab:helike-z} and the weighted  fits with $b Z^n$, performed in the same conditions as in the case of hydrogenlike ions and of the $\Delta n=0$ transitions. 
The coefficients of the $f(Z)=b Z^n$ function as a function of $n$ are plotted in figure \ref{fig:he-wxyz-fit-artemyev-diff-exp-coeff}.

\begin{figure*}[htbp]
\centering
        \includegraphics[width=\textwidth]{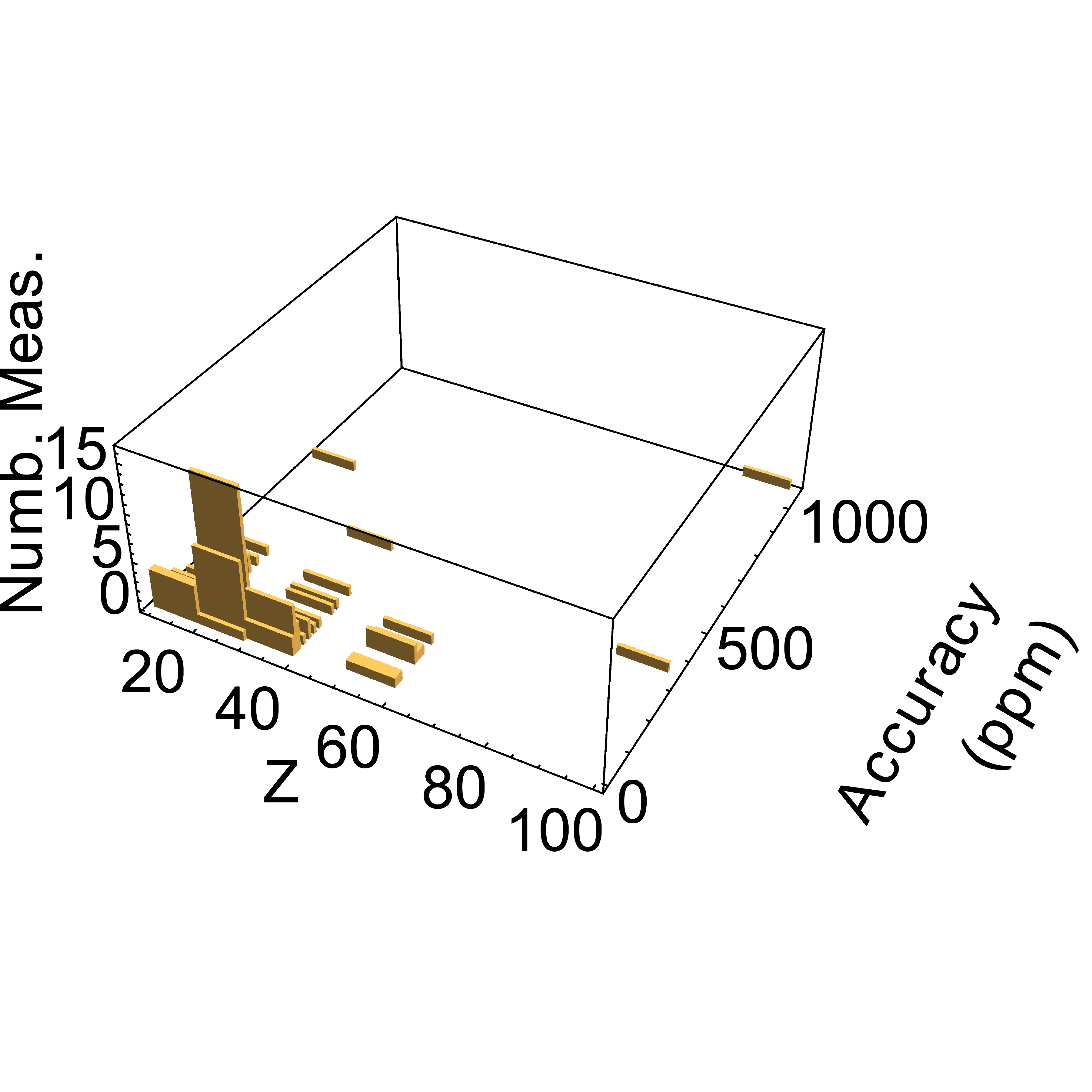}
\caption{
Histogram of the number of measurements for a given accuracy in parts per million and the atomic number $Z$ for the the  w, x, y and z heliumlike lines. All measurements from Tables \protect \ref{tab:helike-w},  \protect \ref{tab:helike-x}, \protect \ref{tab:helike-y} and  \protect \ref{tab:helike-z} are included \cite{dijk2019}.
\label{fig:he-z-vs-ppm-histo}
}
\end{figure*}

\begin{figure*}[htbp]
\centering
        \includegraphics[width=\textwidth]{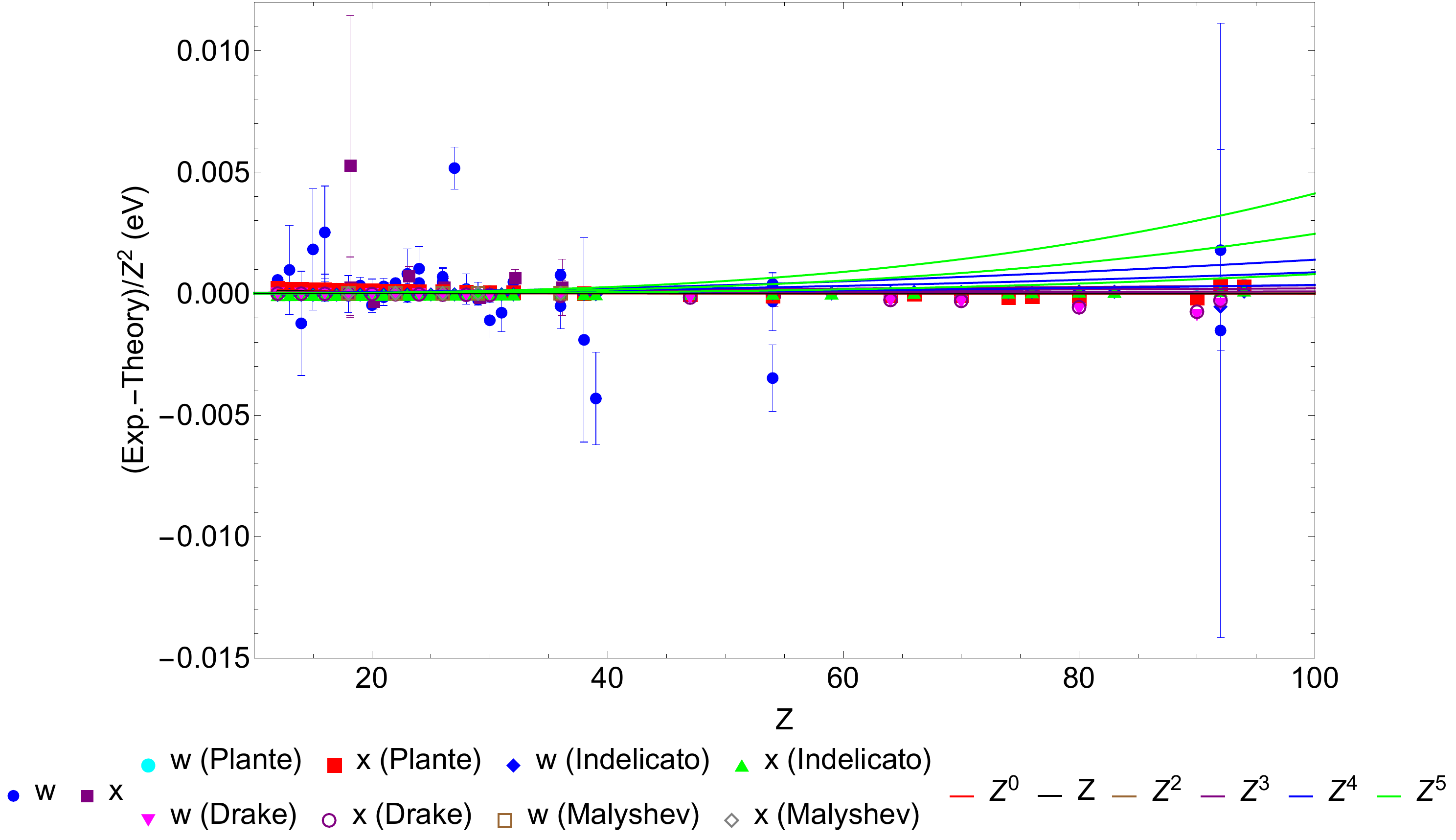}
\caption{
Differences between experiments and  Artemyev \etal \protect\cite{asyp2005}  calculations, compared to the differences between several theoretical results and  \protect \cite{asyp2005}, for the w and x lines. 
(Plante):  \protect \cite{pjs1994}; (Drake):  \protect\cite{dra1988}; (Indelicato):  \protect\cite{ind1988}; (Malyshev):  \protect\cite{mkgt2019}. The fits to the experiment-theory differences, with functions $bZ^n$, $n=0$ to \num{5}, for all four lines are also plotted, together with the error bands. All differences are divided by $Z^2$.
\label{fig:he-like-comp-theories-exp-fits-wx}
}
\end{figure*}

\begin{figure*}[htbp]
\centering
        \includegraphics[width=\textwidth]{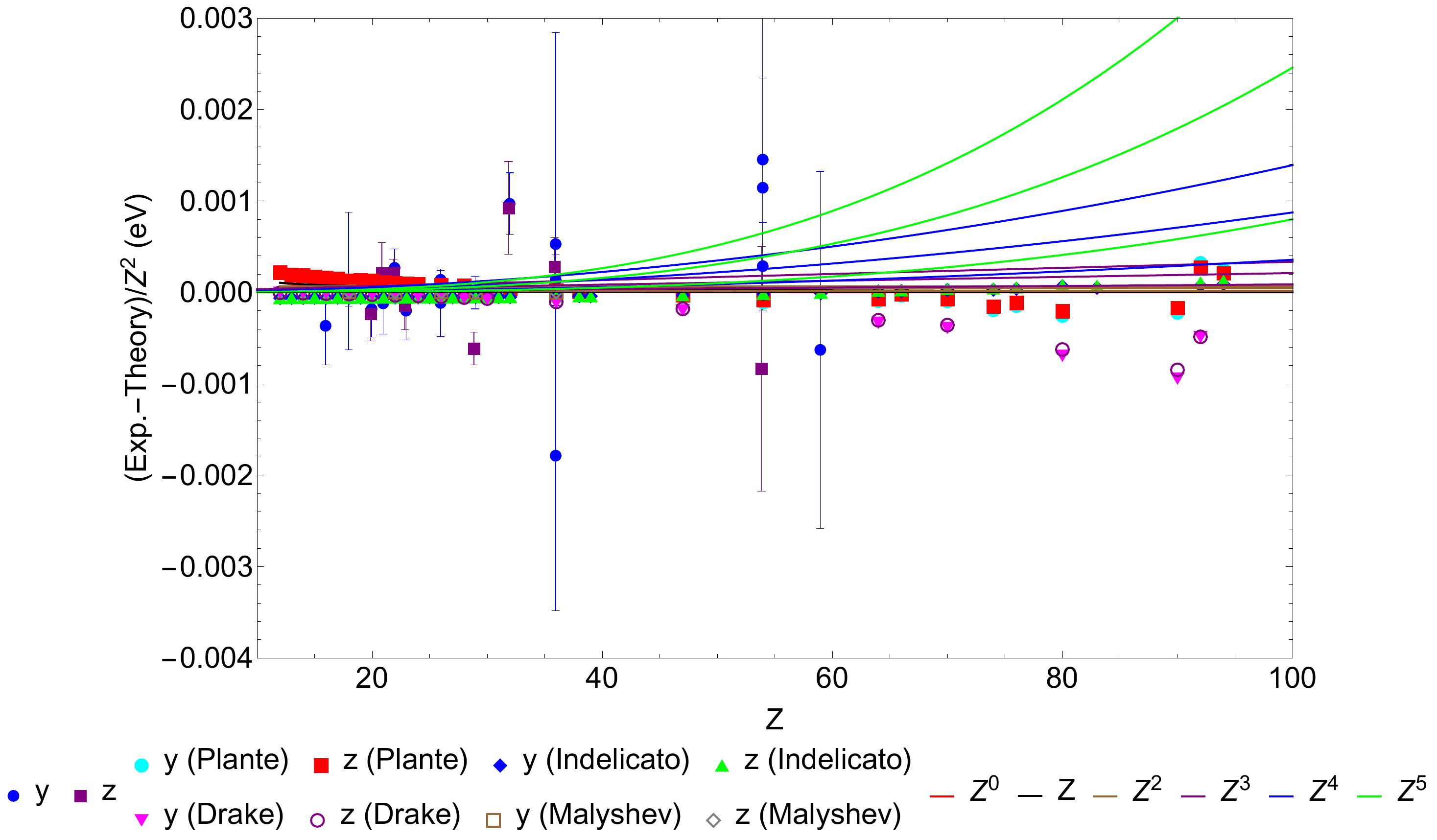}
\caption{Same as figure \ref{fig:he-like-comp-theories-exp-fits-wx}, for the y and z lines.
\label{fig:he-like-comp-theories-exp-fits-yz}
}
\end{figure*}

As in the He-like ion fine structure analysis performed first in this section, the fit error bands are all on the same side of the horizontal axis. This is unlike what was observed in hydogenlike ions. Yet there are no measurements of equivalent relative accuracy.
This is shown in figure \ref{fig:he-z-vs-ppm-histo}, where one can easily see the difference with respect to hydrogenlike ions, with a smaller number of  accurate results at medium and high-$Z$.
To complete this discussion, I have plotted in figures \ref{fig:he-like-comp-theories-exp-fits-wx} and \ref{fig:he-like-comp-theories-exp-fits-yz} the data from figure \ref{fig:he-wxyz-fit-artemyev-diff-exp} and the comparison between the different calculations presented above on figure \ref{fig:he-plante-vs-artemyev}. These figures clearly show that the difference between different theories, some of which do not include second-order QED contributions or use obsolete nuclear sizes, is much smaller than the dispersion among measurements performed for nearby $Z$ or than the error bars for the heaviest ions.

\begin{center}
\begin{longtable}{cD{.}{.}{3}D{.}{.}{3}D{.}{.}{3}l}
\caption{Comparison between experimental and theoretical  energies for the $1s2p\,^1P_1 \to 1s^2\, ^1S_0$ (w) transition in heliumlike ions. Theory is from Plante \etal \protect\cite{pjs1994} for $Z<12$ and from Artemyev \protect\cite{asyp2005}for $Z\geq 12$.}
\label{tab:helike-w}\\
\hline
\hline
	&	\multicolumn{3}{c}{ $1s2p\,^1P_1 \to 1s^2\, ^1S_0$ (w)}					&		\\
$Z$	&	\multicolumn{1}{c}{Exp. (eV)}	&	\multicolumn{1}{c}{Err.}	&	\multicolumn{1}{c}{Exp.-Th.}	& 	\multicolumn{1}{c}{Reference}	\\
\hline	
\hline									
\endfirsthead
\caption{  $1s2p\,^1P_1 \to 1s^2\, ^1S_0$ (w) transition  (continued)}\\
\hline
\hline
	&	\multicolumn{3}{c}{ $1s2p\,^1P_1 \to 1s^2\, ^1S_0$ (w)}					&		\\
$Z$	&	\multicolumn{1}{c}{Exp. (eV)}	&	\multicolumn{1}{c}{Err.}	&	\multicolumn{1}{c}{Exp.-Th.}	& 	\multicolumn{1}{c}{Reference}	\\
\hline									
\endhead
\hline 
\endfoot

\hline 
\endlastfoot
7	&	430.6870	&	0.0030	&	-0.0257	& 	\cite{eal1995}	\\
8	&	573.949	&	0.011	&	-0.032	& 	\cite{eal1995}	\\
11	&	1126.72	&	0.31	&	-0.14	& 	\cite{abzp1974}	\\
12	&	1352.329	&	0.015	&	0.080	& 	\cite{abzp1974}	\\
13	&	1598.46	&	0.31	&	0.16	& 	\cite{abzp1974}	\\
14	&	1864.76	&	0.42	&	-0.24	& 	\cite{abzp1974}	\\
15	&	2152.84	&	0.56	&	0.41	& 	\cite{abzp1974}	\\
16	&	2461.27	&	0.49	&	0.64	& 	\cite{abzp1974}	\\
16	&	2460.69	&	0.15	&	0.06	& 	\cite{aamp1988}	\\
16	&	2460.6300	&	0.021	&	0.001	& 	\cite{kmmu2014}	\\
16	&	2460.6700	&	0.090	&	0.041	& 	\cite{sbbt1982}	\\
18	&	3139.5927	&	0.0080	&	0.0106	& 	\cite{mssa2018}	\\
18	&	3139.5810	&	0.0092	&	-0.0011	& 	\cite{kmmu2014}	\\
18	&	3139.5517	&	0.0366	&	-0.0304	& 	\cite{dbf1984}	\\
18	&	3139.57	&	0.25	&	-0.01	& 	\cite{bmic1983}	\\
19	&	3510.58	&	0.12	&	0.12	& 	\cite{bbgh1989}	\\
20	&	3902.43	&	0.18	&	0.06	& 	\cite{aamp1988}	\\
20	&	3902.19	&	0.12	&	-0.19	& 	\cite{rrag2014}	\\
21	&	4315.54	&	0.15	&	0.13	& 	\cite{bbgh1989}	\\
21	&	4315.35	&	0.15	&	-0.07	& 	\cite{rgtm1995}	\\
22	&	4749.73	&	0.17	&	0.09	& 	\cite{bbgh1989}	\\
22	&	4749.852	&	0.072	&	0.208	& 	\cite{pckg2014}	\\
23	&	5205.592	&	0.546	&	0.427	& 	\cite{aamp1988}	\\
23	&	5205.264	&	0.208	&	0.099	& 	\cite{bbgh1989}	\\
23	&	5205.100	&	0.140	&	-0.065	& 	\cite{cphs2000}	\\
24	&	5682.656	&	0.521	&	0.588	& 	\cite{aamp1988}	\\
24	&	5682.318	&	0.398	&	0.249	& 	\cite{bbgh1989}	\\
26	&	6700.762	&	0.362	&	0.327	& 	\cite{aamp1988}	\\
26	&	6700.725	&	0.201	&	0.291	& 	\cite{bbgh1989}	\\
26	&	6700.441	&	0.049	&	0.006	& 	\cite{kmmu2014}	\\
26	&	6700.90	&	0.25	&	0.47	& 	\cite{btmd1984}	\\
26	&	6700.549	&	0.070	&	0.114	& 	\cite{rbes2013}	\\
27	&	7245.9	&	0.6	&	3.8	& 	\cite{aamp1988}	\\
28	&	7805.75	&	0.49	&	0.14	& 	\cite{aamp1988}	\\
29	&	8391.03	&	0.40	&	0.00	& 	\cite{aamp1988}	\\
29	&	8390.82	&	0.15	&	-0.21	& 	\cite{bab2015}	\\
30	&	8997.53	&	0.65	&	-0.99	& 	\cite{aamp1988}	\\
31	&	9627.45	&	0.75	&	-0.76	& 	\cite{aamp1988}	\\
32	&	10280.70	&	0.22	&	0.48	& 	\cite{mbvk1992}	\\
36	&	13115.45	&	0.30	&	0.98	& 	\cite{itbl1986}	\\
36	&	13114.68	&	0.36	&	0.21	& 	\cite{wbdb1996}	\\
36	&	13114.47	&	0.14	&	0.00	& 	\cite{esbr2015}	\\
36	&	13113.80	&	1.20	&	-0.67	& 	\cite{bitg1984}	\\
38	&	14666.8	&	6.1	&	-2.8	& 	\cite{aamp1988}	\\
39	&	15475.6	&	2.9	&	-6.6	& 	\cite{aamp1988}	\\
54	&	30629.1	&	3.5	&	-1.0	& 	\cite{biss1989}	\\
54	&	30619.9	&	4.0	&	-10.2	& 	\cite{wbbc2000}	\\
54	&	30631.2	&	1.2	&	1.1	& 	\cite{tgbb2009}	\\
92	&	100626	&	35	&	15	& 	\cite{bcid1990}	\\
92	&	100598	&	107	&	-13	& 	\cite{ldhs1994}	\\
\hline
\end{longtable}
\end{center}

\begin{table}
\begin{center}
\caption{Comparison between experimental and theoretical  energies for the  $1s2p\,^3P_2 \to 1s^2\, ^1S_0$ (x) transitions in heliumlike ions. Theory is from Plante \etal \protect\cite{pjs1994} for $Z<12$ and from Artemyev \etal \protect\cite{asyp2005} for $Z\geq 12$.}
\label{tab:helike-x}
\begin{tabular}{cD{.}{.}{4}D{.}{.}{4}D{.}{.}{4}l}
\hline
\hline
	&	\multicolumn{3}{c}{ $1s2p\,^3P_2 \to 1s^2\, ^1S_0$ (x)}					&		\\
$Z$	&	\multicolumn{1}{c}{Exp. (eV)}	&	\multicolumn{1}{c}{Err.}	&	\multicolumn{1}{c}{Exp.-Th.}	& 	\multicolumn{1}{c}{Reference}	\\
\hline									
18	&	3128.0	&	2.0	&	1.7	& 	\cite{dlp1978}	\\
18	&	3126.283	&	0.036	&	-0.007	& 	\cite{dbf1984}	\\
18	&	3126.37	&	0.40	&	0.08	& 	\cite{bmic1983}	\\
20	&	3887.63	&	0.12	&	-0.13	& 	\cite{rrag2014}	\\
21	&	4300.23	&	0.15	&	0.06	& 	\cite{rgtm1995}	\\
22	&	4733.83	&	0.13	&	0.03	& 	\cite{pckg2014}	\\
23	&	5189.12	&	0.21	&	0.38	& 	\cite{cphs2000}	\\
26	&	6682.50	&	0.25	&	0.17	& 	\cite{btmd1984}	\\
29	&	8371.17	&	0.15	&	-0.15	& 	\cite{bab2015}	\\
32	&	10259.52	&	0.37	&	0.64	& 	\cite{mbvk1992}	\\
36	&	13091.17	&	0.37	&	0.30	& 	\cite{wbdb1996}	\\
36	&	13091.2	&	1.5	&	0.33	& 	\cite{bitg1984}	\\
54	&	30594.5	&	1.7	&	0.1	& 	\cite{tgbb2009}	\\
\hline															
\end{tabular}
\end{center}
\end{table}

\begin{table}
\begin{center}
\caption{Comparison between experimental and theoretical  energies for the  $1s2p\,^3P_1 \to 1s^2\, ^1S_0$ (y)  transitions in heliumlike ions. Theory is from Plante \etal \protect\cite{pjs1994} for $Z<12$ and from Artemyev \etal \protect\cite{asyp2005} for $Z\geq 12$.}
\label{tab:helike-y}
\begin{tabular}{cD{.}{.}{4}D{.}{.}{4}D{.}{.}{4}l}
\hline
\hline
	&	\multicolumn{3}{c}{ $1s2p\,^3P_1 \to 1s^2\, ^1S_0$ (y)}					&		\\
$Z$	&	\multicolumn{1}{c}{Exp. (eV)}	&	\multicolumn{1}{c}{Err.}	&	\multicolumn{1}{c}{Exp.-Th.}	& 	\multicolumn{1}{c}{Reference}	\\
\hline									
16	&	2447.05	&	0.11	&	-0.09	& 	\cite{sbbt1982}	\\
18	&	3123.521	&	0.036	&	-0.01	& 	\cite{dbf1984}	\\
18	&	3123.57	&	0.24	&	0.04	& 	\cite{bmic1983}	\\
20	&	3883.24	&	0.12	&	-0.07	& 	\cite{rrag2014}	\\
21	&	4294.57	&	0.15	&	-0.05	& 	\cite{rgtm1995}	\\
22	&	4727.07	&	0.10	&	0.13	& 	\cite{pckg2014}	\\
23	&	5180.22	&	0.17	&	-0.11	& 	\cite{cphs2000}	\\
26	&	6667.5	&	0.25	&	-0.08	& 	\cite{btmd1984}	\\
26	&	6667.67	&	0.07	&	0.09	& 	\cite{rbes2013}	\\
29	&	8346.99	&	0.15	&	0.00	& 	\cite{bab2015}	\\
32	&	10221.79	&	0.35	&	0.99	& 	\cite{mbvk1992}	\\
36	&	13026.8	&	3	&	0.68	& 	\cite{itbl1986}	\\
36	&	13026.29	&	0.36	&	0.17	& 	\cite{wbdb1996}	\\
36	&	13026.15	&	0.14	&	0.03	& 	\cite{esbr2015}	\\
36	&	13023.8	&	2.2	&	-2.32	& 	\cite{bitg1984}	\\
54	&	30209.6	&	3.5	&	3.33	& 	\cite{biss1989}	\\
54	&	30210.5	&	4.5	&	4.23	& 	\cite{wbbc2000}	\\
54	&	30207.1	&	1.4	&	0.83	& 	\cite{tgbb2009}	\\
59	&	36389.1	&	6.8	&	-2.19	& 	\cite{tbcc2008}	\\
\hline															
\end{tabular}
\end{center}
\end{table}

\begin{table}
\begin{center}
\caption{Comparison between experimental and theoretical  energies for the  $1s2p\,^3S_1 \to 1s^2\, ^1S_0$ (z)  transitions in heliumlike ions. Theory is from  Plante \etal \protect\cite{pjs1994} for $Z<12$ and from Artemyev \etal \protect\cite{asyp2005} for $Z\geq 12$.}
\label{tab:helike-z}
\begin{tabular}{cD{.}{.}{4}D{.}{.}{4}D{.}{.}{4}l}
\hline
\hline
	&	\multicolumn{3}{c}{ $1s2p\,^3S_1 \to 1s^2\, ^1S_0$ (z)}					&		\\
$Z$	&	\multicolumn{1}{c}{Exp. (eV)}	&	\multicolumn{1}{c}{Err.}	&	\multicolumn{1}{c}{Exp.-Th.}	& 	\multicolumn{1}{c}{Reference}	\\
\hline									
18	&	3104.161	&	0.008	&	0.012	& 	\cite{asgl2012}	\\
20	&	3861.11	&	0.12	&	-0.09	& 	\cite{rrag2014}	\\
21	&	4271.19	&	0.15	&	0.09	& 	\cite{rgtm1995}	\\
22	&	4702.08	&	0.07	&	0.10	& 	\cite{pckg2014}	\\
23	&	5153.82	&	0.14	&	-0.08	& 	\cite{cphs2000}	\\
29	&	8310.83	&	0.15	&	-0.52	& 	\cite{bab2015}	\\
32	&	10181.33	&	0.52	&	0.95	& 	\cite{mbvk1992}	\\
36	&	12979.63	&	0.41	&	0.36	& 	\cite{wbdb1996}	\\
54	&	30126.70	&	3.90	&	-2.44	& 	\cite{wbbc2000}	\\
\hline															
\end{tabular}
\end{center}
\end{table}

There are other measurements available, performed either by beam-foil spectroscopy or in plasmas, using higher excitations. For  beam foil spectroscopy examples, one can cite the measurement of the $n=4\to n=1$  and $n=3\to n=1$ transitions in He-like iron, performed at the Super-HILAC in Berkeley  \cite{igtb1986}. The $1s 3p\,^1P_1\to 1s^2 \,^1S_0$ line has been observed in potassium and the  $1s np\to 1s^2 \,^1S_0, n=5,\,\ldots 10$ in argon have been observed and measured relative to the Lyman $\alpha_1$ of calcium \cite{saf1985b}. 

Finally, I present in Table \ref{tab:helike-ground-comp} the direct measurements of the ground state energy of He-like ions, performed at the Livermore Super-EBIT  and their comparison with theory. 
The agreement with theory is quite good, with deviations at most equal to $1.5\sigma$. 

\begin{table}
\begin{center}
\caption{Comparison between experimental and theoretical \protect\cite{asyp2005}  energies for the ground state  of heliumlike ions. 
The ionization energies provided in Artemyev \etal \protect\cite{asyp2005} have been transformed into total binding energies by subtracting the $1s$ binding energy of the corresponding ion, using values from Yerokhin \& Shabaev \protect \cite{yas2015}.}
\label{tab:helike-ground-comp}
\begin{tabular}{cD{.}{.}{1}D{.}{.}{1}D{.}{.}{1}l}
\hline
\hline
Z	&	\multicolumn{1}{c}{Exp. (eV)}	&	\multicolumn{1}{c}{Err. (eV)}	&	\multicolumn{1}{c}{Diff. Th.(eV)}	&	\multicolumn{1}{c}{Reference}	\\
\hline
32	&	13556.9	&	1.6	&	-0.5	&	\cite{sem1996}	\\
36	&	17294.4	&	1.7	&	-2.0	&	\cite{mlcn2008}	\\
54	&	40272.7	&	3.5	&	1.0	&	\cite{sem1996}	\\
66	&	61731.6	&	4.3	&	-4.9	&	\cite{sem1996}	\\
74	&	79188	&	15	&	6	&	\cite{sem1996}	\\
76	&	84006	&	20	&	30	&	\cite{sem1996}	\\
83	&	102257	&	14	&	6	&	\cite{sem1996}	\\
92	&	129568.1	&	9.0	&	-2.2	&	\cite{gsbb2004}	\\
\hline
\end{tabular}
\end{center}
\end{table}



\subsubsection{Lithiumlike ions}
\label{sec:lilike}

The   $1s^2 2p \,^2P_{J} \to 1s^2 2s\,^2S_{1/2}$ transitions in Li-like ions are the next logical candidates to test relativistic  few-body effects and QED in few-electron systems. They have been studied both experimentally and theoretically by many authors. One of the reasons is that on the experimental side, it is possible to create and excite the fundamental state $1s^2 2s\,^2S_{1/2}$ in heavy ion sources (EBIT), which has not been possible up to now for one and two-electron ions because of the low rate of production in the heaviest elements \cite{mek1994,mar1996,beos1996}. The second reason is that the energies of the transitions can be measured very accurately with deep UV or low-energy x-ray spectrometers or using dielectronic recombination at storage rings like CRYRING  \cite{zslg1997}, TSR \cite{nlkb2004,kssm2004} or the ESR \cite{bkms2003}. We thus have many more measurements, with higher accuracy, since in particular they are not affected by the Doppler shift when the ion is produced in an ion source. 

Theoretical calculations have been performed with a variety  of methods. Johnson \etal \cite{jbs1988} performed a RMBPT calculation up to third order, but without QED. Later  Seely \cite{see1989} completed the calculations by adding approximate QED obtained with the GRASP MCDF package from Dyall \etal \cite{dgjp1989}.  Indelicato \& Desclaux \cite{iad1990} made a MCDF calculation using an independently developed MCDF code \cite{des1975}, using the Welton method for the self-energy screening.   Kim \etal \cite{kbid1991} combined the RMBPT correlation from  \cite{jbs1988} with the single configuration values obtained from the MCDF code developed in Indelicato \& Desclaux \cite{iad1990}. 
Ynnerman \etal \cite{yjlp1994} implemented a coupled-cluster approach with single and double excitations, and including higher-order Breit contributions and  QED corrections from \cite{plss1993} for uranium. 
Chen \etal  \cite{ccjs1995} also did calculations with the coupled cluster method, using a BSpline basis set for the Dirac equation with higher-order Breit contributions and nuclear deformation. This work was later extended to more elements, with third order MBPT correlations \cite{jls1996}. Safronova \& Shlyaptseva \cite{sas1996} used the $1/Z$ expansion method up to order $1/Z^3$, with relativistic and QED corrections, for $6\leq Z \leq 64$. Few-electron ions, $2p\to2s$ transition energies, from Li-like to Ne-like Bi, Th and U were calculated using the MCDF and RMBPT methods  \cite{smpl1998}.  The Saint Petersburg group made successive calculations with advanced QED corrections, self-energy screening, two-loop diagrams and two-photon exchange diagrams, completed with higher order correlations from many-body techniques \cite{yass2000,yis2006,yas2007,kvag2010}. In particular Yerokhin \etal \cite{yas2007} have calculated the electronic structure part for $4 \leq Z \leq  92$, and final values with loop QED corrections for a few elements. These calculations have been recently improved and extended to other, core-excited Li-like levels  for $6 \leq Z \leq 36$ \cite{yas2012,ysm2017} and up to $Z=92$ more recently \cite{yas2018}.  Gu \cite{gu2005} combined RMBPT and RCI techniques to evaluate transition energies for $1s^2 nl$, $n=2$, to \num{6}.  Cheng \etal \cite{ccs2000} and Sapirstein \& Cheng \cite{sac2001} combined RCI and QED calculations with  Dirac-Kohn-Sham (DKS) potentials. Finally the S-matrix method was used together with the DKS potential to perform a  calculation for all $10\leq Z \leq 100$ \cite{sac2011}. I will use the latter calculation as reference theory.
  
  The possibility of making accurate measurements of the $1s^2 2p \,^2P_{J} \to 1s^2 2s\,^2S_{1/2}$  transitions has also lead to the use of three-electron ions to measure and compare isotopic shifts and deduce the difference in nuclear radii between different isotopes, which theory can predict much more accurately for few-electron ions than in neutral atoms. This method was used for three isotopes of Nd$^{57+}$  \cite{bkhm2008}. The isotope shift between Th and U isotopes was also measured using an EBIT and by comparing transition energies in Li-like, Be-like, B-like and C-like ions \cite{ebc1996}. Calculations of the isotope shift on the $1s^2 2p \,^2P_{J} \to 1s^2 2s\,^2S_{1/2}$ transitions for several elements have been performed in \cite{lngf2012} and \cite{zkst2014}.

\begin{figure*}[htbp]
\centering
  \begin{subfigure}{\textwidth}
        \includegraphics[width=\textwidth]{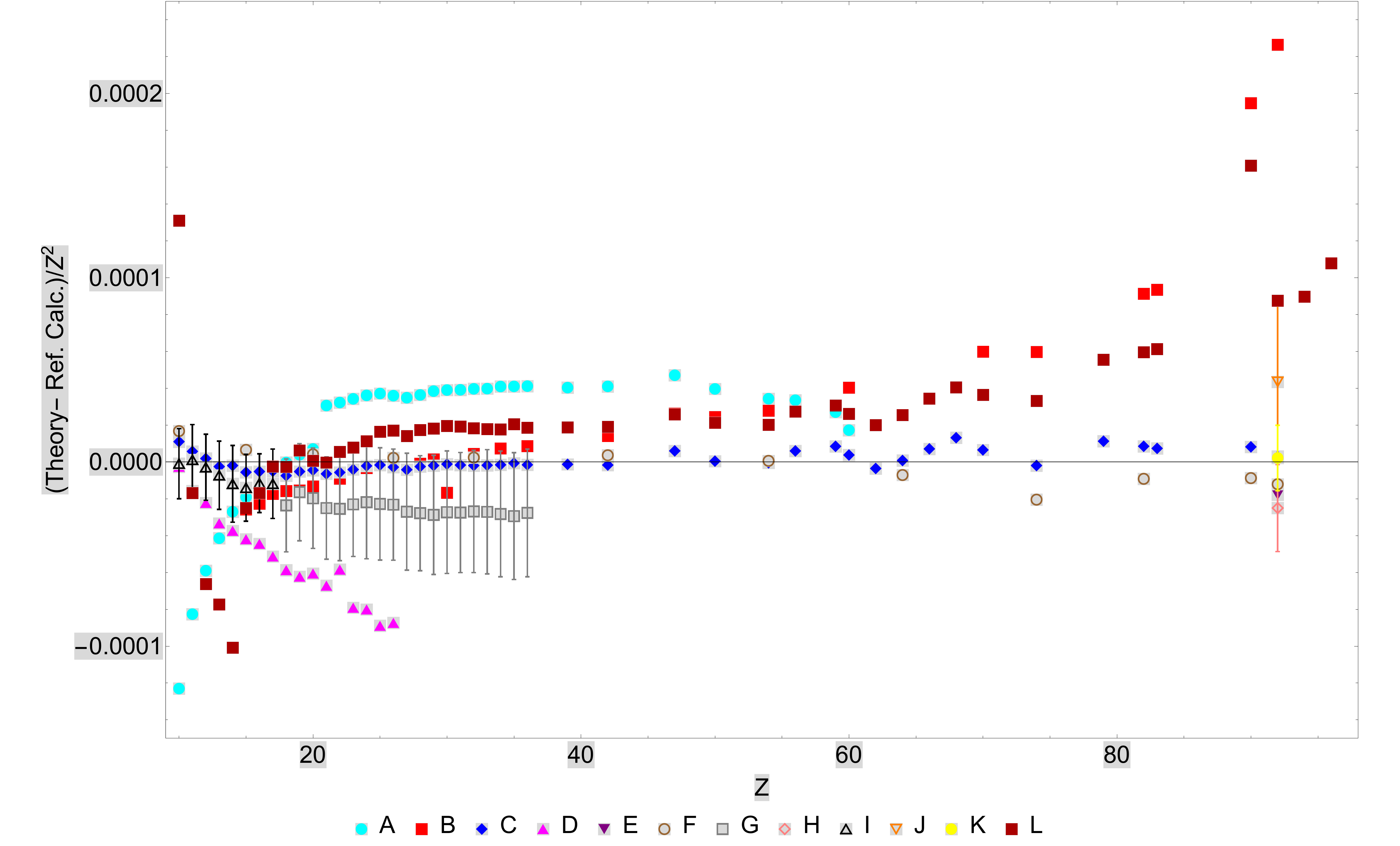}
       \subcaption{Comparison with calculations that do not include second order QED corrections}
    \end{subfigure}
  \begin{subfigure}{\textwidth}
        \includegraphics[width=\textwidth]{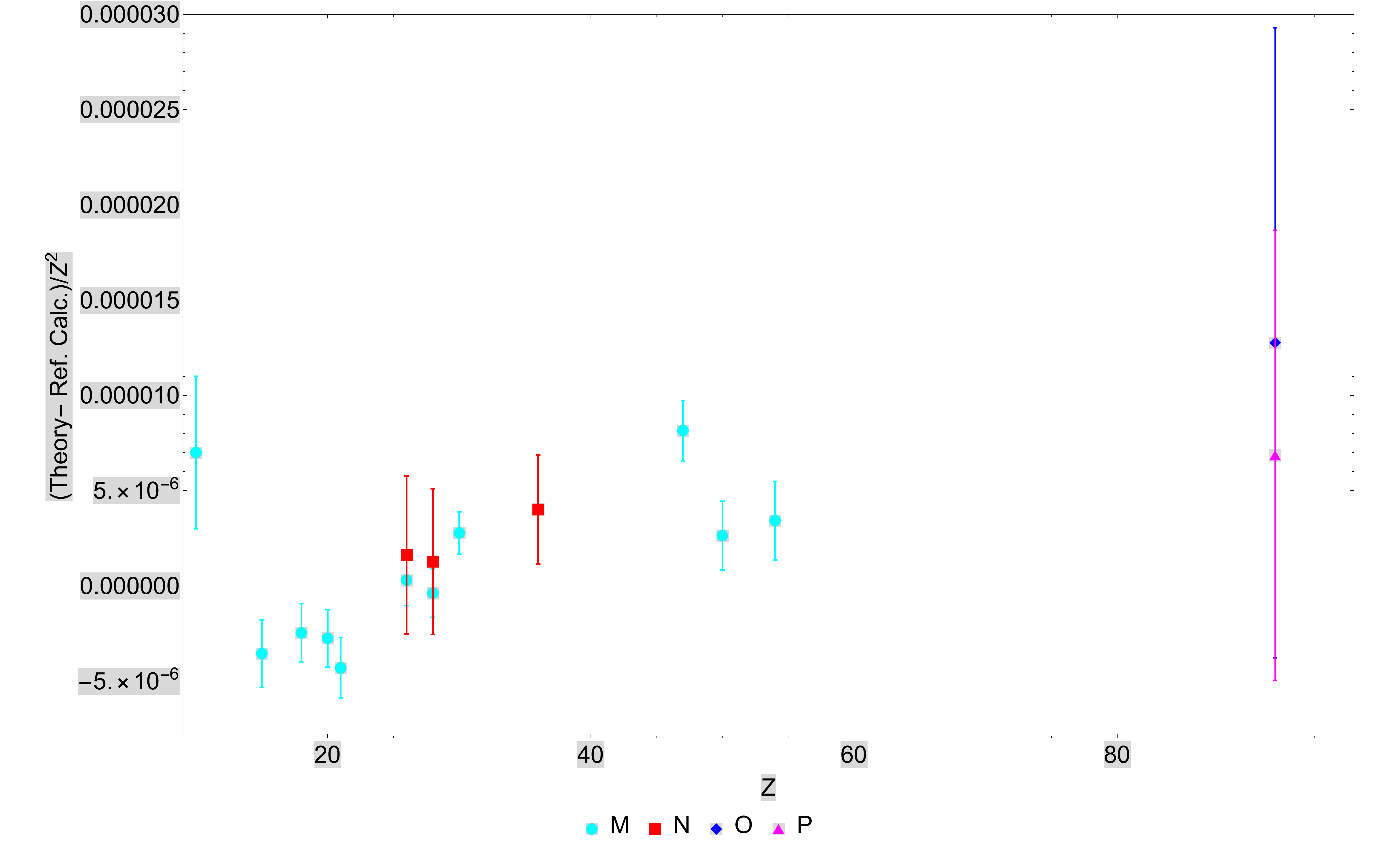}
       \subcaption{Comparison with calculations that include second order QED corrections}
    \end{subfigure}%
\caption{
Comparison between theoretical calculations  and  the work of Sapirstein \& Cheng  \protect \cite{sac2011} for the $1s^2 2p\,^2P_{1/2} \to 1s^2 2s\, ^2S_{1/2}$ transition in lithiumlike ions.
The calculations from  Seely \protect \cite{see1989} have not been plotted as their behavior at high-$Z$ is too different from the more recent results.
References:
A: \protect \cite{gu2005},
B: \protect \cite{iad1990},
C: \protect \cite{kbid1991},
D: \protect \cite{sas1996},
E: \protect \cite{yjlp1994},
F: \protect \cite{ccjs1995},
G: \protect \cite{yas2012},
H: \protect \cite{yass2000},
I: \protect \cite{ysm2017},
J: \protect \cite{bjs1990},
K: \protect \cite{cjs1991},
L: \protect \cite{jls1996}
M: \protect \cite{kvag2010},
N: \protect \cite{yas2007},
O: \protect \cite{yis2006}
P: \protect \cite{kast2008}
\label{fig:li-2s-2p1-theo-sapir}
}
\end{figure*}

\begin{figure*}[htbp]
\centering
  \begin{subfigure}{\textwidth}
        \includegraphics[width=\textwidth]{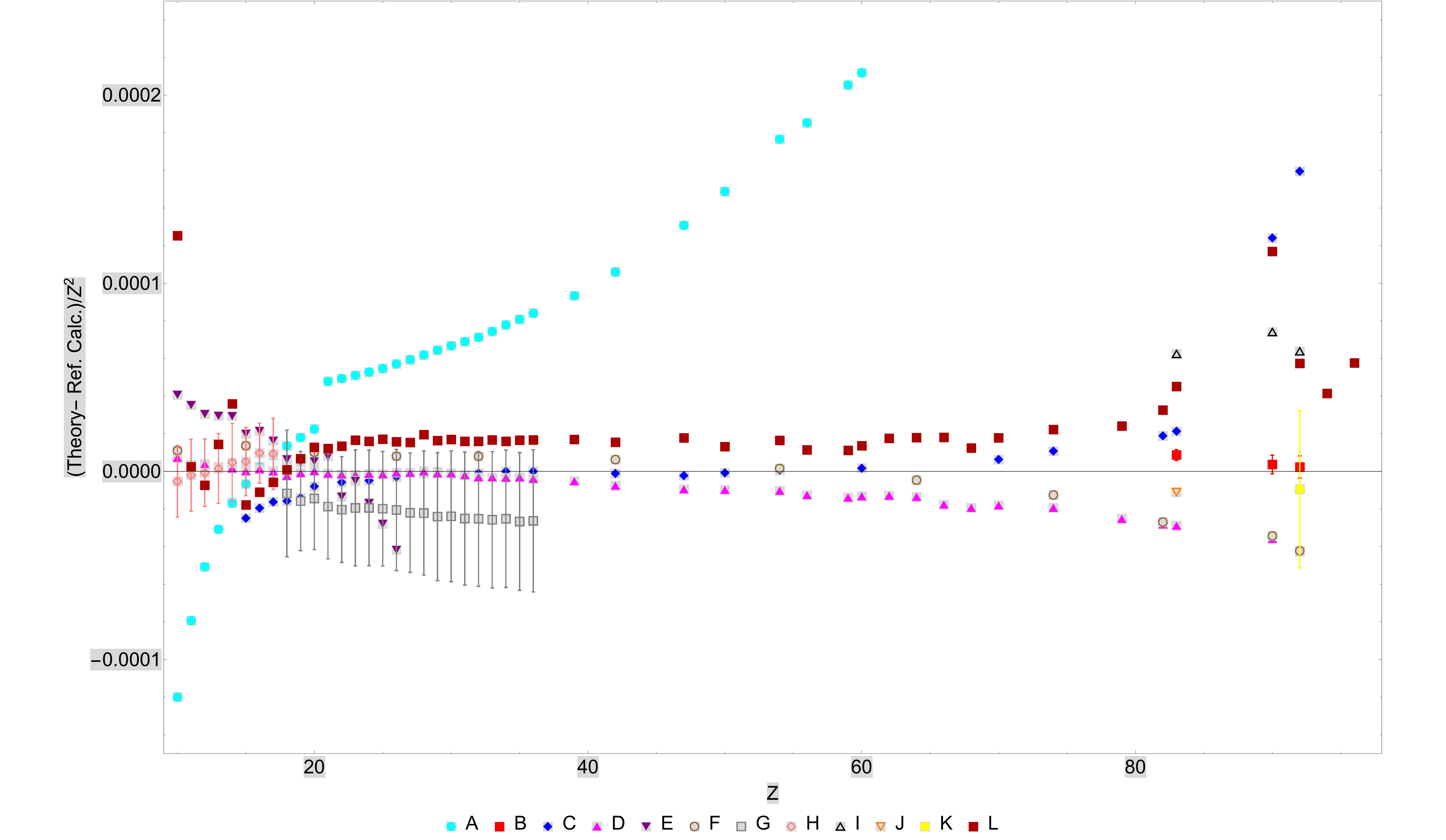}
       \subcaption{Comparison with calculations that do not include second order QED corrections}
    \end{subfigure}
  \begin{subfigure}{\textwidth}
        \includegraphics[width=\textwidth]{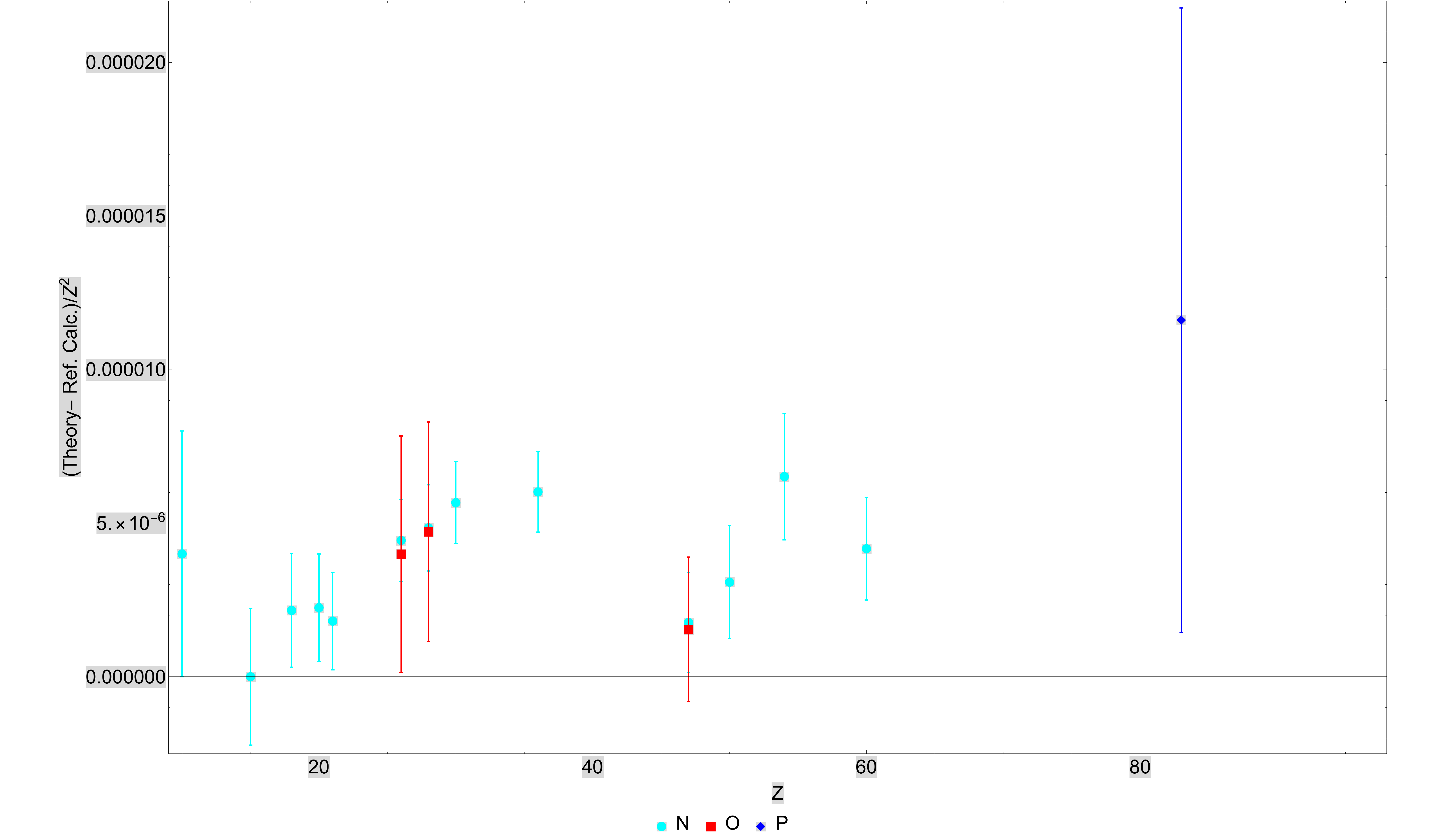}
       \subcaption{Comparison with calculations that include second order QED corrections}
    \end{subfigure}%
\caption{
Comparison between theoretical calculations  and  the work of Sapirstein \& Cheng \protect \cite{sac2011} for the $1s^2 2p\,^2P_{3/2} \to 1s^2 2s\, ^2S_{1/2}$ transition in lithiumlike ions.
The calculations from   \protect \cite{see1989} have not been plotted as their behavior at high-$Z$ is too different from the more recent results.
References:
A: \protect \cite{gu2005},
B: \protect \cite{ccs2000},
C: \protect \cite{iad1990},
D: \protect \cite{kbid1991},
E: \protect \cite{sas1996},
F: \protect \cite{ccjs1995},
G: \protect \cite{sac2011},
H: \protect \cite{yas2012},
I: \protect \cite{ysm2017},
J: \protect \cite{smpl1998},
K: \protect \cite{sac2001},
L: \protect \cite{bjs1990},
M: \protect \cite{jls1996}
N: \protect \cite{kvag2010},
O: \protect \cite{yas2007},
P: \protect \cite{yis2006}
\label{fig:li-2s-2p3-theo-sapir}
}
\end{figure*}

 The comparison between the different calculations presented above is shown in Figs. \ref{fig:li-2s-2p1-theo-sapir} for the  $1s^2 2p \,^2P_{1/2} \to 1s^2 2s\,^2S_{1/2}$ transition and \ref{fig:li-2s-2p1-theo-sapir} for the $1s^2 2p \,^2P_{3/2} \to 1s^2 2s\,^2S_{1/2}$ transition.
 Except for the early calculation by Seely \cite{see1989}, the agreement is globally quite good, with the same limitations due to the uncertainty on the finite nuclear size and its evolution over time, whether some second order QED corrections are included or not, and what part of the electron-electron operator is included in the correlation energy evaluation.

\begin{center}
\begin{longtable}{cD{.}{.}{4}D{.}{.}{4}D{.}{.}{4}l}
\caption{Comparison between experimental and theoretical \protect\cite{sac2011}  energies for the $1s^2 2p\, ^2P_{1/2}\to 1s^2 2s \,^2S_{1/2}$ transition in lithiumlike ions}
\label{tab:lilike-FS2p1-comp}\\
\hline
	&	\multicolumn{3}{c}{$1s^2 2p_{1/2}\to 1s^2 2s_{1/2}$ }					&	\multicolumn{1}{c}{Reference}	\\
Z	&	\multicolumn{1}{c}{Experiment }	&	\multicolumn{1}{c}{Error }	&	\multicolumn{1}{c}{Diff. Th.}	&		\\
\hline									
\endfirsthead
\caption{ $1s^2 2p\, ^2P_{1/2}\to 1s^2 2s \,^2S_{1/2}$ transition  (continued)}\\
\hline
	&	\multicolumn{3}{c}{$1s^2 2p_{1/2}\to 1s^2 2s_{1/2}$ }					&	\multicolumn{1}{c}{Reference}	\\
Z	&	\multicolumn{1}{c}{Experiment }	&	\multicolumn{1}{c}{Error }	&	\multicolumn{1}{c}{Diff. Th.}	&		\\
\hline
\endhead
\hline 
\endfoot

\hline 
\endlastfoot
10	&	15.88870	&	0.00025	&	-0.00110	&	\cite{edl1983}	\\
10	&	15.88881	&	0.00020	&	-0.00121	&	\cite{bhh1963}	\\
11	&	17.86141	&	0.00037	&	-0.00081	&	\cite{edl1983}	\\
11	&	17.8614	&	0.0010	&	-0.0008	&	\cite{nlkb2004}	\\
12	&	19.83938	&	0.00063	&	-0.00118	&	\cite{wap1976}	\\
12	&	19.83896	&	0.00037	&	-0.00076	&	\cite{edl1983}	\\
13	&	21.82271	&	0.00050	&	-0.00081	&	\cite{edl1983}	\\
14	&	23.81015	&	0.00091	&	0.00215	&	\cite{wap1976}	\\
14	&	23.81253	&	0.00037	&	-0.00023	&	\cite{edl1983}	\\
15	&	25.80979	&	0.00149	&	0.00201	&	\cite{edl1983}	\\
16	&	27.81723	&	0.00125	&	0.00257	&	\cite{wap1976}	\\
16	&	27.81871	&	0.00074	&	0.00109	&	\cite{edl1983}	\\
17	&	29.8379	&	0.0015	&	0.0001	&	\cite{edl1983}	\\
18	&	31.86353	&	0.00164	&	0.00457	&	\cite{wap1976}	\\
18	&	31.86642	&	0.00087	&	0.00168	&	\cite{edl1983}	\\
18	&	31.86370	&	0.00057	&	0.00440	&	\cite{brfa2007}	\\
18	&	31.8671	&	0.0012	&	0.0010	&	\cite{pss1984}	\\
20	&	35.9625	&	0.0021	&	-0.0002	&	\cite{wap1976}	\\
20	&	35.9614	&	0.0010	&	0.0009	&	\cite{edl1983}	\\
21	&	38.020	&	0.040	&	0.011	&	\cite{sccf1980}	\\
22	&	40.1159	&	0.0019	&	-0.0035	&	\cite{htdj1989}	\\
22	&	40.11496	&	0.00099	&	-0.00256	&	\cite{edl1983}	\\
22	&	40.1150	&	0.0012	&	-0.0026	&	\cite{pss1984}	\\
24	&	44.3291	&	0.0032	&	-0.0082	&	\cite{wap1976}	\\
24	&	44.3230	&	0.0032	&	-0.0021	&	\cite{htdj1989}	\\
24	&	44.3276	&	0.0037	&	-0.0067	&	\cite{edl1983}	\\
25	&	46.4569	&	0.0035	&	-0.0064	&	\cite{wap1976}	\\
25	&	46.4586	&	0.0050	&	-0.0081	&	\cite{edl1983}	\\
26	&	48.6022	&	0.0038	&	-0.0033	&	\cite{wap1976}	\\
26	&	48.6003	&	0.0038	&	-0.0014	&	\cite{dah1987}	\\
26	&	48.6001	&	0.0019	&	-0.0012	&	\cite{kni1991}	\\
26	&	48.6033	&	0.0019	&	-0.0044	&	\cite{htdj1989}	\\
26	&	48.6012	&	0.0031	&	-0.0023	&	\cite{edl1983}	\\
28	&	52.9395	&	0.0045	&	0.0112	&	\cite{wap1976}	\\
28	&	52.9530	&	0.0045	&	-0.0023	&	\cite{dah1987}	\\
28	&	52.9467	&	0.0038	&	0.0040	&	\cite{sbbh1998}	\\
28	&	52.9496	&	0.0023	&	0.0011	&	\cite{htdj1989}	\\
28	&	52.960	&	0.014	&	-0.009	&	\cite{zllw1988}	\\
29	&	55.1531	&	0.0049	&	0.0027	&	\cite{dah1987}	\\
29	&	55.1595	&	0.0027	&	-0.0037	&	\cite{kni1991}	\\
29	&	55.1543	&	0.0025	&	0.0015	&	\cite{htdj1989}	\\
30	&	57.3839	&	0.0029	&	-0.0018	&	\cite{sbbh1998}	\\
32	&	61.8992	&	0.0062	&	0.0043	&	\cite{dah1987}	\\
32	&	61.9008	&	0.0019	&	0.0027	&	\cite{kni1991}	\\
32	&	61.9023	&	0.0031	&	0.0012	&	\cite{htdj1989}	\\
34	&	66.5294	&	0.0071	&	-0.0098	&	\cite{dah1987}	\\
34	&	66.5294	&	0.0025	&	-0.0098	&	\cite{kni1991}	\\
34	&	66.5240	&	0.0054	&	-0.0044	&	\cite{htdj1989}	\\
36	&	71.241	&	0.011	&	-0.001	&	\cite{dhrs1989}	\\
36	&	71.235	&	0.033	&	0.005	&	\cite{klvb2005}	\\
36	&	71.243	&	0.012	&	-0.003	&	\cite{dah1987}	\\
36	&	71.2430	&	0.0080	&	-0.0031	&	\cite{mlet2002}	\\
36	&	71.284	&	0.016	&	-0.044	&	\cite{mdbb1990}	\\
36	&	71.241	&	0.011	&	-0.001	&	\cite{htdj1989}	\\
39	&	78.5396	&	0.0045	&	-0.0061	&	\cite{stdg2017}	\\
42	&	86.101	&	0.012	&	0.003	&	\cite{dmj1989}	\\
42	&	86.101	&	0.012	&	0.003	&	\cite{htdj1989}	\\
47	&	99.4379	&	0.0072	&	-0.0238	&	\cite{bshs1999}	\\
50	&	107.9109	&	0.0075	&	-0.0139	&	\cite{fbsf2000}	\\
54	&	119.811	&	0.012	&	0.010	&	\cite{tblc2003}	\\
54	&	119.816	&	0.042	&	0.005	&	\cite{bbhk2015}	\\
54	&	119.8204	&	0.0081	&	0.0006	&	\cite{fbsf2000}	\\
56	&	126.112	&	0.013	&	-0.043	&	\cite{rgor2014}	\\
79	&	216.134	&	0.096	&	0.084	&	\cite{bkms2003}	\\
82	&	230.650	&	0.081	&	0.110	&	\cite{bkms2003}	\\
92	&	280.59	&	0.10	&	0.06	&	\cite{sbbc1991}	\\
92	&	280.516	&	0.099	&	0.136	&	\cite{bkms2003}	\\
92	&	280.645	&	0.015	&	0.007	&	\cite{bctt2005}	\\
\hline
\end{longtable}
\end{center}

\begin{center}
\begin{longtable}{cD{.}{.}{4}D{.}{.}{4}D{.}{.}{4}l}
\caption{Comparison between experimental and theoretical \protect\cite{sac2011}  energies for the $1s^2 2p\, ^2P_{3/2}\to 1s^2 2s \,^2S_{1/2}$ transition in lithiumlike ions}
\label{tab:lilike-FS2p3-comp}\\
\hline
	&	\multicolumn{3}{c}{$1s^2 2p_{3/2}\to 1s^2 2s_{1/2}$ }					&	\multicolumn{1}{c}{Reference}	\\
Z	&	\multicolumn{1}{c}{Experiment }	&	\multicolumn{1}{c}{Error }	&	\multicolumn{1}{c}{Diff. Th.}	&		\\
\hline									
\endfirsthead
\caption{ $1s^2 2p\, ^2P_{3/2}\to 1s^2 2s \,^2S_{1/2}$ transition  (continued)}\\
\hline
	&	\multicolumn{3}{c}{$1s^2 2p_{3/2}\to 1s^2 2s_{1/2}$ }					&	\multicolumn{1}{c}{Reference}	\\
Z	&	\multicolumn{1}{c}{Experiment }	&	\multicolumn{1}{c}{Error }	&	\multicolumn{1}{c}{Diff. Th.}	&		\\
\hline
\endhead
\hline 
\endfoot

\hline 
\endlastfoot
10	&	16.09330	&	0.00010	&	-0.00050	&	\cite{bhh1963}	\\
10	&	16.09315	&	0.00035	&	-0.00035	&	\cite{edl1983}	\\
11	&	18.18761	&	0.00053	&	-0.00091	&	\cite{edl1983}	\\
11	&	18.18700	&	0.00100	&	-0.00030	&	\cite{nlkb2004}	\\
12	&	20.33228	&	0.00067	&	-0.00078	&	\cite{wap1976}	\\
12	&	20.33180	&	0.00053	&	-0.00030	&	\cite{edl1983}	\\
13	&	22.54132	&	0.00070	&	-0.00102	&	\cite{edl1983}	\\
14	&	24.82514	&	0.00099	&	0.00066	&	\cite{wap1976}	\\
14	&	24.82635	&	0.00053	&	-0.00055	&	\cite{edl1983}	\\
15	&	27.2050	&	0.0021	&	-0.0024	&	\cite{edl1983}	\\
16	&	29.6847	&	0.0014	&	0.0003	&	\cite{wap1976}	\\
16	&	29.6863	&	0.0011	&	-0.0013	&	\cite{edl1983}	\\
17	&	32.2891	&	0.0021	&	0.0016	&	\cite{edl1983}	\\
18	&	35.0357	&	0.0020	&	0.0014	&	\cite{wap1976}	\\
18	&	35.03803	&	0.00061	&	-0.00093	&	\cite{pss1984}	\\
18	&	35.0369	&	0.0012	&	0.0002	&	\cite{edl1983}	\\
18	&	35.03160	&	0.00059	&	0.00550	&	\cite{brfa2007}	\\
20	&	41.0286	&	0.0027	&	-0.0044	&	\cite{wap1976}	\\
20	&	41.0261	&	0.0014	&	-0.0019	&	\cite{edl1983}	\\
21	&	44.30943	&	0.00020	&	-0.00103	&	\cite{llos2008}	\\
21	&	44.312	&	0.035	&	-0.004	&	\cite{sccf1980}	\\
21	&	44.3107	&	0.0019	&	-0.0023	&	\cite{kssm2004}	\\
22	&	47.8150	&	0.0037	&	-0.0002	&	\cite{htdj1989}	\\
22	&	47.82012	&	0.00074	&	-0.00532	&	\cite{pss1984}	\\
22	&	47.8201	&	0.0014	&	-0.0053	&	\cite{edl1983}	\\
24	&	55.5983	&	0.0050	&	-0.0065	&	\cite{wap1976}	\\
24	&	55.5958	&	0.0050	&	-0.0040	&	\cite{htdj1989}	\\
24	&	55.5936	&	0.0015	&	-0.0018	&	\cite{kni1991}	\\
24	&	55.5992	&	0.0053	&	-0.0074	&	\cite{edl1983}	\\
25	&	59.9247	&	0.0058	&	-0.0105	&	\cite{wap1976}	\\
25	&	59.9275	&	0.0070	&	-0.0133	&	\cite{edl1983}	\\
26	&	64.5617	&	0.0067	&	0.0003	&	\cite{wap1976}	\\
26	&	64.5583	&	0.0067	&	0.0037	&	\cite{dah1987}	\\
26	&	64.5711	&	0.0067	&	-0.0091	&	\cite{htdj1989}	\\
26	&	64.5596	&	0.0030	&	0.0024	&	\cite{kni1991}	\\
26	&	64.5665	&	0.0044	&	-0.0045	&	\cite{edl1983}	\\
28	&	74.9620	&	0.0091	&	-0.0072	&	\cite{dah1987}	\\
28	&	74.9575	&	0.0068	&	-0.0027	&	\cite{sbbh1998}	\\
28	&	74.9620	&	0.0045	&	-0.0072	&	\cite{htdj1989}	\\
28	&	74.976	&	0.011	&	-0.021	&	\cite{bsnb1995}	\\
29	&	80.7694	&	0.0053	&	-0.0062	&	\cite{befs1987}	\\
29	&	80.766	&	0.011	&	-0.003	&	\cite{dah1987}	\\
29	&	80.768	&	0.011	&	-0.005	&	\cite{htdj1989}	\\
29	&	80.7683	&	0.0032	&	-0.0051	&	\cite{kni1991}	\\
30	&	87.0303	&	0.0037	&	-0.0072	&	\cite{sbbh1998}	\\
32	&	101.022	&	0.016	&	0.022	&	\cite{dah1987}	\\
32	&	101.042	&	0.016	&	0.002	&	\cite{htdj1989}	\\
32	&	101.0425	&	0.0049	&	0.0015	&	\cite{kni1991}	\\
34	&	117.298	&	0.022	&	0.019	&	\cite{dah1987}	\\
34	&	117.314	&	0.022	&	0.003	&	\cite{htdj1989}	\\
36	&	136.173	&	0.037	&	0.001	&	\cite{dhrs1989}	\\
36	&	136.246	&	0.045	&	-0.072	&	\cite{mdbb1990}	\\
36	&	136.198	&	0.036	&	-0.024	&	\cite{pgrr2014}	\\
36	&	136.216	&	0.090	&	-0.042	&	\cite{klvb2005}	\\
36	&	136.157	&	0.030	&	0.017	&	\cite{dah1987}	\\
36	&	136.173	&	0.037	&	0.001	&	\cite{htdj1989}	\\
39	&	170.135	&	0.014	&	-0.038	&	\cite{stdg2017}	\\
42	&	211.942	&	0.072	&	0.040	&	\cite{dmj1989}	\\
42	&	211.942	&	0.072	&	0.040	&	\cite{htdj1989}	\\
47	&	303.667	&	0.030	&	0.000	&	\cite{bshs1999}	\\
54	&	492.174	&	0.052	&	0.032	&	\cite{bbhk2015}	\\
74	&	1697.3	&	1.0	&	-1.2	&	\cite{pcbw2009}	\\
74	&	1696.20	&	0.50	&	-0.10	&	\cite{cbbg2011}	\\
82	&	2642.26	&	0.10	&	-0.09	&	\cite{znca2008}	\\
83	&	2788.139	&	0.039	&	-0.099	&	\cite{bosl1998}	\\
90	&	4025.23	&	0.15	&	0.02	&	\cite{boec1995}	\\
92	&	4459.37	&	0.27	&	0.09	&	\cite{bkme1993}	\\
92	&	4460.9	&	2.2	&	-1.4	&	\cite{ntik2013}	\\
\hline
\end{longtable}
\end{center}

\begin{figure*}[htbp]
\centering
        \includegraphics[width=\textwidth]{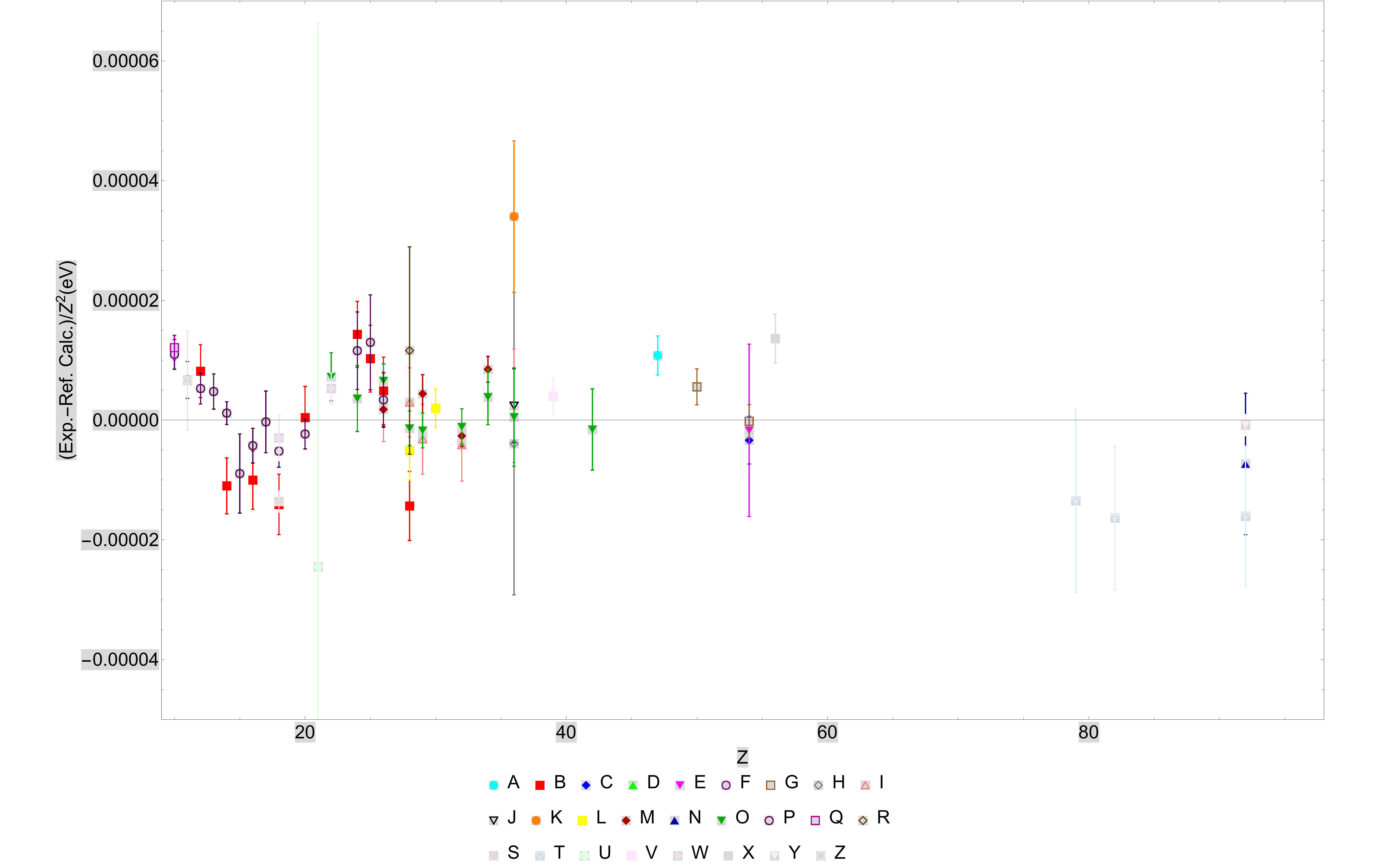}
\caption{
Comparison between available experiments and the work of  Sapirstein \& Cheng \protect \cite{sac2011} for the $1s^2 2p\;^2P_{1/2} \to 1s^2 2s\; ^2S_{1/2}$ transition in lithiumlike ions.
References: 
A: \protect\cite{bshs1999},
B: \protect\cite{wap1976},
C: \protect\cite{tblc2003},
D: \protect\cite{dmj1989},
E: \protect\cite{bbhk2015},
F: \protect\cite{dhrs1989},
G: \protect\cite{fbsf2000},
H: \protect\cite{klvb2005},
I: \protect\cite{dah1987},
J: \protect\cite{mlet2002},
K: \protect\cite{mdbb1990},
L: \protect\cite{sbbh1998},
M: \protect\cite{kni1991},
N: \protect\cite{sbbc1991},
O: \protect\cite{htdj1989},
P: \protect\cite{edl1983},
Q: \protect\cite{bhh1963},
R: \protect\cite{zllw1988},
S: \protect\cite{brfa2007},
T: \protect\cite{bkms2003},
U: \protect\cite{sccf1980},
V: \protect\cite{stdg2017},
W: \protect\cite{pss1984},
X: \protect\cite{rgor2014},
Y: \protect\cite{bctt2005},
Z: \protect\cite{nlkb2004}
Values are from Table \protect \ref{tab:lilike-FS2p1-comp}.
\label{fig:li-2s-2p1-exp-sapir}
}
\end{figure*}

\begin{figure*}[htbp]
\centering
        \includegraphics[width=\textwidth]{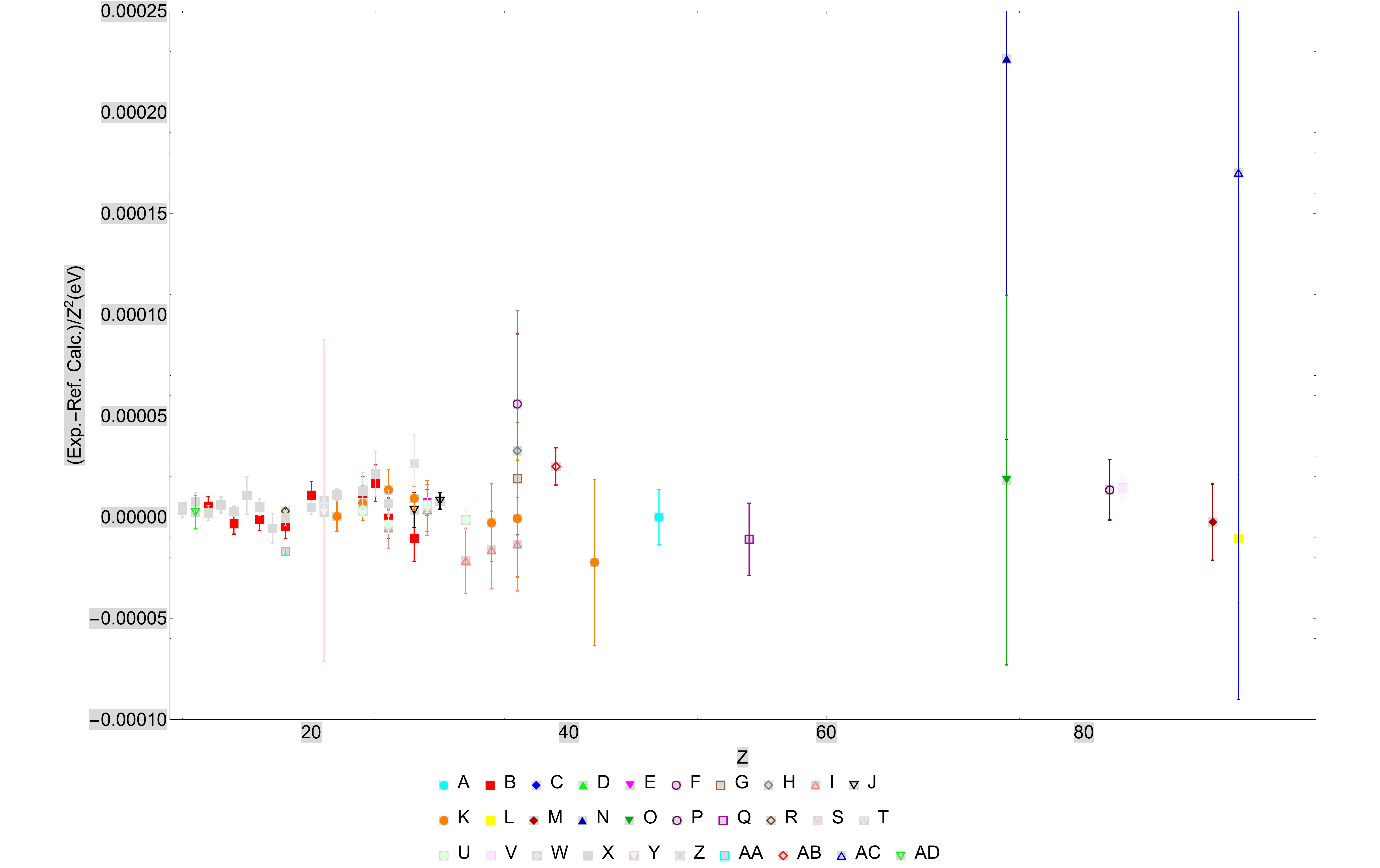}
\caption{
Comparison between available experiments and the work of Sapirstein \& Cheng \protect \cite{sac2011} for the $1s^2 2p\,^2P_{3/2} \to 1s^2 2s\, ^2S_{1/2}$ transition in lithiumlike ions.
References: 
A: \protect\cite{bshs1999},
B: \protect\cite{wap1976},
C: \protect\cite{dmj1989},
D: \protect\cite{dhrs1989},
E: \protect\cite{befs1987},
F: \protect\cite{mdbb1990},
G: \protect\cite{pgrr2014},
H: \protect\cite{klvb2005},
I: \protect\cite{dah1987},
J: \protect\cite{sbbh1998},
K: \protect\cite{htdj1989},
L: \protect\cite{bkme1993},
M: \protect\cite{boec1995},
N: \protect\cite{pcbw2009},
O: \protect\cite{cbbg2011},
P: \protect\cite{znca2008},
Q: \protect\cite{bbhk2015},
R: \protect\cite{pss1984},
S: \protect\cite{sccf1980},
T: \protect\cite{kssm2004},
U: \protect\cite{kni1991},
V: \protect\cite{bosl1998},
W: \protect\cite{bhh1963},
X: \protect\cite{edl1983},
Y: \protect\cite{llos2008},
Z: \protect\cite{bsnb1995},
AA: \protect\cite{brfa2007},
AB: \protect\cite{stdg2017},
AC: \protect\cite{ntik2013},
AD: \protect\cite{nlkb2004}
Values are  from Table \protect \ref{tab:lilike-FS2p3-comp}.
\label{fig:li-2s-2p3-exp-sapir}
}
\end{figure*}

\begin{figure*}[htbp]
\centering
        \includegraphics[width=\textwidth]{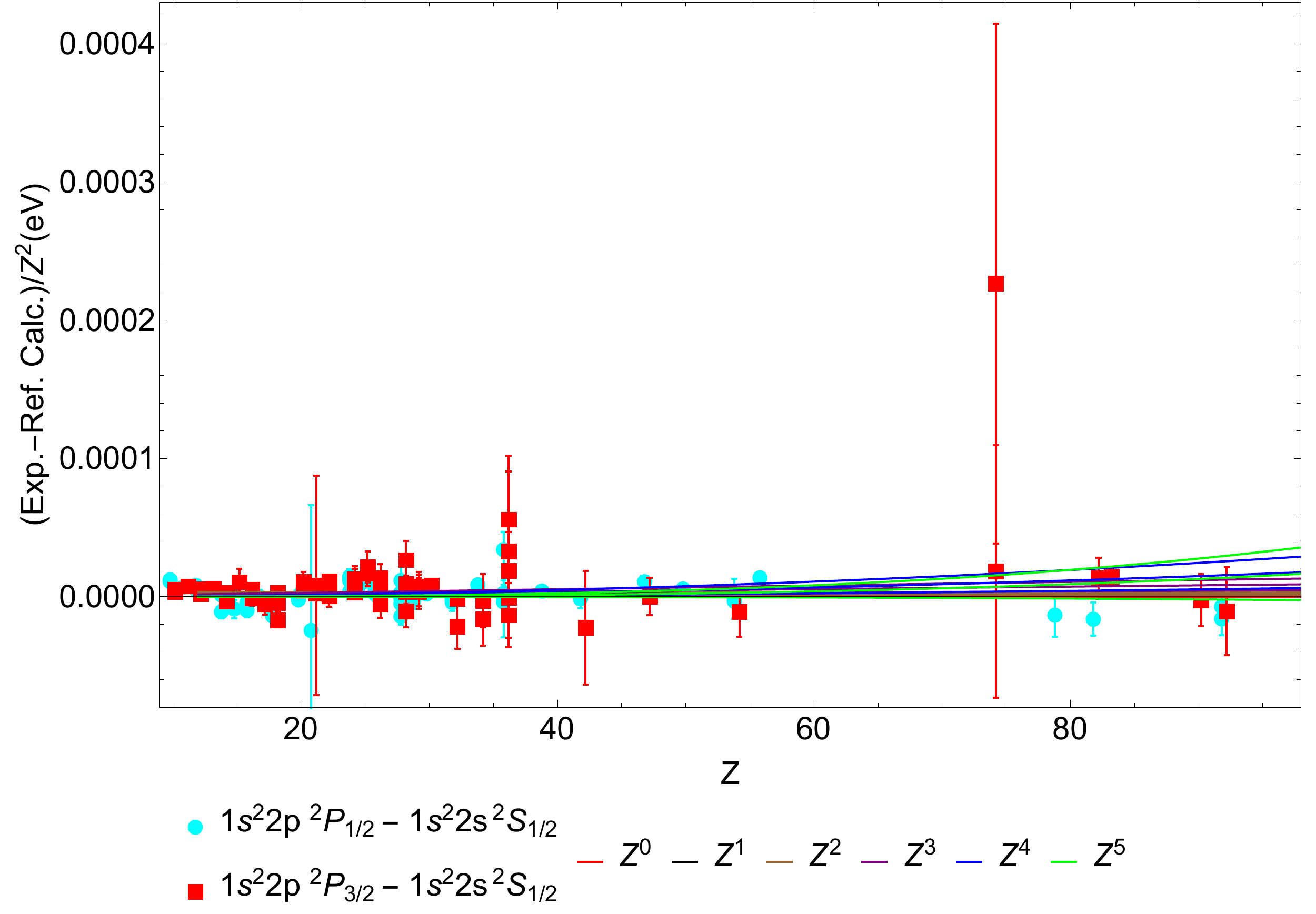}
\caption{
Fit of the  differences between  experiments and the calculations of Sapirstein \& Cheng \protect \cite{sac2011} for the $1s^2 2p\,^2P_{J} \to 1s^2 2s\, ^2S_{1/2}$ transitions in lithiumlike ions with different functions $b Z^n$, $n=0,\ldots,\, 5$.
The error bands corresponding to $(b\pm \delta b) Z^n$, where $\delta b$ is the $1\sigma$  are also shown. The points corresponding to the $1s^2 2p\,^2P_{1/2} \to 1s^2 2s\, ^2S_{1/2}$ and $1s^2 2p\,^2P_{3/2} \to 1s^2 2s\, ^2S_{1/2}$ have been slightly shifted along the $Z$ axis to improve legibility.
\label{fig:li-2s-2pj-fit-sapir-diff-exp}
}
\end{figure*}

\begin{figure*}[htbp]
\centering
        \includegraphics[width=\textwidth]{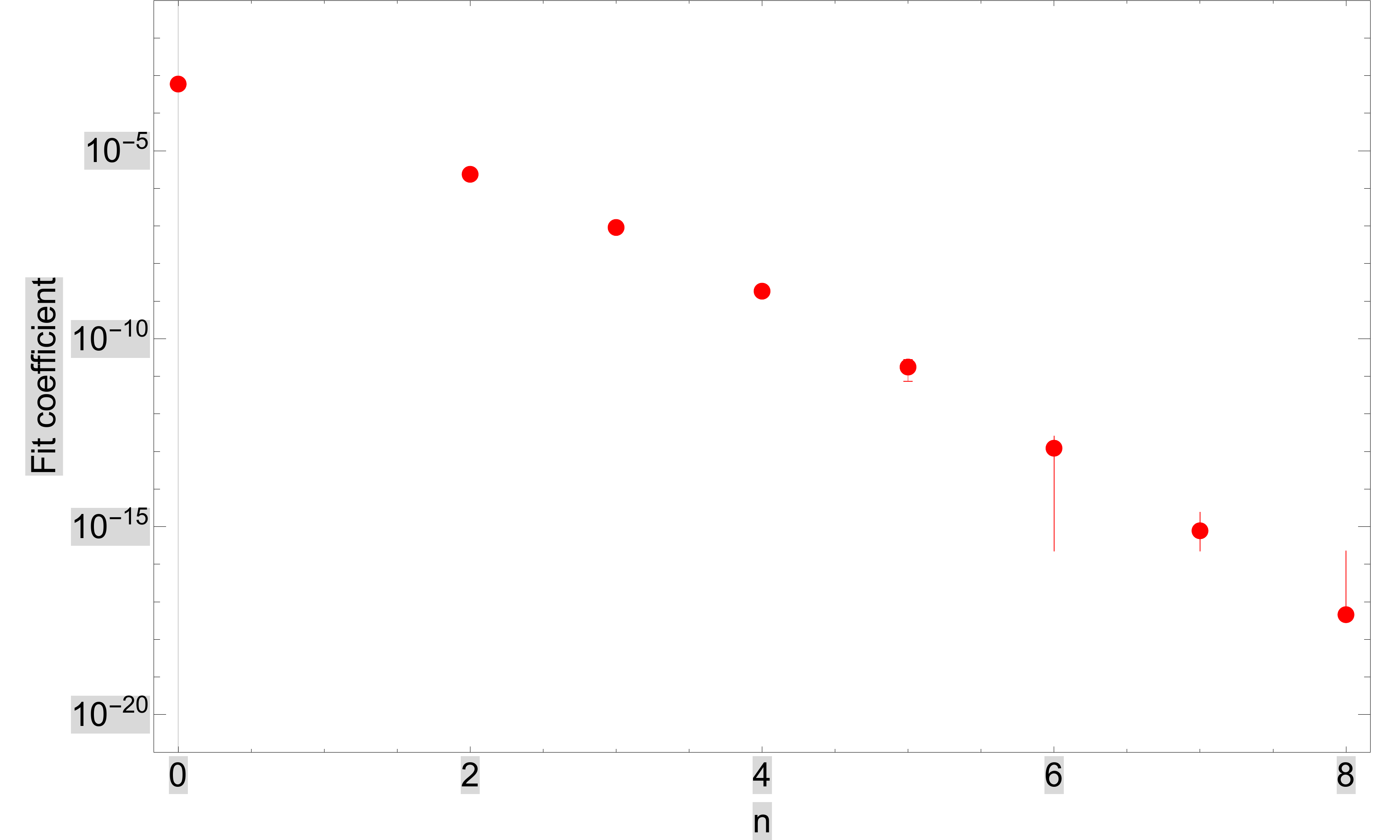}
\caption{
Absolute values of the fit coefficient  $b$ for $n=0,\ldots,\, 8$ with error bars for the differences between  experiments and the calculation of Sapirstein \& Cheng \protect \cite{sac2011} for the $1s^2 2p\,^2P_{J} \to 1s^2 2s\, ^2S_{1/2}$ transitions in lithiumlike ions.
The reduced $\chi^2$ are between \num{0.35} and \num{0.39}. 
\label{fig:li-2s-2pj-fit-sapir-diff-exp-coeff}
}
\end{figure*}

The comparison between the different available experimental results and the theoretical calculations  of Sapirstein \& Cheng \cite{sac2011} is presented in Table \ref{tab:lilike-FS2p1-comp} for the  $1s^2 2p \,^2P_{1/2} \to 1s^2 2s\,^2S_{1/2}$ transition  and Table \ref{tab:lilike-FS2p3-comp} for the  $1s^2 2p \,^2P_{3/2} \to 1s^2 2s\,^2S_{1/2}$ transition. The same results are plotted in Figs. \ref{fig:li-2s-2p1-exp-sapir} and \ref{fig:li-2s-2p3-exp-sapir}. I have performed an analysis of this difference in a manner similar to what was presented in  \ref{sec:hlike} and \ref{sec:helike}.
A sequence of $f(Z)=b Z^n$ functions with $n=0$ to \num{8} have been fitted to the ensemble of all results from Tables   \ref{tab:lilike-FS2p1-comp}  and  \ref{tab:lilike-FS2p3-comp}. The results of the fits are plotted together with the experiment-theory values in figure \ref{fig:li-2s-2pj-fit-sapir-diff-exp}. The fit parameters are plotted on figure \ref{fig:li-2s-2pj-fit-sapir-diff-exp-coeff} together with the statistical error bars.

The $1s^2 2p \;^2P_{1/2} \to 1s^2 2s \;^2S_{1/2}$ transition in lithiumlike uranium has been measured with very good precision. At the same time many-body effects, first and second order QED corrections, as well as finite size, nuclear deformation and recoil corrections have been evaluated with advanced methods. It is thus a complete benchmark of our ability to evaluate precisely transition energies for very-heavy few-electron ions. In table \ref{tab:li-u-contrib}, I present the details of all contributions to this transition energy, following the result of Kozhedub \etal. \cite{kast2008}, which was the first work to provide good agreement between the most accurate theory and experiment, updated using the values from   \cite{kvag2010}.

\begin{table}
\begin{center}
\caption{Comparison between experimental and theoretical energies for the$1s^2 2p \;^2P_{1/2} \to 1s^2 2s \;^2S_{1/2}$ transition in lithiumlike uranium, following  Kozhedub \etal. \protect \cite{kast2008}. 
\label{tab:li-u-contrib}
}
\begin{tabular}{lD{.}{.}{1}D{.}{.}{1}l}
\hline
\hline
Contribution	&	\multicolumn{1}{c}{Value (eV)}	&	\multicolumn{1}{c}{Error (eV)}	&	\multicolumn{1}{c}{Reference}	\\
One-photon exchange 	&	368.83426	&		&	\cite{kvag2010}	\\
One-electron nuclear size 	&	-33.304	&	0.03	&	\cite{kvag2010}	\\
One-electron first-order QED 	&	-42.93	&		&	\cite{mps1998}	\\
Two-photon exchange 	&	-13.371	&		&	\cite{yass2000,yass2001}	\\
Screened QED	&	1.16	&		&	\cite{abps1999,yabp1999}	\\
One-electron second-order QED	&	0.22	&	0.06	&	\cite{yis2006}	\\
Three- and more photon effects	&	0.1370	&	0.050	&	\cite{kvag2010}	\\
Nuclear recoil	&	-0.07	&		&	\cite{asy1995}	\\
Nuclear polarization	&	0.03	&	0.01	&	\cite{pas1995,pas1996,nlps1996}	\\
Total theory	&	280.706	&	0.085	&		\\
Experiment	&	280.645	&	0.015	&	\cite{bctt2005}	\\
Experiment	&	280.52	&	0.10	&	\cite{bkms2003}	\\
Experiment	&	280.59	&	0.10	&	\cite{sbbc1991}	\\

\hline
\hline
\end{tabular}
\end{center}
\end{table}

\begin{figure*}[htbp]
\centering
        \includegraphics[width=\textwidth]{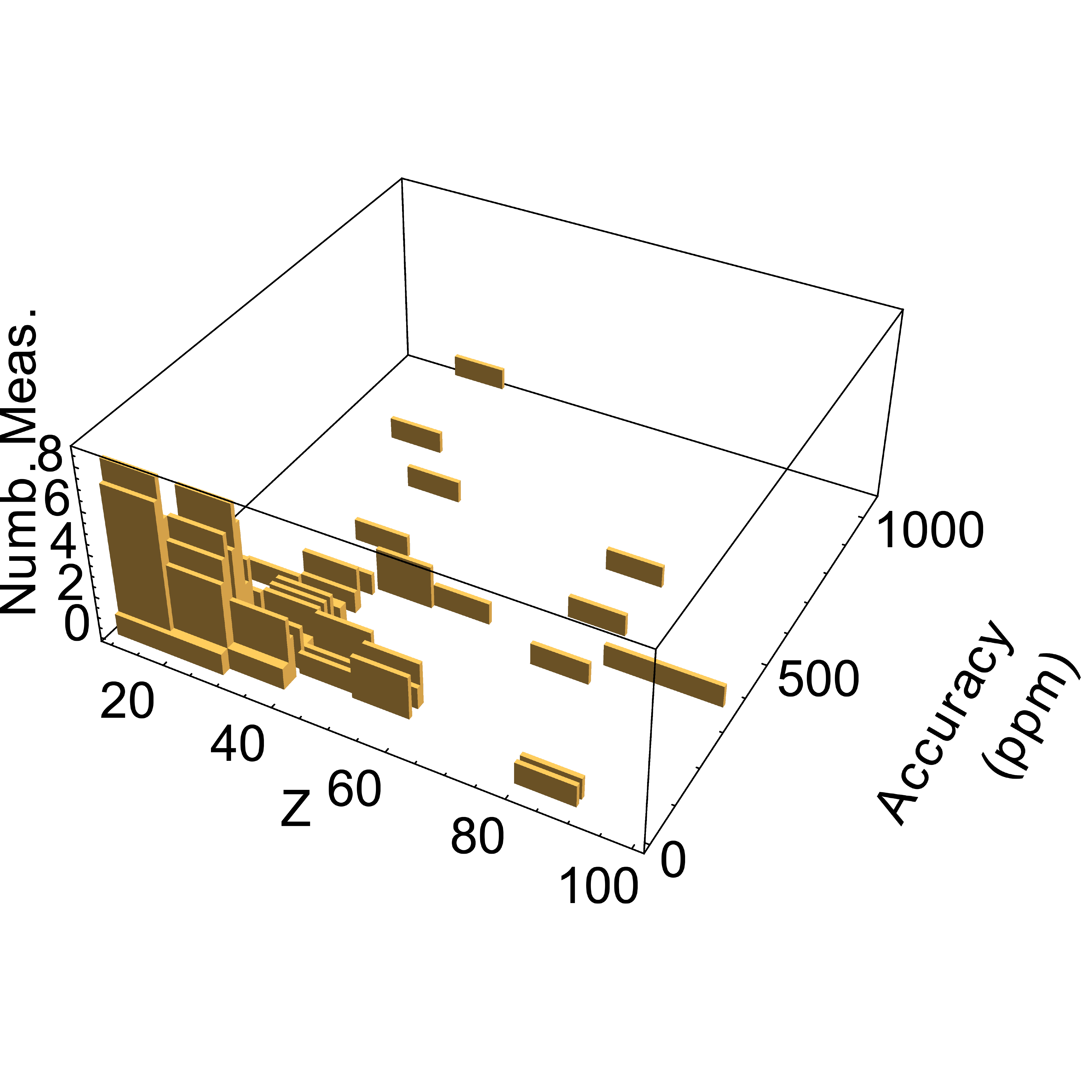}
\caption{
Histogram of the number of measurements for a given accuracy in parts per million and the atomic number $Z$ for the $1s^2 2p\,^2P_{J} \to 1s^2 2s\, ^2S_{1/2}$ transitions in lithiumlike ions. All measurements from Table   \protect \ref{tab:lilike-FS2p1-comp} and \protect \ref{tab:lilike-FS2p3-comp} are taken into account.
\label{fig:li-z-vs-ppm-histo}
}
\end{figure*}

Figure \ref{fig:li-2s-2pj-fit-sapir-diff-exp-coeff}
shows results quite similar to what was presented in figure \ref{fig:lh-ly--fit-yero-diff-exp} for the Lyman $\alpha$ of hydrogenlike ions. The error bands are more symmetrical around the horizontal axis than in the analysis of the $n=2\to n=1$ and $\Delta n=0$ transitions in He-like ions (figures \ref{fig:he-n2-diff-exp-th-fit1} , \ref{fig:he-n2-diff-exp-th-fit2} and \ref{fig:he-wxyz-fit-artemyev-diff-exp}).
It shows that several accurate measurements at medium and high-$Z$ are key for a meaningful comparison with theory. This can be seen on the histogram of the number of experiments with a given accuracy and atomic number in figure \ref{fig:li-z-vs-ppm-histo}: there are many more accurate experiments of good accuracy in the range of $40\leq Z \leq 92$ and in particular several accurate points for bismuth, thorium and uranium.

\section{Operators other than energy}
\label{sec:other-operators}
\begin{figure*}[htbp]
\centering
        \includegraphics[width=0.6\textwidth]{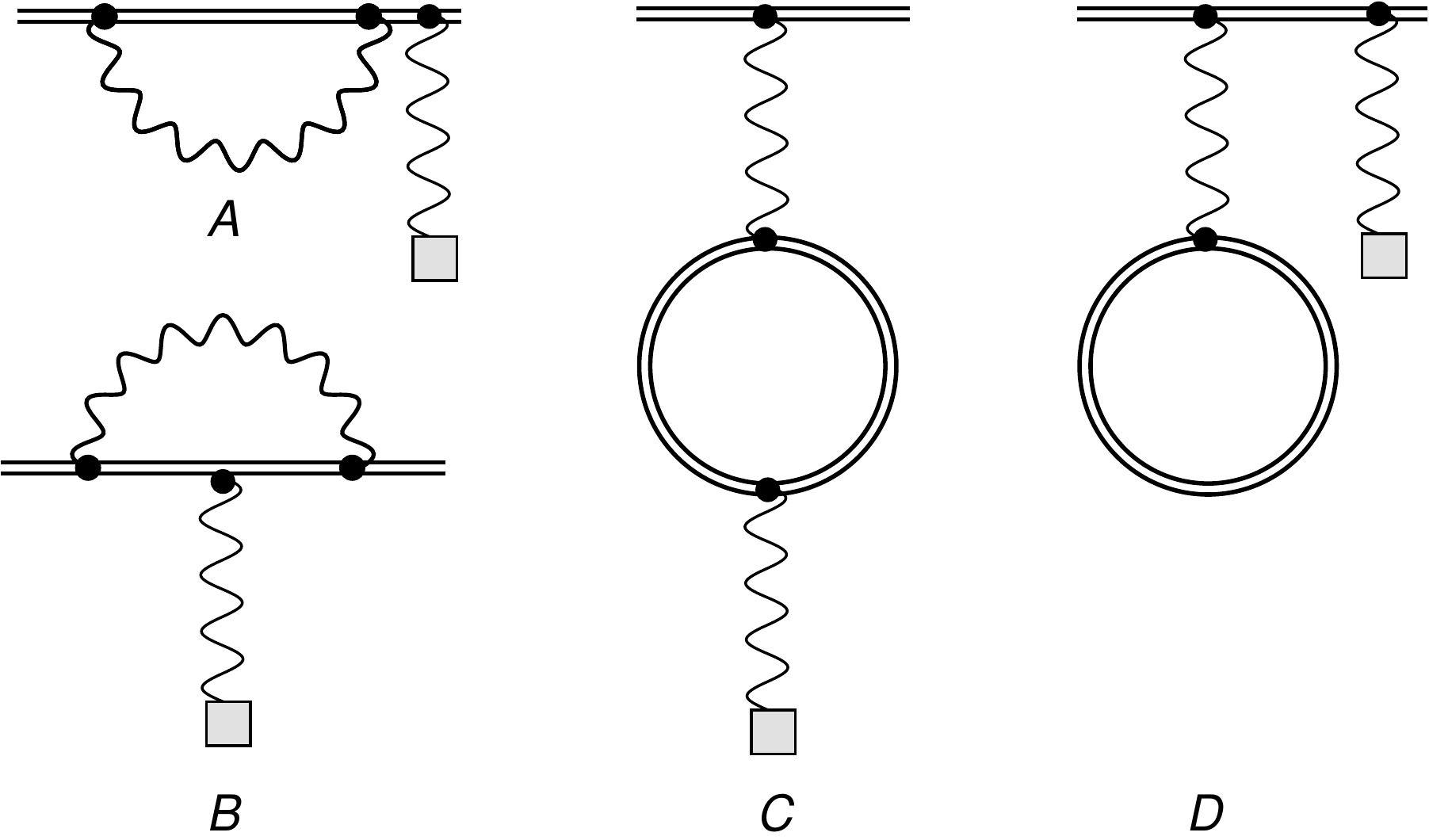}
\caption{
First order QED correction to external interaction corrections like hyperfine structure (nuclear magnetic moment) or Landé $g$-factors (external magnetic field). The square represents the external potential interacting with the electron.
Diagram A represents the self-energy correction to the electron wavefunction. Diagram B represents the vertex correction. Diagram C is the vacuum polarization correction to the external interaction and diagram D is the vacuum polarization correction to the electron wave function.
\label{fig:qed-corr-extern}
}
\end{figure*}

Over the years there have been measurements and extensive theoretical calculations of other quantities than energy, notably the hyperfine structure and the Landé $g$-factors. 
The study of hyperfine structure in few-electron ions allows to test QED in very strong magnetic fields. Because of the very narrow width of transitions between hyperfine sublevels of the ground state of hydrogenlike ions, it has been proposed to use them for highly-charged ion atomic clocks, that would be sensitive to drifts of the fundamental constants \cite{sch2007}.  The  Landé $g$-factor, beyond the realization of QED tests, is also important for its potential application to the determination of fundamental constants like the ratio of the electron to proton mass ratio and of the fine structure constant. The one-loop QED corrections for these operators correspond to the diagrams in figure \ref{fig:qed-corr-extern}. The presence of an extra interaction compared to the equivalent diagrams for energy, increases considerably the number of diagrams to evaluate and their complexity.

\subsection{Hyperfine structure}
\label{sec:hfs}

The study of the hyperfine structure of highly charged ions is also intended as a way to test BSQED, but in a completely different sector: while the main idea behind measurements of transition or level energies is to test BSQED in strong Coulomb fields, studies of the hyperfine structures of the inner shells of highly charged ions provides a test of BSQED in extreme magnetic fields. The magnetic field at the nuclear surface of bismuth, for example, is  \SI{1E10}{\tesla}. In average for the $1s$ orbital due to the $1/r^3$ dependence of the magnetic interaction with a dipole, one finds a value of \SI{1E4}{\tesla}. 
The first accurate experiments on one-electron ions,  performed at GSI with the ESR and Livermore with the super EBIT,  started an intense theoretical activity to evaluate the hyperfine structure on heavy hydrogenlike and later lithiumlike ions taking into account relativistic, QED and nuclear effects.

For a point nucleus, the hyperfine matrix elements between two levels with total atomic angular momentum $J$ and $J'$ are given by
\begin{equation}
\label{eq:hyma1}
	W(J,J')=\langle I,J,F,M_{F}|\bm{M}^{(1)} \cdot \bm{T}^{(1)}|
	I,J',F,M_{F}\rangle ,
\end{equation}
where  $F$ is the  angular momentum of the ion, resulting from adding up the nuclear spin $I$. Equation \eqref{eq:hyma1}) can be transformed as
\begin{eqnarray}
\label{eq:hyma2}
	W_{J,J'} &=& (-1)^{I+J+F} \left\{ \begin{array}{ccc} I & J & F
		\\ J' & I & 1 \end{array} \right\} \nonumber \\ &&
		\times \langle I || \bm{M}^{(1)} || I \rangle \langle J ||\bm{T}^{(1)}|| J'\rangle .
\end{eqnarray}
The nuclear magnetic moment is given by
\begin{equation}
\label{eq:magmoment}
	\mu_{I} \mu_{N}=\langle II | \bm{M}^{(1)}_{0} | II \rangle
	=\left( \begin{array}{ccc} I & 1 & I \\ -I & 0 & I \end{array}
	\right) \langle I || \bm{M}^{(1)} || I \rangle,
\end{equation}
where $\mu_{N}=e\hbar/(m_{p}c)$ is the nuclear magneton $\mu_{N}$  and  the  $3j$ symbol is given by
\begin{equation}
\label{eq:3jsym}
	\left( \begin{array}{ccc} I & 1 & I \\ -I & 0 & I \end{array}
	\right)=\sqrt{\frac{I}{(2I+1)(I+1)}} \, .
\end{equation}
This leads to an energy shift 
\begin{eqnarray}
	W_{J,J} &=&\mu_{N} \left(\frac{ \mu_{I} }{I}\right)\frac{F(F+1)-I(I+1)-J(J+1)}{2} \nonumber \\ 
	&&\qquad \qquad \times \frac{  \langle J ||\bm{T}^{(1)}|| J\rangle}{\sqrt{J(J+1)(J+2)}} \, .
\label{eq:hyma3}
\end{eqnarray}

The  operator $\bm{T}^{(1)}_{q}$ is a sum of one-electron operators $\bm{t}^{(1)}_{q}$:
\begin{eqnarray}
\label{eq:atomoper}
	\bm{T}^{(1)}_{q} &=& \sum \bm{t}^{(1)}_{q} \nonumber \\ &=&
	\sum_{j} -ie \sqrt{\frac{8 \pi}{3}} \, \frac{\boldsymbol{\alpha}_{j} \cdot \bm{ Y}^{(0)}_{1q}\left({\vec {\hat{r}}}_{j}\right)} {r_{j}^{2}},
\end{eqnarray}
where $\bm{\alpha}$ are the Dirac matrices, $\bm{Y}^{(0)}_{1q}\left(\vec {\hat{r}}_{j}\right)$ is a vector 
spherical harmonic and $j$ the $j$th electron of the atom.    The radial part is given by 
\begin{equation}
	\left[r^{k}\right]_{n\kappa\,n'\kappa'}=\int_{0}^{\infty} dr r^{k}
	 \left(
	P_{n \kappa}(r)Q_{n' \kappa'}(r)+Q_{n \kappa}(r)P_{n' \kappa'}(r)
	\right),
	\label{eq:hfsme}
\end{equation}
with $k=-2$.
Detailed derivations and definitions can be found elsewhere \cite{edm1974,lar1974,cac1985}.

These tests are made difficult because of the nuclear structure, which contributes in an even more intricate way than for transition energies. The dependance of the HFS on the nucleus has three parts. The obvious one is that the energy shift in \eqref{eq:hyma3} is proportional to the magnetic moment of the nucleus. These magnetic moments are measured by a number of methods on neutral atoms or on molecules, and the relation between the measured value and the corrected value is complicated. A summary of the values measured over the years for elements which have been used in HFS measurements on one or three-electron ions is shown in Table \ref{tab:mag-moment}. This table includes in particular recent reanalysis of magnetic moment values of Bi, extracted from earlier NMR measurements and on new measurements on specific molecules. One can see that the variations are not negligible.  

\begin{table}
\begin{center}
\caption{Comparison between nuclear magnetic moments $\mu$ (in Bohr magneton units). Measurement method: NMR: Nuclear magnetic resonance, OD: Optical double resonance, AB/D Rev.: Atomic beam magnetic resonance (direct moment measurement), revised in tabulation, CFBLS: Collinear fast beam laser spectroscopy—accelerated beam, OP/RD: Optical pumping with radiative detection. The last line corresponds to a weighted average of the 2018 values }
\label{tab:mag-moment}
\begin{tabular}{ccccD{.}{.}{7}D{.}{.}{7}D{.}{.}{7}ll}
\hline
\hline
Symbol	&	Z	&	A	&	I	&	\multicolumn{1}{c}{$\mu$}	&	\multicolumn{1}{c}{-Err.}	&	\multicolumn{1}{c}{+Err.}	&	\multicolumn{1}{c}{Method}	&	Reference	\\
\hline
Pr	&	59	&	141	&	5/2	&	4.2754	&	0.0005	&	0.0005	&	OD	&	\cite{rag1989,sto2005}	\\
Ho	&	67	&	165	&	7/2	&	4.173	&	0.027	&	0.027	&	AB/D Rev.	&	\cite{rag1989,sto2005}	\\
	&		&		&		&	4.177	&	0.005	&	0.005	&	Rev.	&	\cite{jgs2006}	\\
	&		&		&		&	4.132	&	0.005	&	0.005	&	Rev.	&	\cite{pek1987}	\\
Re	&	75	&	185	&	5/2	&	3.1871	&	0.0003	&	0.0003	&	NMR	&	\cite{rag1989,sto2005}	\\
Re	&	75	&	187	&	5/2	&	3.2197	&	0.0003	&	0.0003	&	NMR	&	\cite{rag1989,sto2005}	\\
Tl	&	81	&	203	&	1/2	&	1.62225787	&	0.00000012	&	0.00000012	&	NMR	&	\cite{rag1989,sto2005}	\\
	&		&		&		&	1.6231	&	0.0013	&	0.0013	&	CFBLS	&	\cite{rag1989,sto2005}	\\
Tl	&	81	&	205	&	1/2	&	1.63821461	&	0.00000012	&	0.00000012	&	NMR	&	\cite{rag1989,sto2005}	\\
Pb	&	82	&	207	&	1/2	&	0.592583	&	0.000009	&	0.000009	&	NMR	&	\cite{rag1989,sto2005}	\\
	&		&		&		&	0.58219	&	0.00002	&	0.00002	&	OP/RD	&	\cite{rag1989,sto2005}	\\
	&		&		&		&	0.59665	&	0.00013	&	0.00013	&	corr. Shielding	&	\cite{jagj2012}	\\
Bi	&	83	&	209	&	9/2	&	4.1106	&	0.0002	&	0.0002	&	NMR	&	\cite{rag1989,sto2005}	\\
	&		&		&		&	4.1103	&	0.0005	&	0.0002	&	NMR (Rev.)	&	\cite{sto2005}	\\
\hline																	
	&		&		&		&	4.0900	&	0.0015	&	0.0015	&	HFS	&	\cite{ssug2018}	\\
	&		&		&		&	4.092	&	0.002	&	0.002	&	BiF$_6^-$	&	\cite{ssug2018}	\\
	&		&		&		&	4.0941	&	0.0024	&	0.0032	&	BiF$_6^-$	&	\cite{arjj2018}	\\
	&		&		&		&	4.0907	&	0.0019	&	0.0028	&	Bi(NO$_3$)$_3$	&	\cite{arjj2018}	\\
	&		&		&		&	4.0919	&	0.0025	&	0.0036	&	Bi(ClO$_4$)$_3$	&	\cite{arjj2018}	\\
	&		&		&		&	4.09122	&	0.00095	&	0.0036	&	average new values	&	This work	\\
		\hline
\end{tabular}
\end{center}
\end{table}

The internal structure of the nucleus has two different effects. The first one comes  from  the modification of the atomic wavefunction when changing from a point to a finite nuclear charge distribution in \eqref{eq:dirac}, leading to a change in the matrix element in \eqref{eq:hfsme}. This is called the Breit-Rosenthal effect \cite{rab1932}.

The second one, called the Bohr-Weisskopf  effect \cite{baw1950}, comes from changes in \eqref{eq:hfsme} when one assumes an extended magnetic moment distribution. Following \cite{mdf1978,fwdm1980}, one can replace 
equation~\eqref{eq:hfsme} by 
\begin{eqnarray}
	\left[r^{k}\right]^{\mbox{\tiny{B.-W.}}}_{n\kappa\,n'\kappa'}&=&
	\int_{0}^{R_{\mbox{\tiny{nuc}}}}ds \,
	s^{2} \rho_{\mbox{\tiny{Mag}}} (s)  \int_{0}^{s} dr \,
	r^{k}\nonumber \\
	&& 
	\times \left(
	P_{n \kappa}(r)Q_{n' \kappa'}(r)+Q_{n \kappa}(r)P_{n' 
	\kappa'}(r)\right) 
	\label{eq:hfsmebw}
\end{eqnarray}
where $R_{\mbox{\tiny{nuc}}}$ is the radius of the nuclear distribution, which can be infinite in the case, \eg, of a Fermi distribution, and $\rho_{\mbox{\tiny{Mag}}}$ is the nuclear magnetic moment density.
Equation \eqref{eq:hfsmebw} was originally obtained to describe  the interaction of the muon magnetic moment to the electronic one in a muonic atom, using the unretarded part of the Breit interaction.
Here it is used to couple the nuclear magnetic moment density to the bound electron or muon.
If one assumes a Fermi distribution for $\rho_{\mbox{\tiny{Mag}}}$ it must be normalized so that
\begin{equation}
	\rho_{\mbox{\tiny{Mag}}}(r)=\frac{1}{\mu_{I}\mu_{N}}\frac{ \rho_{\mathrm{Mag}}^0}{1+\exp{\frac{r-c}{t}}} \; .
	\label{eq:magden}
\end{equation}
The $c$ and $t$ parameter are defined in \eqref{eq:fermi-model}. The nuclear parameters $c$ and $t$ should not be identical to their counterpart in the charge distribution, as the neutrons distribution also impact the magnetic moment distribution. The parameter $\rho_{\mathrm{Mag}}^0$ 
is defined so that the integral of $\rho_{\mbox{\tiny{Mag}}}$ over the nuclear volume is equal to $\mu_{I}\mu_{N}$. The Bohr-Weisskopf effect has been evaluated in a number of works for one and three-electron ions \cite{bai2000,nhn2004}.
In  \cite{nhn2004}, the nuclear part for $^{207}$Pb and $^{209}$Bi is evaluated in a relativistic model,  with Lorentz covariant current, and gives a result within a few \% of the measured values.
Much work has been devoted to calculating the first QED corrections to the energy shift above. The corresponding one-loop QED diagrams are presented on figure \ref{fig:qed-corr-extern}. In   Boucard \& Indelicato \cite{bai2000}, the self-energy correction has been evaluated for the ground state of one and three-electron ions, and all stable isotopes with non-zero nuclear spin, using a $Z\alpha$ expansion from \cite{pac1994,kar1996}. The vacuum polarization correction to the wavefunction has been obtained by including the Uehling potential in the Dirac equation or MCDF equation. Full QED calculations of the self-energy for H-like ions have been performed in several works, very accurately for low-$Z$ ions  \cite{yasp2005,yaj2008,yaj2010}, or medium and large $Z$ values \cite{psgs1996,stka1997,spss1998,yas2001,vgto2008}.

In lithiumlike ions, the screened self-energy contribution to the HFS has been evaluated in \cite{vgst2009}. Both H-like and Li-like ion QED corrections have been evaluated in \cite{vgas2012}. In parallel, calculations were performed for Li-like Bi, in order to help find the line which was investigated by laser spectroscopy at GSI \cite{bai2000,sac2001a}.

Evaluation of the vacuum polarization contribution to the HFS with finite nuclear size corrections were  performed  \cite{sgs1994,stka1997,spss1998,bai2000,yasp2005,vgto2008}. The calculation has been extended to Li-like Bi \cite{bai2000,vgto2008,agvs2012}.

Measurements of the hyperfine structure in one and three electron ions has been performed mainly at the Livermore EBIT and the ESR storage ring at GSI, Darmstadt. The spontaneous emission from a hyperfine transition was first observed in hydrogenlike $^{165}\textrm{Ho}^{65+}$, excited by collisions with the electrons in an EBIT \cite{lbsw1996}. The method was later used to measure the nuclear magnetic moment distribution in hydrogenlike $^{185}$Re$^{74+}$ and  $^{187}$Re$^{74+}$ \cite{lbwb1998}, and $^{203}$Tl$^{80+}$ and $^{205}$Tl$^{80+}$ \cite{buwl2001}. More recently this method was extended to the hyperfine structure of Li-like and Be-like $^{141}$Pr \cite{btbc2014}. At the ESR, laser spectroscopy was used in hydrogenlike \cite{kbef1994,uadg2016,uabd2017} and lithiumlike Bi \cite{ljga2014,uabd2017} and on hydrogenlike Pb \cite{sbde1998}. 
A different type of experiment has been performed on Li-like bismuth by comparing x-ray energies of transitions ending on different hyperfine levels of the ground state $1s^2 2s \, ^2S_{1/2,F}$ \cite{bosl1998}, but the accuracy is not as good as with direct  observation with a UV spectrometer. 

In the last few years the experimental accuracy on H-like and Li-like Bi has made tremendous progress, responding to the challenge of the interpretation of the results with a strong disagreement between theory and experiment  \cite{uadg2016,uabd2017,uabd2017}. 

On the theory side, to tackle the difficulty associated with taking into account the Bohr-Weisskopf effect, several methods have been used. One consists in combining simultaneous calculations with muonic atoms and highly-charged ions, obtaining from the muonic atoms the nuclear parameters to evaluate the Bohr-Weisskopf effect in the ions \cite{esot2005}. Yet the uncertainty of the results is greatly affected by the nuclear model dependence. 
Nuclear calculations were also used to try to predict  the Bohr-Weisskopf effect. The dynamic-correlation model (DCM) \cite{tksh1998,tknb2002} was used to evaluate the HFS of $^{165}$Ho$^{66+}$, $^{185}$Re$^{74+}$, $^{187}$Re$^{74+}$, $^{207}$Pb$^{81+}$ \cite{tksh1998} and $^{203}$Tl$^{80+}$ and $^{205}$Tl$^{80+}$  \cite{tknb2002}. Nuclear wavefunctions evaluated with a Wood-Saxon potential were also proposed to evaluate this effect \cite{stka1997,gfm2000,esot2005}. The dynamical model of hyperfine interaction has been used by Labzowsky \etal \cite{lnps1997}.  The Migdal’s finite Fermi system theory was also used for Pb and Bi in  \cite{sad2002}. One can also mention the model-independent approach, making an analogy between the Bohr-Weisskopf effect and the internal conversion coefficient anomalies \cite{kat2015}.

Table \ref{tab:hlike-hfs-1s} contains the theoretical and experimental values known so far for the ground state of hydrogenlike ions. 

\begin{figure*}[htbp]
\centering
        \includegraphics[width=0.6\textwidth]{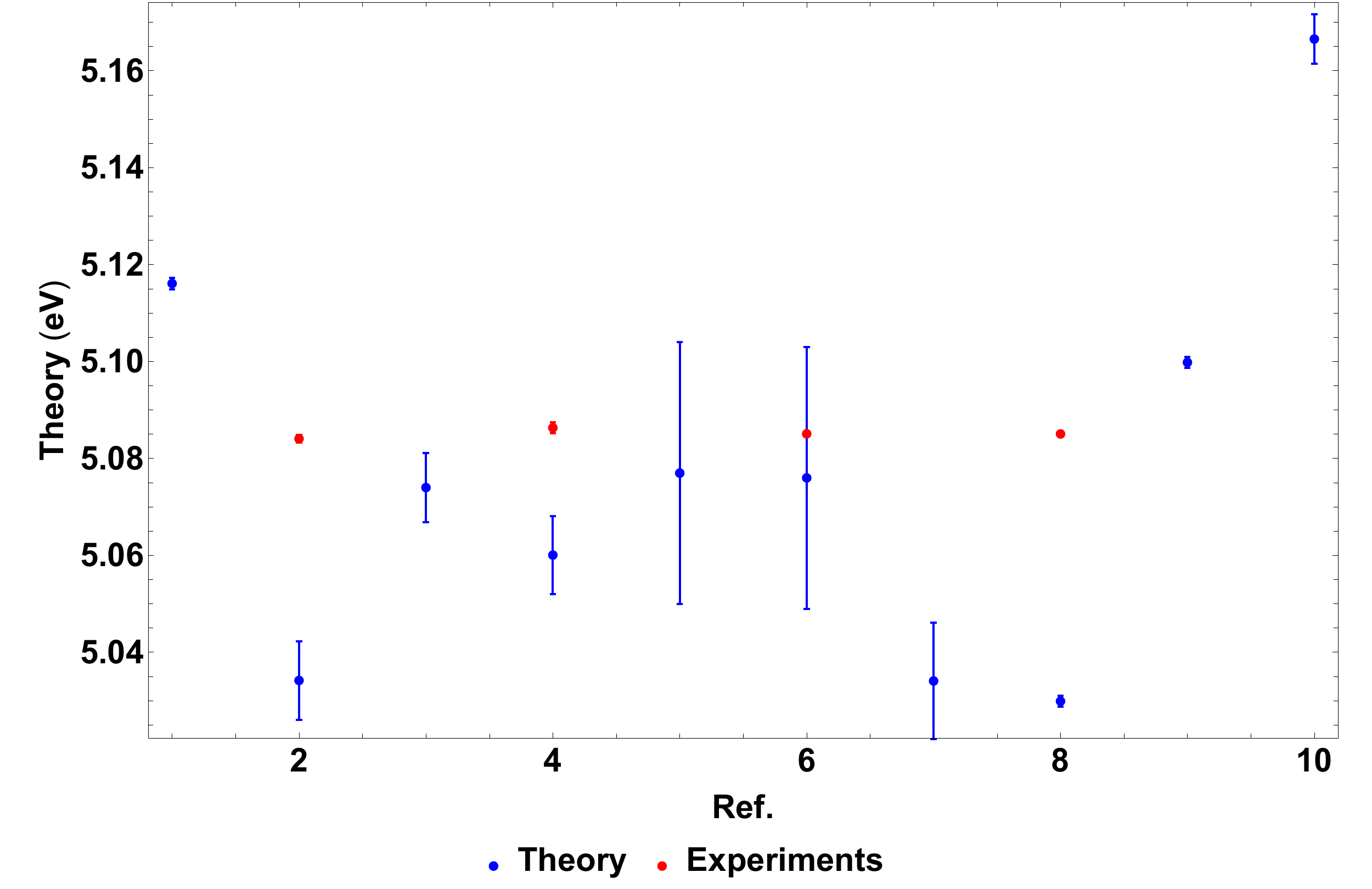}
\caption{
Plot of the experimental and of the corrected theoretical values for the hyperfine structure of $^{209}$Bi$^{82+}$ presented in table \protect \ref{tab:hlike-hfs-1s}.
Experiments: from left to right: \protect \cite{kbef1994,ljga2014,uadg2016,uabd2017}. Theory: from left to right: \protect \cite{sha1994,psgs1996,esot2005,sha1999,stka1997,sazy2000,tksh1998,bai2000}.
\label{fig:bi-hfsexp-th}
}
\end{figure*}

These data are also shown in figure \ref{fig:bi-hfsexp-th}. The theoretical values are scattered, due mostly to the variations in the methods used to treat the Bohr-Weisskopf effect. 

\begin{center}
\begin{longtable}{ccccD{.}{.}{4}D{.}{.}{4}D{.}{.}{4}D{.}{.}{4}D{.}{.}{4}cl}
\caption{ Experimental and theoretical HFS transition energies  for the ground state of H-like ions. The value of the nuclear magnetic moment with which the theoretical calculation have been performed is displayed.
The column QED specifies if the calculation contains QED corrections or not.}
\label{tab:hlike-hfs-1s}\\
\hline
\hline
Symbol	&	Z	&	A	&	I	&	\multicolumn{1}{c}{Exp. (eV)}	&	\multicolumn{1}{c}{Ex. Err. (eV)}	&	\multicolumn{1}{c}{Th.  (eV)}	&	\multicolumn{1}{c}{Th. Err. (eV)}	&	\multicolumn{1}{c}{$\mu_I$ Theo.}	&	QED	&	Ref.	\\
\endfirsthead
\caption{ Experimental and theoretical HFS transition energies  in H-like ions  (continued)}\\
\hline
\hline
Symbol	&	Z	&	A	&	I	&	\multicolumn{1}{c}{Exp. (eV)}	&	\multicolumn{1}{c}{Ex. Err. (eV)}	&	\multicolumn{1}{c}{Th.  (eV)}	&	\multicolumn{1}{c}{Th. Err. (eV)}	&	\multicolumn{1}{c}{$\mu_I$ Theo.}	&	QED	&	Ref.	\\
		
\endhead
\hline 
\endfoot

\hline 
\endlastfoot
Ho	&	67	&	165	&	5/2	&	2.16513	&	0.00057	&		&		&		&		&	\cite{lbsw1996}	\\
	&		&		&		&		&		&	2.1987	&		&	4.1730	&	n	&	\cite{sha1994}	\\
	&		&		&		&		&		&	2.1550	&		&	4.1320	&	y	&	\cite{tksh1998}	\\
	&		&		&		&		&		&	2.1659	&	0.0066	&	4.1320	&	y	&	\cite{stka1997}	\\
	&		&		&		&		&		&	2.189	&	0.007	&	4.1770	&	y	&	\cite{sazy2000}	\\
	&		&		&		&		&		&	2.1602	&		&	4.1320	&	y	&	\cite{bai2000}	\\
	&		&		&		&		&		&	2.1817	&		&	4.1730	&	y	&	\cite{bai2000}	\\
Re	&	75	&	185	&	5/2	&	2.7194	&	0.0018	&		&		&		&		&	\cite{lbwb1998}	\\
	&		&		&		&		&		&	2.7638	&		&	3.1871	&	n	&	\cite{sha1994}	\\
	&		&		&		&		&		&	2.7064	&		&	3.1871	&	y	&	\cite{tksh1998}	\\
	&		&		&		&		&		&	2.749	&	0.01	&	3.1871	&	y	&	\cite{stka1997}	\\
	&		&		&		&		&		&	2.748	&	0.01	&	3.1871	&	y	&	\cite{sazy2000}	\\
	&		&		&		&		&		&	2.7408	&		&	3.1871	&	y	&	\cite{bai2000}	\\
Re	&	75	&	187	&	5/2	&	2.7456	&	0.0018	&		&		&		&		&	\cite{lbwb1998}	\\
	&		&		&		&		&		&	2.7760	&	0.01	&	3.2197	&	y	&	\cite{sazy2000}	\\
	&		&		&		&		&		&	2.7322	&		&	3.2197	&	y	&	\cite{tksh1998}	\\
	&		&		&		&		&		&	2.7683	&		&	3.2197	&	y	&	\cite{bai2000}	\\
Tl	&	81	&	203	&	1/2	&	3.21351	&	0.00025	&		&		&		&		&	\cite{buwl2001}	\\
	&		&		&		&		&		&	3.2440	&		&	1.6223	&	n	&	\cite{sha1994}	\\
	&		&		&		&		&		&	3.2129	&		&	1.6217	&	n	&	\cite{tknb2002}	\\
	&		&		&		&		&		&	3.229	&	0.018	&	1.6223	&	y	&	\cite{stka1997}	\\
	&		&		&		&		&		&	3.229	&	0.017	&	1.6217	&	y	&	\cite{gfm2000}	\\
	&		&		&		&		&		&	3.2239	&		&	1.6223	&	y	&	\cite{bai2000}	\\
	&		&		&		&		&		&	3.22	&	0.02	&		&		&	\cite{esot2005}	\\
Tl	&	81	&	205	&	1/2	&	3.24410	&	0.00029	&		&		&		&		&	\cite{buwl2001}	\\
	&		&		&		&		&		&	3.2748	&		&	1.6382	&	n	&	\cite{sha1994}	\\
	&		&		&		&		&		&	3.2390	&		&	1.6372	&	n	&	\cite{tknb2002}	\\
	&		&		&		&		&		&	3.261	&	0.018	&	1.6382	&	y	&	\cite{stka1997}	\\
	&		&		&		&		&		&	3.261	&	0.018	&	1.6379	&	y	&	\cite{gfm2000}	\\
	&		&		&		&		&		&	3.2549	&		&	1.6382	&	y	&	\cite{bai2000}	\\
	&		&		&		&		&		&	3.238	&	0.009	&		&		&	\cite{esot2005}	\\
Pb	&	82	&	207	&	1/2	&	1.21589	&	0.00024	&		&		&		&		&	\cite{sbde1998}	\\
	&		&		&		&		&		&	1.2191	&		&	0.5870	&	n	&	\cite{sha1994}	\\
	&		&		&		&		&		&	1.2347	&		&	0.5926	&	y	&	\cite{psgs1996}	\\
	&		&		&		&		&		&	1.2105	&	0.0024	&	0.5822	&	y	&	\cite{tksh1998}	\\
	&		&		&		&		&		&	1.215	&	0.005	&	0.5926	&	y	&	\cite{stka1997}	\\
	&		&		&		&		&		&	1.215	&	0.005	&	0.5926	&	y	&	\cite{sazy2000}	\\
	&		&		&		&		&		&	1.2208	&		&	0.5822	&	y	&	\cite{bai2000}	\\
	&		&		&		&		&		&	1.2427	&		&	0.5926	&	y	&	\cite{bai2000}	\\
	&		&		&		&		&		&	1.222	&		&	0.5925	&	y	&	\cite{sad2002}	\\
Bi	&	83	&	209	&	9/2	&	5.08403	&	0.00083	&		&		&		&		&	\cite{kbef1994}	\\
	&		&		&		&	5.0863	&	0.0011	&		&		&		&		&	\cite{ljga2014}	\\
	&		&		&		&	5.08505	&	0.00013	&		&		&		&		&	\cite{uadg2016}	\\
	&		&		&		&	5.085027	&	0.000091	&		&		&		&		&	\cite{uabd2017}	\\
	&		&		&		&		&		&	5.1403	&		&	4.1106	&	n	&	\cite{sha1994}	\\
	&		&		&		&		&		&	5.058	&	0.0080	&	4.1106	&	y	&	\cite{psgs1996}	\\
	&		&		&		&		&		&	5.098	&	0.0070	&	4.1106	&		&	\cite{esot2005}	\\
	&		&		&		&		&		&	5.102	&	0.027	&	4.1106	&		&	\cite{sha1999}	\\
	&		&		&		&		&		&	5.101	&	0.027	&	4.1106	&	y	&	\cite{stka1997}	\\
	&		&		&		&		&		&	5.100	&	0.027	&	4.1106	&	y	&	\cite{sazy2000}	\\
	&		&		&		&		&		&	5.06	&	0.01	&	4.1106	&	y	&	\cite{tksh1998}	\\
	&		&		&		&		&		&	5.054	&		&	4.1106	&	y	&	\cite{bai2000}	\\
	&		&		&		&		&		&	5.124	&		&	4.1106	&	n	&	\cite{ljss1995}	\\
	&		&		&		&		&		&	5.111	&	0.0050	&	4.110	&	y	&	\cite{sad2002}	\\	
	&		&		&		&		&		&	5.089	&	0.002	&	4.092	&	y	&	\cite{ssug2018}	\\
								\hline
\end{longtable}
\end{center}

\begin{figure*}[htbp]
\centering
        \includegraphics[width=0.6\textwidth]{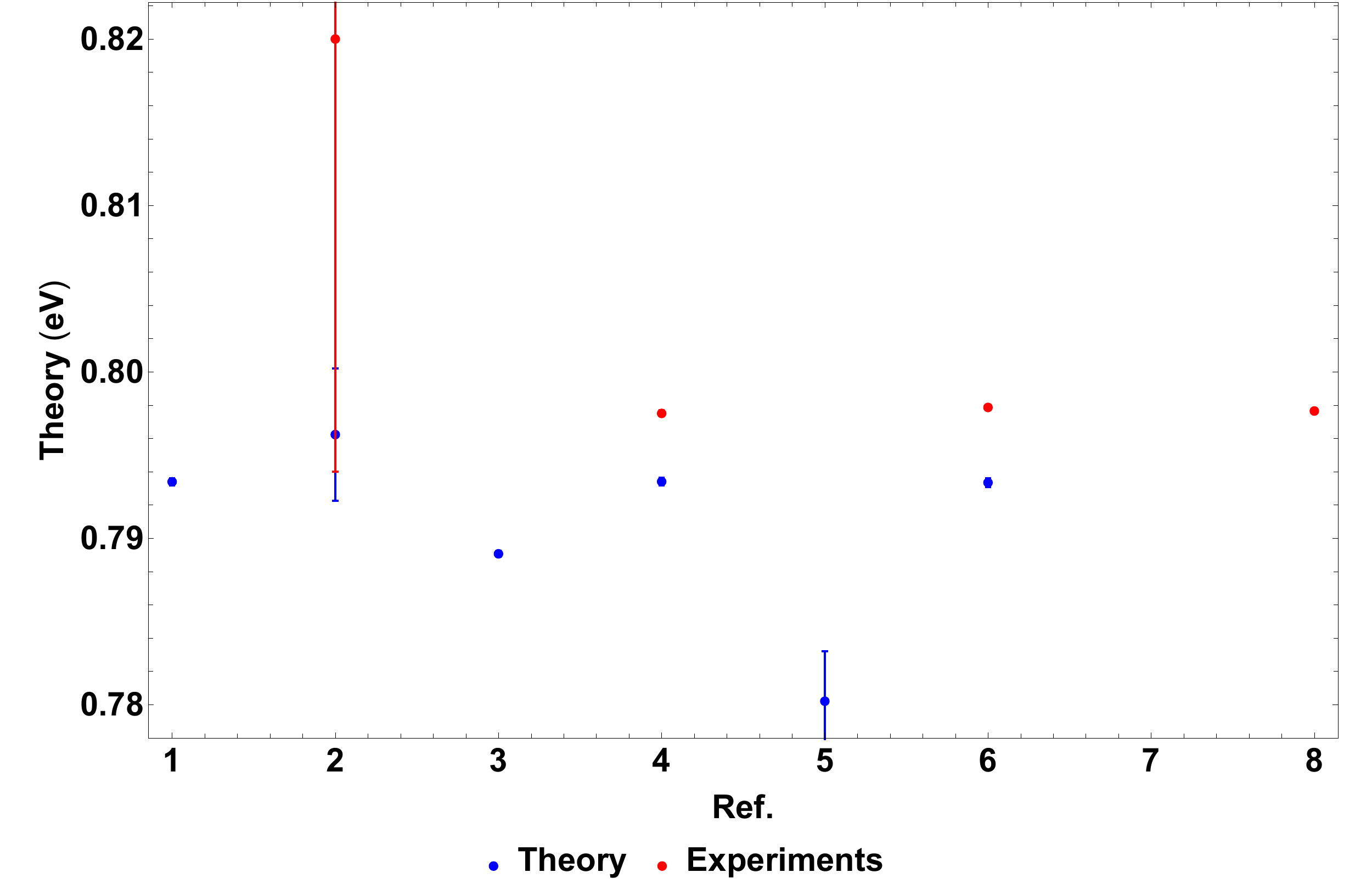}
\caption{
Plot of the experimental and theoretical values for the hyperfine structure of $^{209}$Bi$^{80+}$ presented in table \protect \ref{tab:lilike-hfs-2s}.
Experiments: from left to right: \protect \cite{bosl1998,ljga2014,slja2017,uabd2017}. Theory: from left to right: \protect \cite{sac2001a,ssty1998,bai2000,vgas2012,tfkw2000,sazy2000}.
\label{fig:bi-li-hfsexp-th}
}
\end{figure*}

Table \ref{tab:lilike-hfs-2s} contains the theoretical and experimental values for the ground state hyperfine structure of lithiumlike ions.  These data are also shown in figure \ref{fig:bi-li-hfsexp-th}.

In order to perform QED tests without measuring the nuclear magnetic moment distribution or linking it to the internal nuclear structure, another method  has been proposed which consists in combining the measurements of  H-like and Li-like ions of the same element to eliminate the Bohr-Weisskopf contribution \cite{ssty1998a,ssty1998,sayz2001}. The so-called \emph{specific difference} employed in this case is
\begin{equation}
\Delta E_{\mathrm{HFS}}=E_{\mathrm{HFS}}^{2s}-\xi E_{\mathrm{HFS}}^{1s}
\label{eq:specific-diff-hfs}
\end{equation}
where $\xi$ has been evaluated as $\xi=0.16886$ for Bi. This approximation is based on the fact that inside the nucleus, the binding energy of an electron is very small compared to the nuclear Coulomb potential. The $1s$ and $2s$ wavefunctions inside the nucleus are thus weakly dependent on the respective binding energies. It was thus shown in  \cite{sha1994}, that the integrals over the atomic wavefunction for the $1s$ and $2s$, which are used to evaluate the Bohr-Weisskopf correction (see, \eg equations (4) to (8) in  \cite{ssty1998}) differ only by an overall factor, leading to \eqref{eq:specific-diff-hfs}.

Using the new  magnetic moment and theoretical values obtained in \cite{ssug2018} and the corresponding HFS energy, one obtains $\Delta E_{\mathrm{HFS}}=\SI{0.061043+-0.000005}{\electronvolt}(30)$, while the theoretical value from Volotka \etal \cite{vgas2012}, rescaled with the new magnetic moment value is \SI{-0.061042+-0.000064}{\electronvolt} and the experimental value \SI{0.061012+-0.000005}(21) \cite{uabd2017}. All three values are thus in excellent agreement.

\begin{table}
\begin{center}
\caption{Experimental and theoretical HFS transition energies  for the ground state of Li-like ions. The value of the nuclear magnetic moment with which the theoretical calculation have been performed is displayed.
The column QED specifies if the specific calculation contains QED corrections or not.}
\label{tab:lilike-hfs-2s}
\begin{tabular}{ccccD{.}{.}{4}D{.}{.}{4}D{.}{.}{4}D{.}{.}{4}D{.}{.}{4}cl}
\hline
\hline
Symbol	&	Z	&	A	&	I	&	\multicolumn{1}{c}{Exp. (eV)}	&	\multicolumn{1}{c}{Ex. Err. (eV)}	&	\multicolumn{1}{c}{Th.  (eV)}	&	\multicolumn{1}{c}{Th. Err. (eV)}	&	\multicolumn{1}{c}{$\mu_I$ Theo.}	&	QED	&	Ref.	\\
\hline
Pr	&	59	&	141	&	5/2	&	0.19650	&	0.00120	&		&		&		&		&	\cite{btbc2014}	\\
	&		&		&		&		&		&	0.1974	&		&	4.2754	&	y	&	\cite{ssty1998}	\\
	&		&		&		&		&		&	0.1975	&		&	4.2754	&	y	&	\cite{bai2000}	\\
Bi	&	83	&	209	&	9/2	&	0.820	&	0.026	&		&		&		&		&	\cite{bosl1998}	\\
	&		&		&		&	0.79750	&	0.00018	&		&		&		&		&	\cite{ljga2014}	\\
	&		&		&		&	0.79786	&	0.00015	&		&		&		&		&	\cite{slja2017}	\\
	&		&		&		&	0.7976456	&	0.0000015	&		&		&		&		&	\cite{uabd2017}	\\
	&		&		&		&		&		&	0.79715	&	0.00013	&	4.1106	&	y	&	\cite{sac2001a}	\\
	&		&		&		&		&		&	0.8000	&	0.0040	&	4.1106	&	y	&	\cite{ssty1998}	\\
	&		&		&		&		&		&	0.7928	&		&	4.1106	&	y	&	\cite{bai2000}	\\
	&		&		&		&		&		&	0.79716	&	0.00014	&	4.1106	&	y	&	\cite{vgas2012}	\\
	&		&		&		&		&		&	0.7839	&	0.0030	&	4.1106	&	y	&	\cite{tfkw2000}	\\
	&		&		&		&		&		&	0.79710	&	0.00020	&	4.1106	&	y	&	\cite{sazy2000}	\\
	&		&		&		&		&		&	0.7983	&	0.0004	&	4.0920	&	y	&	\cite{ssug2018}	\\
					\hline
\end{tabular}
\end{center}
\end{table}

Although the latest experiments and calculations have led to increased precision and understanding, we are still far from being able to make truly accurate tests of QED with HFS. While the weighted QED correction to the $1s$ HFS energy in hydrogenlike  Bi is $\xi \Delta E^{1s}_{\mathrm{QED}}=\SI{5.088}{\milli\electronvolt}$ and  is $\Delta E^{2s}_{\mathrm{QED}}=\SI{5.052}{\milli\electronvolt}$ in lithiumlike  Bi, the value of the difference is only   $\delta E_{\mathrm{QED}}=\Delta E^{2s}_{\mathrm{QED}}-\xi \Delta E^{1s}_{\mathrm{QED}}=\SI{0.036}{\milli\electronvolt}$. This value is thus \num{1700} times smaller than the weighted difference of energies $\delta E_{\mathrm{Tot.}}$, and \num{880} times smaller than the weighted differences of Dirac values or  many-body corrections  \cite{vgas2012}. The latter contributes only to the Li-like energy. This value is also 5 times smaller than the screened QED correction which is the main QED contribution to the weighted difference. It should also be noted that the ratio of the QED correction to the transition energy in hydrogenlike Bi is  5 times larger than $\delta E_{\mathrm{QED}}/\delta E_{\mathrm{Tot.}}$.

So while the new calculations and measurements of the $^{209}$Bi  magnetic moments \cite{ssug2018,arjj2018}  enabled to solve the \emph{hyperfine structure puzzle} found in the experiment \cite{uabd2017},  it does not improve our capacity to test QED. 
New experiments are planed to work on this aspect, like a measurement of the hyperfine structure of $^{208}$Bi at the ESR to be able to have system with identical atomic properties and different nuclear corrections. In preparation for this experiment, the magnetic moment of  $^{208}$Bi has been recently remeasured at ISOLDE \cite{sbbb2018}.

\subsection{Landé $g$-factors}
\label{sec:lande}
In this section, I will describe recent progress in theory and experimental measurements of the bound-electron g-factor in hydrogenlike and lithiumlike ions.

The bound electron $g$-factor connects the electron dipole magnetic moment and its total angular momentum
\begin{equation}
\bm{\mu}_J =- g^{(e)}_J \mu_B \frac{\bm{J}}{\hbar},
\label{eq:g-fact-def}
\end{equation}
 where $\mu_\mathrm{B}$ is the Bohr magneton and $\bm{J}$ the total angular momentum of the electron. The energy of the atom in a magnetic field is then
\begin{equation}
\Delta E = - \langle a | \bm{\mu}_J \cdot \bm{B} | a \rangle.
\label{eq:g-fact-bfield}
\end{equation}

One can then evaluate
\begin{equation}
g^{(e)}=-\frac{\left\langle J M_J| \mu_z^{(e)}| J M_J\right\rangle}{\mu_\mathrm{B}M_J}
\label{eq:g-fact}
\end{equation}
where  $\mu_z^{(e)}$ is the $z$-component of the electron magnetic moment,  and $M_J$ the projection of  $\bm{J}$.

For a point nucleus, the Dirac solution for the Landé $g$-factor was derived by Breit \cite{bre1928}. He found for the $1s$ of a one-electron atom
\begin{equation}
g^{(e)}_{\mathrm{Dirac}} = 2 + \frac{4}{3}\left[\sqrt{1-(Z\alpha)^2}-1\right].
\label{eq:dirac-g-fact}
\end{equation}
Breit already noted at the time the strong decrease of the $g$-factor for high-$Z$. Higher-order terms, from the $Z\alpha$ expansion in QED were provided in \cite{gro1970,gah1971}.
When adding QED, the Landé $g$-factor can be written as
\begin{eqnarray}
g^{(e)} & =& 2\left[\frac{g^{(e)}_{\mathrm{Dirac}}}{2} + \left(\frac{\alpha}{\pi}\right)C^{(2)} + \left(\frac{\alpha}{\pi}\right)^2C^{(4)} + \left(\frac{\alpha}{\pi}\right)^3C^{(6)}+ \cdots \right],
\label{eq:g-fact-qed}
\end{eqnarray}
where the $C^{(n)}$ coefficients can be written 
\begin{equation}
C^{(2n)}=A_1^{(2n)}+ \textrm{bound QED effects}
\end{equation}
where $A_1^{(2n)}$ is the contribution to the free electron anomalous magnetic moment of order $(\alpha/\pi)^n$ (see, \eg  \cite{blps2000}). This expression is also sensitive to finite nuclear size corrections.
A complete description of the different theoretical methods and contributions can be found in \cite{bei2000,sgpv2015}.

There is a close connection between the Landé factor and the HFS described in  the previous section. Shabaev \cite{sha1998b} has shown that there exists a link between the HFS energy and transition probability and the bound-electron $g$-factor 
\begin{equation}
\omega_{F_i\to F_f}^{\mathrm{HFS}}=\frac{\alpha}{3}\frac{\left(\Delta E_{F_i\to F_f}^{\mathrm{HFS}}\right)^3}{\hbar\left(m_ec^2\right)^2}\frac{I}{2I+1}\left[g^{(e)}_J-\frac{m_e}{m_p}\mu_{I}\right]
\label{eq:hfs-g}
\end{equation}
where $m_p$ is the proton mass and $\mu_{I}$ the nuclear magnetic moment as defined in \eqref{eq:magmoment}. 
In this equation, $g^{(e)}_J$ is the sum of the Breit contribution \eqref{eq:dirac-g-fact}, and of the QED and finite nuclear size corrections. The transition energy $\Delta E_{F_i\to F_f}^{\mathrm{HFS}}$ contains the corresponding QED and finite size corrections.
The magnetic dipole approximation is very good as the HFS transition wavelength is large compared to the size of the ions. The radiative corrections to the transitions rate, however are not included. 
This allows to extract an experimental value for the $g$-factor for the elements for which the transition probability between hyperfine levels has been measured, like $^{209}$Bi$^{82+}$. In this case one finds \SI{49.5+-6.5}{\milli\second} \cite{sbde1998} leading to $g^{(e)}=$\num{1.78+-.12}.
It would require an improvement of the transition probability by two orders of magnitude to be sensitive to the QED corrections through \eqref{eq:hfs-g}.

In the recent years there has been a strong push to measure very accurately the $g$-factor. It can be connected to  $\nu_{\mathrm{L}}$, the Larmor frequency, and to $\nu_{\mathrm{c}}$, the cyclotron frequency of an ion trapped in a Penning trap by
\begin{equation}
g^{(e)} = 2 \frac{q}{e}\frac{m_e}{M}\frac{\nu_{\mathrm{L}}}{\nu_{\mathrm{c}}}\, ,
\label{eq:g-fact-freq-exp}
\end{equation}
where $q$ is the ion charge state, $e$ the unit charge, $m_e$ is the electron mass and  $M$ is the nuclear mass. With the progress made in the handling of Penning traps over the years, which allows to measure Landé factors and nuclear masses with extreme precision, and theoretical advances, which allow to evaluate $g^{(e)}$ very accurately provided the fundamental  constants are well measured,  equation  \eqref{eq:g-fact-freq-exp} allows one to deduce the electron to proton mass ratio.
In parallel, for heavier elements, one can think of using \eqref{eq:dirac-g-fact} completed with all the QED corrections, which also depend on the fine structure constant $\alpha$, to derive it accurately. Since the dependance is in $Z\alpha$ it is of course better to do it for large values of $Z$ although in this case the nuclear size corrections start to be large enough to limit the accuracy. More details on both possibilities and experimental methods can be found in recent reviews  \cite{vog2009,svkq2017}.
When nuclear corrections become an issue, the St Petersburg group has proposed a way to limit the effect of nuclear corrections, in the same spirit as for the hyperfine structure, \ie to combine the $g$-factor of the ground states of hydrogenlike ions and ions with more electrons.
The difference between the $g$-factor of the $1s$ level in hydrogenlike  ions and the one of the $2s$ level in lithiumlike ions has thus been shown to be relatively independent of the nuclear corrections \cite{ybht2016,ybht2016a}. In the same way, a comparison of H-like and B-like $g$-factors  measured in the same spinless isotope of lead \cite{sgov2006} allows to reduce the effect of the nuclear size correction uncertainties. 

The evaluation of the different QED corrections to the $g$-factor are relatively similar to the hyperfine structure, and are represented by identical diagrams as in figure \ref{fig:qed-corr-extern}. The nuclear recoil contribution has also been evaluated in a number of papers \cite{sha2001,say2002,klm2005,pac2008,msgt2017}.  The QED diagrams resemble the ones for the free-electron anomalous magnetic moment evaluation, with the free electron propagator replaced by the bound one. It thus suffers from the same complexity, with many more diagrams to evaluate at a given order than for energy.  All-order one-loop radiative corrections have been evaluated beyond the Breit and Grotch term  in  several works \cite{pssl1997,blps2000,yps2002,yis2002,yis2002a,sha2002,yis2004,yaj2008,pcjy2005,yaj2010,yah2017}. The nuclear size correction to the one-loop QED diagrams have also been calculated  \cite{gas2002,ykh2013}. Coefficients for the two-loop QED correction expansion in $Z\alpha$, beyond the free electron value \cite{som1957}  have been evaluated \cite{pcjy2005,cmy2000,cdps2018}. More recently all-order two-loop radiative corrections have been calculated as well \cite{yah2013}. The changes between the known $Z\alpha$ expansion results and the new all-order results are very large.

Recoil corrections are important and have been evaluated in the QED framework for  H-like ions\cite{sha2001,say2002,msgt2017}. The effect of the nuclear magnetic moment on the $g$-factor for nuclei with non-zero spin has also been evaluated \cite{mosb2004}. This could allow to measure the nuclear magnetic moment via a measurement of the $g$-factor.

Screened QED corrections for Li-like ions have also been calculated \cite{gstv2004,vgst2009,vgst2014} as well as recoil corrections \cite{sgmt2017,sgmt2018}. Recent reviews of most contributions are available in  \cite{bisy2003} and \cite{sgpv2015}.
As in energy calculations, the $g$-factor evaluation is affected by nuclear corrections beyond the finite nuclear size correction. In particular, nuclear polarization has been studied \cite{nps2002} to see what limitations there could be in going to higher-$Z$.
It was found that the nuclear polarization effects did set a limit of around \num{1E-9} for medium-$Z$ and \num{1E-6} for the heaviest elements leading to the same kind of limitations to test QED or derive fundamental constants  than for transition energies.

The continuous Stern and Gerlach effect \cite{deh1986} on hydrogenlike ions was used to measure $g^{(e)}$ on $^{12}$C$^{5+}$  \cite{hbhk2000,skzw2014,kskw2015}. Over time, the trap technology was improved, as well as the ion creation techniques.
Recent experiments used a triple trap design with a precision trap, an analysis trap and an \emph{in situ} miniature EBIT \cite{skzw2014}. There were also measurements on hydrogenlike oxygen $^{16}$O$^{7+}$  \cite{vdsv2004} and $^{28}$Si$^{13+}$   \cite{swsz2011,sswa2012,swkq2013}.  The result of these measurements, together with the different contributions to the calculation can be found in tables \ref{tab:h-like-gfac} (for He, C, and O) and \ref{tab:h-like-gfac2} for Si and Ca.
By comparing the theoretical and experimental value in Si, it was possible to estimate the next order missing correction and improve the theoretical value for C  \cite{skzw2014,kskw2015}. The improved semi-theoretical value is \num{2.001 041 590 179 8 +-0.0000000000047} to be compared with the fully theoretical value from table  \ref{tab:h-like-gfac} of \num{2.001041590176 +-0.000000000006}.
Assuming this improved value of $g^{(e)}$  it was possible to obtain the electron mass as $m_e=$\SI{0.000 548 579 909 069 4 +-0.0000000000000155}{u} \cite{hbhk2000,skzw2014,kskw2015}, a 13-fold improvement compared to the 2010 CODATA value \cite{mtn2012}. The current 2014 CODATA value is \SI{0.000548579909070+-0.000000000000016}{u}, clearly dominated by this improved result \cite{mnt2016}. In 2017, improved theory slightly shifted the electron mass to  \SI{0.000548579909065+-0.000000000000016}{u} \cite{zsks2017}. This work also proposed new measurements to improve the result, like in He$^+$. The table shows that at the moment, for light to medium-$Z$ elements, the nuclear contribution is limited to the finite nuclear size correction, and will not constitute a limitation for improving fundamental constants.
In the future the new setup ALPHATRAP should help improve the results even more \cite{swb2013,svkq2017}.

Lithiumlike  $^{28}$Si$^{11+}$  \cite{wskg2013} and $^{40}$Ca$^{17+}$  \cite{zsks2017} were then measured. A comparison between two doubly magic isotopes of Ca,  $^{40}$Ca$^{17+}$ and  $^{48}$Ca$^{17+}$, was also performed  \cite{kbbc2016}, enabling a better understanding of the nuclear effects.

\begin{center}
\begin{table}
\caption{Theoretical contributions to the Landé $g$-factor and comparison with experiment for He, C and O. The value corresponding to \cite{skzw2014,kskw2015} has been derived from equations (3), (53), (55) and (56) in \cite{kskw2015}.
}
\label{tab:h-like-gfac}
\footnotesize
\begin{tabular}{llD{.}{.}{14}D{.}{.}{13}D{.}{.}{11}l}
\hline
\hline
Z	&		&	2		&	6		&	8		&	Ref.	\\
A	&		&	4		&	12		&	16		&		\\
nuclear radius(fm)	&		&	1.681		&	2.4703		&	2.7013		&		\\
\hline														
Contribution	&	order	&			&			&			&		\\
\hline
Dirac	&		&	1.99985798882537		&	1.9987213543921		&	1.99772600306		&	\cite{bre1928}	\\
Finite nuclear size	&		&	0.00000000000230		&	0.0000000004074		&	0.00000000155		&	\cite{zokh2012}	\\
One loop QED	&	$(Z\alpha)^0$	&	0.00232281946485		&	0.0023228194649		&	0.00232281946		&	\cite{mnt2016,sch1948}	\\
	&	$(Z\alpha)^2$	&	0.00000008246219		&	0.0000007421597		&	0.00000131940		&	\cite{gro1970}	\\
	&	$(Z\alpha)^4$	&	0.00000000197670		&	0.0000000934220		&	0.00000024007		&	\cite{pcjy2005}	\\
	&	h.o. SE	&	0.00000000003542		&	0.0000000082826		&	0.00000003443		&	\cite{yis2002,yis2004,pcjy2005}	\\
	&	SE-FS	&	0.00000000000000		&	-0.0000000000007		&	0.00000000000		&	\cite{ykh2013}	\\
	&	h.o. VP-EL	&	0.00000000000252		&	0.0000000005559		&	0.00000000224		&	\cite{ykh2013,kis2001}	\\
	&	VP-EL FS	&	0.00000000000000		&	0.0000000000002		&	0.00000000000		&	\cite{ykh2013}	\\
	&	h.o. VP-ML	&	0.00000000000016		&	0.0000000000381		&	0.00000000016		&	\cite{ykh2013,kis2001}	\\
	&	h.o. VP-ML FS	&	0.00000000000000		&	0.0000000000000		&	0.00000000000		&	\cite{lmtk2005,kam2002}	\\
Two-loop QED	&	$(Z\alpha)^0$	&	-0.00000354460449		&	-0.0000035446045		&	-0.00000354460		&	\cite{ykh2013,lmtk2005}	\\
	&	$(Z\alpha)^2$	&	-0.00000000012584		&	-0.0000000011325		&	-0.00000000200		&	\cite{pet1957,som1958}	\\
	&	$(Z\alpha)^4$ (w/o LBL)	&	0.00000000000241		&	0.0000000000601		&	0.00000000008		&	\cite{gro1970}	\\
	&	LBL at $(Z\alpha)^4$	&	-0.00000000000039		&	-0.0000000000315		&			&	\cite{pcjy2005,pjy2004}	\\
	&	$\ge (Z\alpha)^5$ S(VP)E	&	0.00000000000000		&	0.0000000000000		&			&	\cite{cas2016}	\\
	&	$\ge (Z\alpha)^5$ SEVP	&	0.00000000000003		&	0.0000000000069		&			&	\cite{yah2013}	\\
	&	$\ge (Z\alpha)^5$ VPVP	&	0.00000000000003		&	0.0000000000055		&			&	\cite{yah2013}	\\
	&	$\ge (Z\alpha)^5$ SESE (estimate)	&	0.00000000000000		&	-0.0000000000012		&	0.00000000005		&	\cite{yah2013,jen2009}	\\
Three-loop QED	&	$(Z\alpha)^0$	&	0.00000002949795		&	0.0000000294980		&	0.00000002950		&	\cite{say2002}	\\
	&	$(Z\alpha)^2$	&	0.00000000000105		&	0.0000000000094		&	0.00000000000		&	\cite{gro1970}	\\
Recoil	&	$m/M$ all-orders	&	0.00000002920251		&	0.0000000877251		&	0.00000011710		&	\cite{say2002}	\\
	&	$(m/M)^2$ order $(Z\alpha)^2$	&	0.00000000001201		&	-0.0000000000281		&	-0.00000000013		&	\cite{pac2008}	\\
	&	Radiat Recoil	&	-0.00000000002261		&	-0.0000000000679		&			&	\cite{bei2000,gro1970}	\\
Other corrections	&	Nuclear polarizability	&	0.00000000000000		&	0.0000000000000		&			&	\cite{nps2002}	\\
	&	Nuclear susceptibilty	&	0.00000000000000		&	0.0000000000000		&			&	\cite{jcpy2006}	\\
	&	Weak interaction $(Z\alpha)^0$	&	0.00000000000006		&	0.0000000000001		&			&	\cite{mnt2016,ckm1996}	\\
			\hline	
	&	Hadronic effects $(Z\alpha)^0$	&	0.00000000000347		&	0.0000000000035		&			&	\cite{nat2013,klms2014,prv2010}	\\
Total Theory	&		&	2.00217740673570	(87)	&	2.0010415901650	(6)	&	2.00004702036	(11)	&		\\
Theory corrected	&		&	2.00217740671168	(87)	&	2.0010415901652	(51)	&			&		\\
Experiment	&		&			&	2.0010415964	(45)	&	2.0000470254	(46)	&		\\
Exp. Reference	&		&			&	\multicolumn{1}{c}{\cite{hbhk2000}}		&	\multicolumn{1}{c}{\cite{vdsv2004}}		&		\\
Experiment	&		&			&	2.001041590180	(56)	&			&		\\
Exp. Reference	&		&			&	\multicolumn{1}{c}{\cite{skzw2014,kskw2015}}		&			&		\\
		\hline	

\end{tabular}
\normalsize
\end{table}
\end{center}

\begin{center}
\begin{table}
\caption{Theoretical contributions to the Landé $g$-factor and comparison with experiment for Si and Ca. 
}
\label{tab:h-like-gfac2}
\begin{tabular}{llD{.}{.}{13}D{.}{.}{13}l}
\hline
\hline
Z	&		&	14		&	20		&	Ref.	\\
A	&		&	28		&	40		&		\\
nuclear radius(fm)	&		&	3.1223		&	3.4764		&		\\
\hline											
Contribution	&	order	&			&			&		\\
\hline											
Dirac	&		&	1.993023571557		&	1.9857232037		&	\cite{bre1928}	\\
Finite nuclear size	&		&	0.000000020468		&	0.0000001130		&	\cite{zokh2012}	\\
One loop QED	&	$(Z\alpha)^0$	&	0.002322819465		&	0.0023228195		&	\cite{mnt2016,sch1948}	\\
	&	$(Z\alpha)^2$	&	0.000004040647		&	0.0000082462		&	\cite{gro1970}	\\
	&	$(Z\alpha)^4$	&	0.000001244596		&	0.0000025106		&	\cite{pcjy2005}	\\
	&	h.o. SE	&	0.000000542856		&	0.0000031077		&	\cite{yis2002,yis2004,pcjy2005}	\\
	&	SE-FS	&	-0.000000000068		&	0.0000000000		&	\cite{ykh2013}	\\
	&	h.o. VP-EL	&	0.000000032531		&	0.0000001727		&	\cite{ykh2013,kis2001}	\\
	&	VP-EL FS	&	0.000000000022		&	0.0000000000		&	\cite{ykh2013}	\\
	&	h.o. VP-ML	&	0.000000002540		&	0.0000000146		&	\cite{ykh2013,kis2001}	\\
	&	h.o. VP-ML FS	&	-0.000000000001		&	0.0000000000		&	\cite{lmtk2005,kam2002}	\\
Two-loop QED	&	$(Z\alpha)^0$	&	-0.000003544604		&	-0.0000035446		&	\cite{ykh2013,lmtk2005}	\\
	&	$(Z\alpha)^2$	&	-0.000000006166		&	-0.0000000125		&	\cite{pet1957,som1958}	\\
	&	$(Z\alpha)^4$ (w/o LBL)	&	-0.000000001318		&	-0.0000000109		&	\cite{gro1970}	\\
	&	LBL at $(Z\alpha)^4$	&	-0.000000000933		&			&	\cite{pcjy2005,pjy2004}	\\
	&	$\ge (Z\alpha)^5$ S(VP)E	&	0.000000000009		&			&	\cite{cas2016}	\\
	&	$\ge (Z\alpha)^5$ SEVP	&	0.000000000458		&			&	\cite{yah2013}	\\
	&	$\ge (Z\alpha)^5$ VPVP	&	0.000000000315		&			&	\cite{yah2013}	\\
	&	$\ge (Z\alpha)^5$ SESE (estimate)	&	-0.000000000082		&	0.0000000041		&	\cite{yah2013,jen2009}	\\
Three-loop QED	&	$(Z\alpha)^0$	&	0.000000029498		&	0.0000000295		&	\cite{say2002}	\\
	&	$(Z\alpha)^2$	&	0.000000000051		&	0.0000000000		&	\cite{gro1970}	\\
Recoil	&	$m/M$ all-orders	&	0.000000206100		&	0.0000002974		&	\cite{say2002}	\\
	&	$(m/M)^2$ order $(Z\alpha)^2$	&	-0.000000000060		&	-0.0000000003		&	\cite{pac2008}	\\
	&	Radiat Recoil	&	-0.000000000159		&			&	\cite{bei2000,gro1970}	\\
Other corrections	&	Nuclear polarizability	&	0.000000000000		&			&	\cite{nps2002}	\\
	&	Nuclear susceptibilty	&	0.000000000000		&			&	\cite{jcpy2006}	\\
	&	Weak interaction $(Z\alpha)^0$	&	0.000000000000		&			&	\cite{mnt2016,ckm1996}	\\
	&	Hadronic effects $(Z\alpha)^0$	&	0.000000000003		&	0.0000000000		&	\cite{nat2013,klms2014,prv2010}	\\
Total Theory	&		&	1.995348957722	(71)	&	1.9880569507	(100)	&		\\
Theory corrected	&		&	1.995348957708	(156)	&			&		\\
Experiment	&		&	1.99534895910	(81)	&			&		\\
Exp. Reference	&		&	\multicolumn{1}{c}{\cite{swkq2013}}	&			&		\\
		\hline	
\end{tabular}
\end{table}
\end{center}

\section{Conclusion}
\label{sec:conc}

In this review, I have analyzed four decades worth of experimental and theoretical results on transition energies, hyperfine structure splitting and Landé $g$-factors in few-electron ions.

A complete analysis of the experiment-theory differences show that there are no significant deviations between experiment and theory as a function of $Z$ for the $2p\to1s$ transitions in one and two-electron systems, nor in the $\Delta n=0$, $n=2$ transitions or splitting in heliumlike and lithiumlike ions. The measurements of the ground state energy in two- and three-electron systems are also in good agreement with theory. Yet it is clear that there is a lack of measurements in the range of atomic numbers $36\leq Z \leq 80$. Concomitantly the distribution of experimental accuracies for one- and two-electron ions is asymmetric, with few accurate  measurements in the range $10 \leq Z \leq 36$ (accuracies lower than \SI{20}{ppm}) and almost none above. In that sense, the  $1s^2 2p\to1s^2 2s$ transitions in three-electron ions represent the more coherent set of available data for few-electron ions, even though the theory is obviously more difficult to handle. There is thus a clear need for both improved measurements with ion sources whenever possible, performed with reference-free methods. Such measurements are mostly immune to the difficulties connected to the Doppler effect determination on storage rings. One can hope to see improved EBIT with higher intensity currents and higher energies for the electron beam. There is exploratory work to be performed on the x-ray emission of plasmas with the new generation of high-frequency, supraconducting ECRIS \cite{zsgz2018}. The FAIR facility \cite{gsl2015,dijk2019}, some sections of which have become operational recently, in particular the CRYRING storage ring \cite{laab2016}, will provide new possibilities to perform more accurate measurements on highly charged ions.
The installation of CRYRING, which  allows to decelerate highly charged ions to much lower energies than possible in the ESR, together with improved measurements of the ions' energy using voltage measurement devices calibrated by the PTB, which have already been used in the latest generation of HFS measurements, will undoubtedly allow for more accurate measurements. The development of methods like laser spectroscopy on fast ions in the FAIR synchrotron SIS, using the Doppler effect to shift the laser energy to several \si{\kilo\electronvolt}, or like RCE on higher-energy beams, will provide new ways to measure transition energies in lithiumlike ions in particular. 

The HITRAP beam line \cite{hbek2006,abbb2009,habc2015}, which allows to decelerate fast ion beams from the ESR, will also bring new possibilities. In this system, one can cool and trap ions in two different trap systems, one oriented towards   measurements using the continuous Stern-Gerlach method to measure heavy ion masses and Landé factors, and the other one towards laser spectroscopy of ultra-cold highly charged ions. HITRAP can be used, for example, for measuring the mass of successive charge states differing by one electron, including bare ions, permitting a direct measurement of binding energies. Measurements with accuracies approaching \num{1E-11} are already in progress using the PENTATRAP high-precision traps \cite{sdrb2017,dbbe2018}. This will provide mass differences with a few \si{electronvolt} accuracy, compatible with the best available measurements by photon spectroscopy.  

The analysis of the theoretical contributions to hydrogenlike ion transition energies shows that at some point the contribution from the nucleus, through the finite nuclear size correction, nuclear deformation and nuclear polarization, is going to be the limitation to our understanding of strong field QED. This is also true for hyperfine structure measurements and to a lesser extent for $g$-factor measurements. There is thus a clear need in the long run to improve our understanding of nuclear structure and its interaction with the ion's electrons. 
This will probably require combining measurements performed with muonic atoms like projected in the \emph{muX} collaboration \cite{kir2017} at the Paul Scherrer institute with those of electronic atoms. The use of pionic or antiprotonic atoms \cite{tjlh2001,grfi2008} could  complement our understanding of the nuclear structure.

Besides bringing a better comprehension of QED, one of the four fundamental interactions in physics,  in strong electrostatic and magnetic fields, this field of research is also of major interest for other areas of fundamental physics. Obtaining improved accuracy on the electron mass is a major issue in the determination of fundamental constants. Search for new physics with high precision measurements in atoms and ions is also an important issue \cite{ksls2018,sbdk2018}. The determination of the fine structure constant by comparing three independent methods based on different sectors of physics is a very promising direction of research, which has already set strong constraints on new physics. One can combine  the use of one and three electron ions'  Landé $g$-factor, which uses strong field QED, with the free electron anomalous magnetic moment measurements \cite{hfg2008,hhg2011}, and calculations of higher-order QED \cite{akn2018}, including weak and strong interaction contributions, with direct measurements with cold atom interferometry which do not depend on QED \cite{bcgn2011,bcgn2013,pyze2018}. 

Better knowledge of transition energies in few-electron atoms can also help confirm or inform observation of unknown effects that could be linked to the identification of dark matter or new particles. For example, there has been discussion about the possible discovery of dark matter through the observation of x-ray spectra with energies around \SI{3.5}{\keV}  by the XMM-Newton ( X-ray Multi-Mirror-Newton) space x-ray telescope \cite{bmfs2014,brif2014}. But spectroscopy performed at the Livermore EBIT has allowed to attribute these spectra to a set of lines from highly-charged sulfur ions  \cite{sdbs2016}. Later observation by the high-resolution x-ray spectrometer of the HITOMI satellite  did not confirm the original observations \cite{aaaa2017}. In the same way, there are now several propositions to check for the possible existence of a \emph{new light neutral boson}, which may explain a signal observed in the nuclear spectroscopy of $^{8}$Be \cite{kccg2016}. Several works propose to study the effect of this boson on muonic atoms \cite{jan2018} or in the isotopic shift \cite{ffps2017,dffs2017,bbdf2108}. 

\ack
Laboratoire Kastler Brossel is Unit{\'e} Mixte de Recherche du CNRS n$^{\circ}$ 8552. All Feynman diagrams have been drawn using Jaxodraw 2.1 \cite{bckt2009}.
P. Indelicato is a member of the Allianz Program of the Helmholtz Association, contract n$^{\circ}$ EMMI HA-216 ``Extremes of Density and Temperature: Cosmic Matter in the Laboratory'' and of the Stored Particle Collaboration (SPARC).
Part of this work was prepared in the framework of the Extreme Matter Institute \emph{Task force} on the $1s$ Lamb shift that convened in Jena in September 2018.
I wish to thank Dr. P.J. Mohr  for years of collaboration and support for several stays at NIST in which part of this work has been completed. I also thank Dr. J.P. Desclaux for many years of fruitful discussions and work on the subjects presented here.
I wish also to thank Prof. J.P. Santos for his kind hospitality and helpful discussion during many stays at the New University of Lisbon. Part of this work has been performed during stays financed by the \emph{Programme Hubert Curien} PESSOA 38028UD and program PAUILF 2017-C08.
I am also very grateful to Dr. Nancy Paul for  her very careful reading of the manuscript and many very useful suggestions and ameliorations.

\section*{References}
\bibliography{ref2019.bib}

\end{document}